\documentclass[12pt]{article}

\usepackage{fancyhdr,titlesec,geometry,ragged2e}
\usepackage[dvipsnames]{xcolor}
\usepackage[many]{tcolorbox}
\usepackage{layout}
\usepackage{lipsum}
\usepackage{hyperref}
\hypersetup{
  colorlinks,
  citecolor=red,
  linkcolor=blue,
  linktoc=all
}
\usepackage[T1]{fontenc}                            % Font Styling
\usepackage{lmodern,mathrsfs}

\usepackage[shortlabels]{enumitem}
\usepackage{mathtools,amssymb,amsfonts,amsthm,bm,mathrsfs}   % Math Presets
\usepackage{array,tabularx,booktabs}                % Table Presets
\usepackage{graphicx,wrapfig,float,caption,epsfig}         % Figure Presets
\usepackage{setspace,multicol}                      % Text Presets
\usepackage{tikz,tikz-cd,physics,cancel}            % Physics Presets

\usepackage{esint}
\allowdisplaybreaks

\usepackage{parskip}
\DeclareGraphicsExtensions{.pdf,.png,.jpg}
\usepackage{geometry}
\renewcommand{\thefootnote}{\arabic{footnote}}

\titleformat{\section}[hang]{\normalfont\large\bfseries}{\thesection.}{3mm}{}
\titlespacing{\section}{0mm}{15mm}{4mm}

\titleformat{\subsection}[hang]{\normalfont\bfseries}{\thesubsection}{2mm}{}
\titlespacing{\subsection}{0mm}{10mm}{4mm}

\titleformat{\subsubsection}[hang]{\normalfont\bfseries}{\thesubsubsection}{5mm}{}
\titlespacing{\subsubsection}{0mm}{5mm}{1mm}

\newtheoremstyle{thmstyle}
{5pt} % Space above
{3pt} % Space below
{\itshape} % Body font
{} % Indent amount
{\bfseries} % Theorem head font
{.} % Punctuation after theorem head
{0.5em} % Space after theorem head
{} % Theorem head spec (can be left empty, meaning `normal')

\theoremstyle{remark}{

}

\theoremstyle{thmstyle}{
\newtheorem{definition}{Definition}[section]
\newtheorem{theorem}{Theorem}[section]
\newtheorem{proposition}[theorem]{Proposition}
\newtheorem{lemma}[theorem]{Lemma}
\newtheorem{thmcorollary}{Corollary}[theorem]

}

\numberwithin{equation}{section}
\renewcommand{\theequation}{\thesection.\arabic{equation}}
\newlength{\extraspace}
\newlength{\extraspaces}
\newcommand{\be}{\begin{equation}
\addtolength{\abovedisplayskip}{\extraspaces}
\addtolength{\belowdisplayskip}{\extraspaces}
\addtolength{\abovedisplayshortskip}{\extraspace}
\addtolength{\belowdisplayshortskip}{\extraspace}}
\newcommand{\ee}{\end{equation}}
\newcommand{\ba}{\begin{eqnarray}
\addtolength{\abovedisplayskip}{\extraspaces}
\addtolength{\belowdisplayskip}{\extraspaces}
\addtolength{\abovedisplayshortskip}{\extraspace}
\addtolength{\belowdisplayshortskip}{\extraspace}}
\newcommand{\ea}{\end{eqnarray}}
\newcommand{\bas}{\begin{eqnarray*}
\addtolength{\abovedisplayskip}{\extraspaces}
\addtolength{\belowdisplayskip}{\extraspaces}
\addtolength{\abovedisplayshortskip}{\extraspace}
\addtolength{\belowdisplayshortskip}{\extraspace}}
\newcommand{\eas}{\end{eqnarray*}}

\newcounter{subequation}[equation]
\makeatletter

\expandafter\let\expandafter
\reset@font\csname reset@font\endcsname

\def\subeqnarray{\arraycolsep1pt
    \def\@eqnnum\stepcounter##1{\stepcounter{subequation}%
        {\reset@font\rm(\theequation\alph{subequation})}}
\jot5mm     \eqnarray}

\def\subarray{\arraycolsep1pt
    \def\@eqnnum\stepcounter##1{\stepcounter{subequation}%
        {\reset@font\rm(\alph{subequation})}}
\jot5mm     \eqnarray}

\makeatother

\newcommand{\pp}{{\partial}}
\renewcommand{\dd}{{\partial}}

\newcommand{\ssum}[2]{\sum_{#1}^{#2}}
\newcommand{\exx}[1]{e^{#1}}

\renewcommand{\bra}{\langle}
\renewcommand{\ket}{\rangle}
\newcommand{\ra}{\rightarrow}

\newcommand{\nspace}{\!\!\!\!\!\!\!\!\!\!}

\newcommand{\nonum}{\nonumber \\[1.5mm]}
\newcommand{\is }{&\!\!=\!\!&} % for eqnarray
\newcommand{\isa}{ & = } % for align

\renewcommand{\ip}[2]{\left \langle #1\,\middle|\, #2 \right \rangle}

\DeclarePairedDelimiterX\set[1]\lbrace\rbrace{\def\given{\;\delimsize\vert\;}#1}

\newcommand{\inv}{^{-1}}

\newcommand{\1}{\mbox{1\hspace{-.8ex}1}}

\newcommand{\nrm}[1]{\norm{#1}}
\newcommand{\nrmm}{\norm{\cdot}}

\newcommand{\lb}{\lambda}

\newcommand{\Om}{\Omega}

\newcommand{\vp}{\varphi}

\newcommand{\eps}{\epsilon}
\newcommand{\munu}{{\mu\nu}}
\renewcommand{\th}{\theta}

\newcommand{\mcB}{{\mathcal{B}}}

\newcommand{\mcD}{{\mathcal{D}}}
\newcommand{\mcE}{{\mathcal{E}}}

\newcommand{\mcH}{{\mathcal{H}}}

\newcommand{\mcL}{{\mathcal{L}}}

\newcommand{\mcO}{{\mathcal{O}}}

\newcommand{\mcS}{{\mathcal{S}}}

\newcommand{\mcV}{{\mathcal{V}}}

\newcommand{\cD}{{\cal D}}
\newcommand{\cE}{{\cal E}}

\newcommand{\bbC}{{\mathbb{C}}}
\newcommand{\C}{\mathbb{C}}

\newcommand{\bbF}{{\mathbb{F}}}

\newcommand{\bbN}{{\mathbb{N}}}
\newcommand{\N}{\mathbb{N}}

\newcommand{\bbR}{{\mathbb{R}}}
\newcommand{\R}{\mathbb{R}}

\newcommand{\mff}{{\mathfrak{f}}}

\newcommand{\mfh}{{\mathfrak{h}}}

\newcommand{\mfu}{{\mathfrak{u}}}

\newcommand{\mfB}{{\mathfrak{B}}}

\newcommand{\mfD}{{\mathfrak{D}}}

\newcommand{\mfH}{{\mathfrak{H}}}

\newcommand{\wg}[0]{{\rm g}}

\def\fnum@figure{\textbf{\figurename\nobreakspace\thefigure}}
\def\fnum@table{\textbf{\tablename\nobreakspace\thetable}}

\newcommand{\vtest}{%
  \mathord{% ensure math mode and grouping
    \renewcommand{\arraystretch}{0}%
    \begin{array}[t]{@{}c@{}l@{}}
      \\[-11pt]
    \rightharpoonup\\[-1pt]
    \mfD
    \end{array} 
    \kern\scriptspace
  }%
}

\newcommand{\cvtest}{%
  \mathord{% ensure math mode and grouping
    \renewcommand{\arraystretch}{0}%
    \begin{array}[t]{@{}c@{}l@{}}
    \mfD\\[0.8pt]
    \rightharpoondown
    \end{array} 
    \kern\scriptspace
  }%
}
\newcommand{\vlp}[1]{%
  \mathord{% ensure math mode and grouping
    \renewcommand{\arraystretch}{0}%
    \begin{array}[t]{@{}c@{}l@{}}
    \\[-10.8pt]
    \rightharpoonup\\[-2.1pt]
    L^{#1} & 
    \end{array} 
    \kern\scriptspace
  }%
}

\newcommand{\cvlp}[1]{%
  \mathord{% ensure math mode and grouping
    \renewcommand{\arraystretch}{0}%
    \begin{array}[t]{@{}c@{}l@{}}
    L^{#1}\\[-0.1pt]
    \rightharpoondown
    \end{array} 
    \kern\scriptspace
  }
}

\newcommand{\dth}[0]{\Delta_\theta}

\begin{document}

\begin{titlepage}

%footnotesymbols others than numbers
\renewcommand{\thefootnote}{\arabic{footnote}}
\makebox[1cm]{}
\vspace{2mm}

% The title
\begin{center}
\hypersetup{linkcolor=black}
\mbox{{\large \bf Analytic semigroups approaching a Schr\"{o}dinger group}}
%\\[4mm]
%\mbox{{\large \bf a Schr\"{o}dinger group}}
\\[4mm]
\mbox{{\large \bf on real foliated metric manifolds}} 
\vspace{2.3cm}

{\sc Rudrajit Banerjee}\footnote[1]{email:
\ttfamily\href{mailto:rudrajit.banerjee@oist.jp}{\textcolor{black}{rudrajit.banerjee@oist.jp}}} 
{\sc and} 
{\sc Max Niedermaier}\footnote[2]{email:
\ttfamily\href{mailto:mnie@pitt.edu}{\textcolor{black}{mnie@pitt.edu}}}
\\[8mm]
{\small\sl $^1$Okinawa Institute of Science and Technology Graduate University,}\\
{\small\sl 1919-1, Tancha, Onna, Kunigami District}\\ 
{\small \sl Okinawa 904-0495, Japan}
\\[5mm]
{\small\sl $^2$Department of Physics and Astronomy}\\
{\small\sl University of Pittsburgh, 100 Allen Hall}\\
{\small\sl Pittsburgh, PA 15260, USA}
\vspace{16mm}

{\bf Abstract} \\[4mm]
\begin{quote}
On real metric manifolds admitting a co-dimension one foliation, sectorial operators are introduced that interpolate between the generalized Laplacian and the d'Alembertian. This is used to construct a one-parameter family of analytic semigroups that remains well-defined into the near Lorentzian regime. In the strict Lorentzian limit we identify a sense in which a well-defined Schr\"{o}dinger evolution group arises. For the analytic semigroups we show in addition that: (i) they act as integral operators with kernels that are jointly smooth in the semigroup time and both spacetime arguments. (ii) the diagonal of the kernels admits an asymptotic expansion in (shifted) powers of the semigroup time whose coefficients are the Seeley-DeWitt coefficients evaluated on the complex metrics.
\end{quote} 
\end{center}

\vfill
\setcounter{footnote}{0}
\end{titlepage}
%%%%%%%%%%%%%%%%%%%%%%%%%%%%%%%%%%%%%%%   
% ----------------------------------------------------------------------
%		Table of contents page
% ----------------------------------------------------------------------
\thispagestyle{empty}
\makebox[1cm]{}

\vspace{-23mm}
\begin{samepage}
{\hypersetup{linkcolor=black}
\tableofcontents}
\end{samepage}

\nopagebreak

\newpage
%%%%%%%%%%%%%%%%%%%%%%%%%%%%%%%%%%%%%%%%%%%%%%%%%%%%%%%%%%%%%%%%

%%%%%%%%%%%%%%%%%%%
%	New section 
%%%%%%%%%%%%%%%%%%%

\setcounter{equation}{0}

%%%%%%%%%%%%%%%%%%%%%%%%%%%%%%%%%%%%%%%%%%%%%%%%%%%%%%%%%%%%%%%%%%%%%

\section{Introduction} 

%%%%%%%%%%%%%%%%%%%%%%%%%%%%%%%%%%%%%%%%%%%%%%%%%%%%%%%%%%%%%%%%%%%%%

Let $M$ be a real, smooth, $(1\!+\!d)$-dimensional manifold without boundary
admitting a co-dimension one foliation, which is equipped with a
Euclidean signature metric $g^+$ and a Lorentzian signature metric $g^-$
{of the form (\ref{i1a}) below}. For a non-negative, smooth,
bounded potential $V$, we denote by $\cD_+ = - \nabla_{+}^2 + V$ the
generalized Laplacian associated with $g^+$ and by $\cD_- =
- \nabla_{-}^2 + V$ the generalized d'Alembertian associated with $g^-$.
Here $\nabla_\pm^2$ are the Laplace-Beltrami operators associated to
$g^\pm$. In terms of these data we define the one-parameter family of
interpolating operators  
\be
\label{i0} 
\Delta_{\th} = - \sin \th\, \cD_+ - i \cos \th \, \cD_-, \quad
\th \in (0,\pi)\,,
\ee
initially on $C_c^{\infty}(M)$, where in general $[\cD_+, \cD_-] \neq 0$.
The analysis of this family and associated constructions will be the main
focus of our paper.

We briefly elaborate on the geometrical setting.
%The metric tensors
%$g^\pm$ cannot be prescribed arbitrarily and are related as follows.
The co-dimension one foliation structure may be made manifest by a
diffeomorphism $\bbR\times \Sigma \xrightarrow{} M$; neither $M$ nor
$\Sigma$ are required  to be closed or compact. Then,
  in any local coordinate chart $\bbR\times U\ni (t,q)\mapsto
  (t,x^a)\in \bbR\times \bbR^d$ with $U\subset \Sigma$ open,
  the metrics $g^\pm$  are assumed to be {smooth} and of the 
  Arnowitt-Deser-Misner (ADM) forms 
\begin{equation} 
\label{i1a} 
g^{\pm} =
\pm  N^2 dt^2 + \wg_{ab}(dx^a + N^a dt) (dx^b + N^b dt),
\end{equation}
where $N,N^a, \wg_{ab}$  are the {\it{shared}} ADM component fields
(we refer to~\cite{Gbook} for details on the ADM formulation). 
These metrics and the complex combination in  \eqref{i0} arise
naturally when a generic Lorentzian metric of ADM form is subjected
to a phase rotation in the lapse field, $N \mapsto e^{-i \th} N$, while
all coordinate atlases remain real.
In the above local coordinates $(t,x^a)$, $a=1,\ldots,d$,
the complex metrics arising thereby are of the form
\begin{equation}
\label{i1} 
g^{\th} =
- e^{ -2 i \th} N^2 dt^2 + \wg_{ab}(dx^a + N^a dt) (dx^b + N^b dt).
\end{equation}
The operators (\ref{i0})
can also be viewed as a generalized Laplacian associated with the
complex metrics $g^{\th}$. The operators $\Delta_{\th}$  are
manifestly covariant under diffeomorphisms that preserve the
initial foliation and can be shown to be covariant also under
diffeomorphism that change the foliation (i.e.~mix the component
fields $N,N^a,\wg_{ab})$ while preserving the ADM form of the metric \cite{wickfoli}.
For short, we shall refer to the resulting tensor as the
  {\it lapse-Wick-rotated metric.}
Moreover, the complex metrics (\ref{i1}) are admissible
in the sense of \cite{KSWick,VisserWick2} in the initial as well as
any other foliation and can be viewed as a rank one
deformation of the given Lorentzian metric akin to \cite{Candelas},
but with a metric-dependent co-vector field. For coherence of 
exposition we relegate these differential geometric aspects to a
separate note \cite{wickfoli}.

For $\th =\pi/2$ the operator (\ref{i0}) is minus the standard  {elliptic}
Laplacian plus potential, $\mcD_+\geq 0$, which naturally extends to the
self-adjoint Dirichlet Laplacian on $L^2(M)$ {(denoted momentarily
by the same symbol)}. As such, it generates
the familiar heat semigroup $\R_+ \ni s \mapsto
e^{-s \mcD_+}$, whose existence is typically ensured using the spectral
resolution of the Dirichlet Laplacian. The heat semigroup is
well-known to be smoothening and to act as an integral operator with
a jointly smooth kernel (the heat kernel), see
e.g.~\cite{Daviesheatkbook,Grigorbook,Getzlerbook,Grigor1}. Further, the
heat kernel has an   asymptotic expansion for small semigroup time $s$,
which is all-important in applications
\cite{Avramidibook,Morettiheatk,Getzlerbook,Parkerbook}.

For $\theta=0$ on the other hand,  (\ref{i0}) reduces to $-i\mcD_-$,
where $\mcD_-$ is the d'Alembertian plus potential. The interplay between
the (essential) self-adjointness of $\mcD_-$ and the properties of the
underlying Lorentzian manifold is much more subtle than in the  Riemannian
case \cite{Nonselfad,taira,taira2,vasyesa}. Once self-adjointness has
been achieved, the existence of an associated Schr\"{o}dinger group
$\bbR\ni s\mapsto \exx{-is\mcD_-}$ follows by Stone's Theorem. The
transition from $\exx{-s\mcD_+}$ to $\exx{-is\mcD_-}$ involves shifting
from a heat equation to a Schr\"{o}dinger equation, and from the elliptic
$\mcD_+$ to the hyperbolic $\mcD_-$. Even for $\mcD_+$, the heat and
Schr\"{o}dinger (semi)groups  have markedly different behavior. In
particular, the existence of a non-distributional kernel or its  asymptotic
small time expansion is not automatic \cite{FullingGreens,taira3}.
The counterpart of a Schr\"{o}dinger group and kernel for a hyperbolic
$\mcD_-$ does not appear to have been studied in the mathematics literature.

A key motivation for our work is the implicit desideratum in the
  theoretical physics literature \cite{Parkerbook, heatkoffdiag1} that the
  properties of the  Schr\"{o}dinger kernel for $\mcD_-$ may be obtained
  from the heat kernel expansion of $\mcD_+$ by ad-hoc substitutions
  ($s\mapsto i \tilde{s},\,\tilde{s} \in \R$, and $g^+\mapsto g^-$). The
  resulting formal series are an important computational tool in quantum
  field theory on curved spacetime and not without rationale; c.f.~the
  remarks at the end of Section \ref{Sec3_3}.
Clearly, it is desirable to replace aspects of this by a {valid}
mathematical construction.

The interpolating operator \eqref{i0} is not symmetric for
$\theta\neq 0,\pi/2,\pi$, and is not readily amenable to spectral
analysis  as a tool to establish the existence of an associated semigroup.
The first part of our results, nevertheless, ensures the existence of
the semigroup $\zeta \mapsto e^{\zeta \Delta_{\th}}$ associated with
$\Delta_{\th}$ for $\theta\in (0,\pi)$, with properties mirroring to
a large extent those in the Riemannian setting.
For convenient orientation we summarize these results here in
a slightly cursory form. The precise definitions and conditions can
be found in the corresponding theorems (Theorems \ref{thsec1},
\ref{thsmooth1}, \ref{thsmooth2}).  
\begin{itemize}
\item[(a)] $\Delta_{\th}$ is an unbounded operator on a
  Sobolev domain dense in $L^2(M)$. The same holds for its adjoint
  $\Delta_{\th}^*$, and $\Delta_{\th}^* = \Delta_{\pi - \th}$, including
  domains.  
\item[(b)] The spectrum of $\Delta_{\th}$ is contained in a wedge
  of the left half plane, $|{\rm Arg} \lb| \geq \pi/2 + \tilde{\th}$, with $\tilde{\th}:=\min\set{\theta,\pi-\theta}$.  
\item[(c)] The resolvent $[z - \Delta_{\th}]^{-1}\!$,
 $| {\rm Arg}(z)| < \pi/2 \!+ \!\tilde{\th}$,
   obeys norm bounds that qualify $\Delta_{\th}$ as the generator of
  an analytic semigroup, $\zeta \mapsto e^{\zeta \Delta_{\th}}$, with
  $|{\rm Arg} \zeta | < \tilde{\th}$. 
\item[(d)] $\zeta \mapsto e^{ \zeta \Delta_{\th}}$   
  is a bounded analytic semigroup on $L^2$, which is uniquely determined
  by $\Delta_{\th}$ (including domain)  
  and contractive, $\Vert e^{ \zeta \Delta_{\th}} \Vert \leq 1$,
  $|{\rm Arg}(\zeta)| < \tilde{\th}$ ($\zeta > 0$ in particular).   
\item[(e)]  The operator $e^{\zeta \Delta_{\th}}$, $|{\rm Arg}(\zeta) |
  < \tilde{\th}$,
  acts as an integral operator on $L^2$
  functions with a kernel $K_{\zeta}^{\th}(t,x;t',x')$ that is jointly smooth
  in $(\zeta,t,x,t',x')$, and obeys $K_{\zeta}^{\th}(t,x;t',x')^* =
  K_{\zeta^*}^{\pi - \th}(t',x';t,x)$. 
\end{itemize} 
The second part of our paper addresses aspects not covered by analytic
semigroup theory.

As mentioned, for computational applications, the asymptotic expansions
for small heat time are crucial. With this in mind, we study the existence
of such an asymptotic expansion for our generalized kernel
$K_{\zeta}^{\th}(t,x;t',x')$. The widely used proof strategies use a
parametrix ansatz invoking the Synge function associated to a real
metric tensor (i.e.~one half of the geodesic distance between nearby points).
The constructions of the Synge function are through the solution of
a Hamilton-Jacobi system (the geodesic equation) on the manifold, see
e.g.~\cite{Synge1,Synge2,Avramidibook}. In the present context such
constructions would require a complex manifold to go hand-in-hand with
the complex metric \eqref{i1}. This route, however, comes with its own
limitations and we seek to retain our default real manifold setting.

To sidestep this hurdle we utilize a double expansion
in coordinate differences $(t\!-\!t', x\!- \!x')$ and shifted
powers of $\zeta$ to arrive at a result
for at least the diagonal of the kernel: the diagonal kernel
admits an asymptotic expansion of the form
\be
\label{i3} 
K_{\zeta}^{\th}(t,x;\!t,x) \asymp \frac{ (-i e^{ i\th})^{\frac{d-1}{2}}}%
{( 4 \pi \zeta)^{\frac{d+1}{2}}} 
\sum_{n \geq 0} A_n^{\th}(t,x) ( i e^{- i \th} \zeta)^n,
\ee 
where $A_n^{\th}$ are the standard heat kernel coefficients evaluated
on $g^{\th}$.

Finally, we examine the strict Lorentzian limit, $\th \ra 0^+$, of
our semigroups. Assuming that the generalized d'Alembertian $\cD_-$
is essentially selfadjoint on $C_c^{\infty}(M)$ (and hence
has a unique self-adjoint closure $\overline{\cD}_-$ generating a
Schr\"{o}dinger evolution group $e^{- i s \overline{\cD}_-},
s \in \R)$ we show: for any trace-class operator $T$ on $L^2(M)$  
\ba
\label{i4} 
{\rm tr}[ T e^{s \Delta_{\th}}] \;\xrightarrow{\th \ra 0^+}\;
{\rm tr}[ T e^{ -i s \overline{\cD}_{-}}]\,,
\quad
{\rm tr}[ T e^{s \Delta_{\pi - \th}}] \;\xrightarrow{\th \ra 0^+}\;
{\rm tr}[ T e^{ + i s \overline{\cD}_{-}}]\,,
\ea
for all $s \geq 0$. That is, convergence holds in the weak-star
topology on the bounded operators on $L^2(M)$. Since $\Delta_{\pi -\th}$ is the adjoint of $\Delta_{\th}$ the
second half of the statement accounts for the extension of a
semigroup to a group in $s$. 

The paper is organized as follows. In Section \ref{Sec3_1} we
collect some background material on analytic semigroups and their
generators, after which Section \ref{Sec3_2}  addresses items (a)--(d)
in the above list. The existence of a smooth kernel is proven in
Section \ref{Sec3_3} while Section \ref{Sec5} establishes its asymptotic
expansion on the diagonal. Finally, the strict Lorentzian limit is studied
in Section \ref{Sec6}. For convenient reference we
detail the functional analytical arena in Appendix \ref{App_dist}:
test functions, distributions and Sobolev spaces.
Appendix \ref{apploc} provides a concise summary of a
local regularity theory adapted to our complex metrics, as needed for
Section \ref{Sec3_3}.

%%%%%%%%%%%%%%%%%%%%%%%%%%%%%%%%%%%%%%%%%
%new section
%%%%%%%%%%%%%%%%%%%%%%%%%%%%%%%%%%%%%%%%%

\newpage 
\section{The analytic semi-group generated by $\mathbf{\Delta_{\th}}$}
\label{Sec3} 

The goal of this section is to establish $\Delta_{\th}$ as the generator
of an  {\itshape analytic semigroup} via suitable resolvent estimates.
In the introductory tabulation this corresponds to the results (a)--(d).

%%%%%%%%%%%%%%%%%%%%%%%%%%%%%%%%%%%%%%%%%%%%%%%%%%%%%%%%%%%%%%%%%%%%%%%%%%%%%%%%%%%%%%%

\subsection{Background: analytic semigroups}
\label{Sec3_1}

For readability's sake we include a brief summary of the definitions and 
results needed, see e.g. \cite{Semigroupbook} for a detailed exposition. 

\begin{definition}[Bounded analytic semigroup]\label{asgroupdef}	
Let $\mfH$ be a complex Hilbert space. Given $\alpha\in (0,\pi]$, the ``sector'' $\Sigma_\alpha\subseteq \bbC$ is 
defined by\footnote{``Arg'' is  defined in $(-\pi,\pi)$.}
\begin{align}\label{sec3}
	\Sigma_\alpha:=\big\{\zeta\in \bbC\setminus\{0\}\,\big|\,|{\rm Arg}(\zeta)|<\alpha\big\}\,.
\end{align} 
Then, a family of bounded operators $(T(\zeta))_{\zeta\in \Sigma_\delta\cup \{0\}}$ in $\mfB(\mfH)$ (i.e.~the space of bounded endomorphisms of $\mfH$) is called a {\itshape bounded analytic semigroup} of angle $\delta\in (0,\pi/2]$ if:
\begin{enumerate}[leftmargin=8mm, rightmargin=0mm, label=(\roman*)]
  \item $T(0)=\1$ and $T(\zeta_1+\zeta_2)=T(\zeta_1)T(\zeta_2)$ for all $\zeta_1,\,\zeta_2\in \Sigma_\delta$.
  \item The mapping $\zeta\mapsto T(\zeta)$ is analytic in $\Sigma_\delta$, i.e.~there is a powerseries expansion about every $\zeta\in \Sigma_\delta$ that converges in operator norm with nonzero radius of convergence.
  \item For all $\psi\in \mfH$ and $0<\delta'<\delta$ we have strong continuity, i.e. $\lim_{\Sigma_{\delta'}\ni \zeta\to 0}T(\zeta)\psi =\psi$.
  \item For every $0<\delta'<\delta$, $\norm{T(\zeta)}_{\rm op}$ is uniformly bounded in $\Sigma_{\delta'}$.
\end{enumerate}
\end{definition}

{\bfseries Remarks.}

(i) Restricting the domain of the bounded analytic semigroup to
$\bbR_{\geq 0}\subseteq \Sigma_{\delta}\cup \{0\}$ yields a strongly
continuous semigroup $(T(s))_{s\geq0}$.

(ii) A closed densely defined operator $A:D(A)\to \mfH$,
$D(A)\subseteq \mfH$,  is the {\itshape generator} of the analytic
semigroup $(T(\zeta))_{\zeta\in \Sigma_\delta\cup \{0\}}$ {\itshape iff} 
\begin{align}\label{sec4}
\forall\,\psi\in D(A):\quad \lim_{s\to 0^+}\frac{1}{s}(T(s)\psi-\psi)=A\psi\,,
\end{align}
i.e.~the difference quotient converges in norm (as $\R_+ \ni s\to 0^+$)
to $A\psi$ for all $\psi\in D(A)$. The defining property  of an analytic
semigroup's generator is its {\itshape sectoriality}.
\medskip

\begin{definition}
\label{secdef}
A densely defined closed operator $A:D(A)\to \mfH$ is called sectorial of
angle $\delta\in (0,\pi/2]$ if:
\begin{enumerate}[leftmargin=8mm, rightmargin=0mm, label=(\roman*)]
  \item  $\Sigma_{\pi/2+\delta}\subseteq \rho(A)$, where $\rho(A):=\big\{\lb \in \bbC\,\big|\,\lb -A \text{ is bijective and }(\lb-A)\inv \text{ is bounded}\,\big\}$ is the resolvent set of $A$.
\item The resolvents are uniformly bounded over sectors, i.e. for every $\varepsilon\in (0,\delta)$ there exists  $M_{\varepsilon}\geq 1$ such that 
\begin{align}
  \label{sec5}
  \norm{(\lb-A)\inv}_{\rm op}\leq \frac{M_\varepsilon}{|\lb|}\,,\quad \lb \in \overline{\Sigma_{\pi/2+\delta-\varepsilon}}\setminus\{0\}\,.
\end{align}
\end{enumerate}
\end{definition}
The exponentiation of a sectorial operator (which need not be normal,
c.f.~the operator $\dth$ in \eqref{Adecomp} for generic $\theta$) may
be defined via the holomorphic functional calculus.
\begin{theorem}\
\label{prpsec1}
Let $(A,D(A))$ be a sectorial operator of angle $\delta\in (0,\pi/2]$ in a Hilbert space $\mfH$. Consider the following family of operators in $\mfB(\mfH)$,  
\begin{align}\label{sec6}
T(0):=\1,\quad 	T(\zeta):=\frac{1}{2\pi i}\int_{\gamma}\!\exx{\zeta\lb}(\lb-A)\inv d\lb\,,\quad \zeta\in \Sigma_\delta\,,
\end{align}
where, given $\zeta\in \Sigma_\delta$, $\gamma$ is any piecewise smooth curve in $\Sigma_{\pi/2+\delta}$ going from $\infty \exx{-i(\pi/2+\delta')}$ to $\infty \exx{i(\pi/2+\delta')}$ for some $|{\rm Arg}(\zeta)|<\delta'<\delta$. 

Then the family $(T(\zeta))_{\zeta\in \Sigma_{\delta}\cup\set{0}}$  is a bounded analytic semigroup with generator $(A,D(A))$, and is furthermore uniquely determined  by the generator. 

If, in addition, there is $\varphi\in (0,\delta)$ such that there is the uniform resolvent bound
\begin{align}\label{sec6a}
	\forall\,\lb\in \Sigma_{\varphi}:\quad \nrm{(\lb-A)\inv}_{\rm op}\leq \frac{1}{|\lb|}\,,
\end{align}
then $T(\zeta)$ is contractive in  $\Sigma_{\varphi}\cup\set{0}$, i.e. $\forall\,\zeta\in \Sigma_{\varphi}\cup\set{0}:\,\nrm{T(\zeta)}_{\rm op}\leq 1$.
\end{theorem}
The existence of the analytic semigroup and the contractivity result is a straightforward  combination of the Lumer-Phillips Theorem and the generation theorem for analytic semigroups (resp.~Thms.~3.15 and 4.6 in Ch.~II of \cite{Semigroupbook}), while the uniqueness follows from the analyticity of the mapping $\Sigma_\delta\ni \zeta \to T(z)\in \mfB(\mfH)$, together with the fact that the strongly continuous semigroup $(T(s))_{s\geq 0}$ obtained by restricting to $[0,\infty)$ is uniquely determined by its generator $(A,D(A))$ (c.f.~Thm.~1.4  in Ch.~II of \cite{Semigroupbook}). We omit the proof, and instead note the following.

{\bfseries Remarks.}

(i) The use of the Cauchy integral formula \eqref{sec6}, and the fact that the
semigroup is uniquely determined by its generator, motivates the intuitive
notation $\exx{\zeta A}\equiv T(\zeta)$. For brevity we shall refer to the
semigroup $T(\zeta)$ as the {\it analytic semigroup generated by $A$}
and often write $\exx{\zeta A}$ for it.  

(ii) The map  $\Sigma_{\delta}\ni \zeta\mapsto T(\zeta)\in \mfB(H)$ is analytic, and the  integral formula \eqref{sec6}, together with the closedness of $(A,D(A))$, implies   for all $\zeta\in \Sigma_{\delta}$
\begin{align}\label{sec6c}
	\frac{d^n}{d\zeta^n} T(\zeta)\psi=A^nT(\zeta)\psi\,,\quad n\in \bbN_0,\,\psi\in \mfH\,.
\end{align}
In particular, this entails boundedness of the $A^nT(\zeta)$, and that
$\text{ran}(T(\zeta))\subseteq \cap_{n=1}^\infty D(A^n)$.
 
(iii) Moreover,  a straightforward adaptation of the arguments in \cite{Semigroupbook}  shows that for all $\delta'\in (0,\delta)$ there is a $C_{\delta'}>0$ such that 
\begin{align}
  \forall\, \zeta\in\Sigma_{\delta'}:\quad \nrm{AT(\zeta)}_\text{op}\leq
  \frac{C_{\delta'}}{|\zeta|}\,,
\end{align}
Observing that $A^n T(\zeta)=(AT(\zeta/n))^n$, it follows that for all
$n\in \bbN$,
\begin{align}\label{sec6c2}
\forall\, \zeta\in\Sigma_{\delta'}:\quad	\nrm{A^n T(\zeta)}_\text{op}\leq \Big(\frac{nC_{\delta'}}{|\zeta|}\Big)^n\,,
\end{align}
which will be needed later on.
Finally, for any $\zeta_0\in \Sigma_\delta$, the (convergent) Taylor expansion $T(\zeta)=\ssum{k=0}{\infty}\frac{(\zeta-\zeta_0)^k}{k!} T^{(k)}(\zeta_0)$, together with \eqref{sec6c} implies that for any $\psi\in \mfH$
\begin{align}\label{sec6d}
  A^n T(\zeta)\psi=\ssum{k=0}{\infty}\frac{(\zeta-\zeta_0)^k}{k!}
  A^{n+k}T(\zeta_0)\psi\,,\quad n\in \bbN_0\,,
\end{align}
with the sum converging in  norm within the disk of convergence.

%%%%%%%%%%%%%%%%%%%%%%%%%%%%%%%%%%%%%%%%%%%%%%%%%%%%%%%%%%%%%%%%%%%%%%%%%%%%%%%%%%%%%

\subsection{Spectrum and sectoriality of $\mathbf{\Delta_{\th}}$}
\label{Sec3_2}

We now return to the operator (\ref{i0}) and summarize our
standing assumptions.

{{\bf Standing assumptions.}}
Throughout $M$ is a real, smooth, manifold without boundary admitting a co-dimension one foliation.
As such it is diffeomorphic to $\R \times \Sigma$; we do not require $\Sigma$ or $M$ to be closed or compact. 
Further, we assume $M$ to carry {smooth} metrics of Euclidean signature
and Lorentzian signature of the ADM form, with {{\it shared}} ADM fields. 
Specifically, in adapted local coordinates $(t,x^a), a=1,\ldots, d$, the
metric tensors read $g^{\pm} = \pm N^2 dt^2 + \wg_{ab}(dx^a + N^a dt)
(dx^b + N^b dt)$, where
$N>0$ is the lapse, $N^a \dd/\dd x^a$ is the shift, and
$\wg_{ab} dx^a dx^b$ is the non-degenerate Riemannian metric on
the leaves of the foliation. We also write $g^{\pm} = g^{\pm}_{\mu\nu}
dy^{\mu} dy^{\nu}$ in local coordinates and
$g_{\pm} = g_{\pm}^{\mu\nu} \dd/\dd y^{\mu} \dd/\dd y^{\nu}$,
for the inverse. The volume form is signature independent
and denoted alternatively by $d\mu_g = d^{1\!+\!d} y |g|^{1/2} =
dt d^dx N \sqrt{\wg}$. {It enters in particular the inner product
$\bra u\vert v\ket_{L^2(M)} := \int\! d\mu_g \,u^* v$ on $L^2(M)$.} 
The scalar Laplacian and d'Alembertian associated with $g^{+}$ and
$g^{-}$ are
$\nabla_{\pm}^2 = |g|^{-1/2} \dd_{\mu} ( |g|^{1/2} g_{\pm}^{\mu\nu}
\dd_{\nu})$.  We allow for a potential $V\in C^\infty(M)$ that is
non-negative and bounded and set $\mcD_\pm :=-\nabla_{\pm}^2+V$. 
With these specifications we define
\be
\label{Adecomp}
\dth :=  - \sin \th\, \cD_+ - i \cos \th \,\cD_-\,, \quad 
\th \in (0,\pi)\,,
\ee
with $C_c^{\infty}(M)$ as initial domain. In particular,
$\mcD_+$ defined on $C_c^\infty(M)$ is a positive operator
while $\cD_-$ is indefinite. As mentioned, $\Delta_{\th}$
can be interpreted as a generalized Laplacian associated with the
covariantly defined metric $g^{\th}$, dubbed the {\itshape lapse-Wick-rotated}
metric. These covariance aspects are secondary to the present development
and are omitted.

Our immediate concern it to establish the sectoriality of $\Delta_{\th}$
on a suitable domain $D(\dth)$ dense in the Hilbert space $L^2(M)$.
It is tempting to continue to use $C_c^\infty(M)$, which is densely
contained in $L^2(M)$. However, this domain is unsuitable for
sectoriality as even $\1-\dth$ fails to be surjective. Instead, we
sacrifice classical differentiability and define the domain $D(\dth)$
to be the following subset of the Sobolev space $H_0^1(M)$
(see Appendix \ref{App_dist}) 
\begin{align}\label{sec8}
\forall\,\theta\in (0,\pi):\,\,	D(\dth):=\big\{u\in H^1_0(M)\,\big|\,\dth u\in L^2(M)\big\}\,,
\end{align}
which is dense in $L^2(M)$. 

{\bfseries Remarks.}

(i) Recall that  $u\in H^1_0(M)
\overset{\rm dense}{\hookrightarrow}L^2(M)$ defines a distribution
$\widetilde{u}\in \mfD'(M)$ on the space of test functions
$\mfD(M)\equiv C_c^\infty(M)$, and further the distributional action of
the  operator $\dth$ determines a distribution $\widetilde{\dth u}\in
\mfD'(M)$, c.f. Appendix \ref{App_dist}. Then the condition
$\dth u\in L^2(M)$ in \eqref{sec8} means that there exists
$h\in L^2(M)$ such that 
\begin{align}\label{sec8a}
\forall\,w\in C_c^\infty(M):\,\int\!d\mu_g\,u\dth w=\int\!\,d\mu_g\,hw\,,
\end{align}
with $d\mu_g$  the volume form introduced above. Once \eqref{sec8a} holds,
one {\itshape defines} $\dth u:=h$. 
 
% (ii) Clearly $C_c^\infty(M)\subseteq D(\dth)\subseteq L^2(M)$,
%so $D(\dth)$ is% dense in $L^2(M)$. 
 
(ii) Although $D(\dth)$ contains non-classically differentiable functions,
one may still integrate by parts: for all $u,v\in D(\dth)$
 \begin{align}\label{sec9}
	\int \! d\mu_g\,u(\sin\theta\nabla_+^2+i\cos\theta \nabla_-^2)v =-\int \! d\mu_g\, (\sin \theta g^{\alpha \beta}_++i\cos\theta g^{\alpha \beta}_-)\pp_\alpha u\pp_\beta v\,.
\end{align}
This follows readily from approximation with $C_c^\infty(M)$ functions in $H^1$-norm, together with the definitions of the distributional gradient and the distributional action of  $\sin\theta\nabla_+^2+i\cos\theta \nabla_-^2$; we omit a detailed proof.
\medskip

We now state and prove the main result of this section.
\begin{theorem}
\label{thsec1}
Let $(M,g^{\pm})$ be as above, $\theta\in (0,\pi)$, and
$\tilde{\theta}:=\min\set{\theta,\pi-\theta}$. For
$V\in C^\infty(M)$ non-negative and bounded, let 
\begin{align}\label{sec8a0}
 \dth:=-\sin\theta\mcD_+ -i\cos \theta\mcD_-\,,\quad
 \mcD_\pm :=-\nabla_\pm^2+V\,,
\end{align}
have domain \eqref{sec8} determined by its distributional action.
 Then $(\dth,D(\dth))$ generates a unique  bounded analytic semigroup $(\exx{\zeta\dth})_{\zeta\in \Sigma_{\tilde\theta} \cup \set{0}}$ that is contractive, i.e.~$|\hspace{-1pt}|\exx{\zeta\dth} |\hspace{-1pt}|_\text{op}\leq 1$ for all $\zeta\in \Sigma_{\tilde\theta}\cup\set{0}$.
 
Further, the adjoint semigroup is given by 
 \begin{align}\label{sec8b}
   \forall\,\zeta\in \Sigma_{\tilde\theta}:\,\,(\exx{\zeta\dth})^\ast
   =\exx{\zeta^\ast \Delta_{\pi-\theta}}\,.
\end{align}
\end{theorem}
We prepare the following lemma, whose proof is postponed to the end of the section.

\begin{lemma}
\ \label{lmsec1}
For $\theta\in (0,\pi)$, and $\tilde{\theta}:=\min\set{\theta,\pi-\theta}$,  the resolvent set $\rho(\dth)$ contains the sector $\Sigma_{\pi/2+\tilde\theta}$, i.e. $\Sigma_{\pi/2+\tilde\theta}\subseteq \rho(\dth)$.
Equivalently, the spectrum  $\sigma(\Delta_\theta)$  is contained in the (closed) set $\bbC\setminus \Sigma_{\pi/2+\tilde\theta}$. Moreover, for each  $\tilde{\theta}'\in (0,\tilde{\theta})$ there is a constant $C_{\tilde{\theta}'}\geq 1$ such that 
\begin{align}\label{sec9d}
\forall\,\lb\in \Sigma_{\pi/2+\tilde{\theta}'}:\quad 	\norm{(\lb-\dth)\inv}_{\rm op}\leq \frac{C_{\tilde{\theta}'}}{|\lambda|}\,,
\end{align}
and further
\begin{align}\label{sec9d0}
\forall\,\lb\in \Sigma_{\tilde{\theta}}:\quad	\norm{(\lb-\dth)\inv}_{\rm op}\leq \frac{1}{|\lambda|}\,.
\end{align}
\end{lemma}

{\itshape Proof of Theorem \ref{thsec1}.} Lemma \ref{lmsec1} entails that $(\dth,D(\dth))$ is sectorial of angle $\tilde\theta$, and accordingly generates a unique bounded analytic semigroup $(\exx{\zeta\dth})_{\zeta\in \Sigma_{\tilde\theta}\cup\set{0}}$  by Theorem \ref{prpsec1}. Moreover, the resolvent bound \eqref{sec9d0} of Lemma \ref{lmsec1} for all $\lb\in \Sigma_{\tilde{\theta}}$ implies $(\exx{\zeta\dth})_{\zeta\in \Sigma_{\tilde\theta}\cup\set{0}}$ is a family of contractive operators.

Next, the adjoint result \eqref{sec8b} is an immediate corollary of the fact
\begin{align}
  \label{sec15}
\Delta_\theta^\ast = \Delta_{\pi-\theta}\,,\quad D(\Delta_{\theta}^\ast)=D(\Delta_{\pi-\theta})\,,
\end{align}
which we now prove.
Since the all elements of $D(\Delta_\theta)$ and $D(\Delta_{\pi-\theta})$ are Sobolev $H_0^1$-functions, integration by parts (c.f. \eqref{sec9}) gives  $\Delta_{\pi-\theta}\subseteq \Delta_\theta^\ast$ (i.e. $D(\Delta_{\pi-\theta})\subseteq D(\Delta_\theta^\ast)$ and $\Delta_\theta^\ast|_{D(\Delta_{\pi-\theta})}=\Delta_{\pi-\theta})$. In order to show equality, let  $\lb>0$ and define the operator $B_\theta:=\lb-\Delta_\theta$  with domain $D(B_\theta):=D(\Delta_\theta)$. By definition of the adjoint, $f\in D(B^\ast_\theta)$ {\itshape iff} there exists a unique $g\in L^2(M)$ such that for all $h\in D(B_\theta)$ we have  $\ip{f}{B_\theta h}_{L^2(M)}=\ip{g}{h}_{L^2(M)}$,  
and by definition $B_\theta^\ast f:=g$.
On the other hand, Lemma \ref{lmsec1} implies that $B_{\pi-\theta}\inv=(\lb -\Delta_{\pi-\theta})\inv\in \mfB(L^2(M))$, so there is $f':= (\Delta_{\pi-\theta}-\lb)\inv g\in D(\Delta_{\pi-\theta})$. Hence, for any $h\in D(B_\theta)$
\begin{align}\label{sec16}
	\ip{B_\theta^\ast f}{h}_{L^2(M)}=\ip{(\Delta_{\pi-\theta}-\lb)f'}{h}_{L^2(M)}= \ip{f'}{(\Delta_{\theta}-\lb)h}_{L^2(M)}\,,
\end{align}
where the second equality arises from integrating by parts as $f',\,h\in H^1_0(M)$ (c.f. \eqref{sec9}). Comparing $f$ and $f'$, it  follows that for all $h\in D(B_\theta)$ we have the result $\ip{f-f'}{(\Delta_{\theta}-\lb)h}_{L^2(M)}=0$,  or equivalently $f-f'\in \text{ran}(\lb -\Delta_\theta)^\perp =L^2(M)^\perp$, and hence  
 $f=f'\in D(\Delta_{\pi-\theta})$. So $B_\theta^\ast =(\lb -\Delta_\theta)^\ast = \lb -\Delta_{\pi-\theta}$ holds for all $\lb>0$, which implies $\Delta_\theta^\ast = \Delta_{\pi-\theta}$. 
\qed

\bigskip

{\itshape Proof of Lemma \ref{lmsec1}.} Fixing an arbitrary $\theta\in (0,\pi)$, we begin by establishing $\Sigma_{\pi/2+\tilde\theta}\subseteq \rho(\dth)$. Namely, given $\lb\in \Sigma_{\pi/2+\tilde\theta}$, it is to be shown that $\lb-\dth:D(\dth)\to L^2(M)$ has a bounded inverse.

{\itshape Surjectivity:}  This follows through an application of the Lax-Milgram lemma (Ch.~6 Theorem 6 in \cite{Laxbook}). Fixing $f\in L^2(M)$, one defines the antilinear bounded map $\Psi_f:H_0^1(M)\to \bbC$ by
\begin{equation} 
	\Psi_f(w):=\int\!d\mu_g \,fw^\ast\,,
\end{equation} 
and the sesquilinear form $\mcE:H_0^1(M)\times H_0^1(M)\to \bbC$  mapping $u,w\in H^1_0(M)$ to
\begin{align}
	\mcE(u,w) :=\int\!d\mu_g\,\Big[ (\lb+i\exx{-i\theta}V) uw^\ast  +(\sin\theta g_+^{\alpha \beta}+i\cos\theta g_-^{\alpha \beta})\pp_\alpha u\pp_\beta w^\ast \Big]\in \bbC \,.
\end{align}
Here {\itshape $\mcE$ is bounded}, i.e., 
$\exists\,c>0:\,\forall\, u,w\in H_0^1(M):\,|\mcE(u,w)|\leq c\nrm{u}_{H^1(M)}\nrm{w}_{H^1(M)}$. This follows from an application of the Cauchy-Schwarz inequality, the fact that the potential function $V$ is bounded on $M$, and the identity $g_-^{\mu\sigma}g_{\sigma\rho}^+g_{-}^{\rho \nu}=g_+^\munu$ (which is easily verified from the ADM forms \eqref{i1a}).
Once  $\mcE$ is known to be  also {\itshape  coercive}, i.e., 
$\exists\,\alpha>0:\,\forall\,u\in H^1_0(M):\,|\mcE(u,u)|\geq \alpha\nrm{u}_{H^1}$,
the Lax-Milgram lemma implies the existence of a unique $v\in H_0^1(M)$
such that 
\begin{align}\label{sec9d0a}
	\forall\,w\in H^1_0(M):\,\,\mcE(v,w)=\Psi_f(w)\,.
\end{align}
To establish the coercivity bound, consider the restriction of $\mcE$ to the unit sphere $\mcS_{H_0^1}:=\set{w\in H_0^1(M)\given \nrm{w}_{H^1(M)}=1}$. Then, we may express 
\begin{align}\label{sec9e0}
	\mcE(w,w)=(1-e(w))\lambda -\Big[-\sin\theta\, e(w)-i\cos \theta \,\ell(w)-i \exx{-i\theta} \int\!d\mu_g\,V|w|^2\Big]\,,
\end{align}
with 
\begin{align}\label{sec9e}
	e(w):=\int\!d\mu_g\,g_+^{\alpha \beta}\pp_\alpha w\pp_\beta w^\ast\,,\quad \ell(w):=\int\!d\mu_g\,g_-^{\alpha \beta}\pp_\alpha w\pp_\beta w^\ast\,.
\end{align}
Clearly,  $e(w)\geq 0$, $\ell(w)\in \bbR$ and the bound $|\ell(w)|\leq e(w)$ is easily seen from the ADM forms \eqref{i1a}; hence the two terms on the right hand side of \eqref{sec9e0} are elements of disjoint sets, 
\begin{align}
	&(1-e(w))\lb \in \Sigma_{\pi/2+\tilde\theta}\,,
	\nonum 
	&-\sin\theta \,e(w)-i\cos \theta\, \ell(w)-i \exx{-i\theta} \int\!d\mu_g\,V|w|^2\in \bbC\setminus\Sigma_{\pi/2+\tilde \theta}\,.
\end{align}
It then follows that these terms are bounded 
 away from each other uniformly over $\mcS_{H_0^1}$, and so there exists $\alpha>0 $ such that for all $w\in \mcS_{H_0^1}:\,|\cE(w,w)|\geq \alpha$. Rescaling yields the requisite coercivity bound.
 
 Having established  coercivity, $\Psi_f$ is uniquely realized by $\mcE$  through the relation \eqref{sec9d0a}. In particular, for any $w\in C_c^\infty(M)\subseteq H_0^1(M)$, integrating \eqref{sec9d0a} by parts yields 
\begin{align}
	\int\!d\mu_g\,fw^\ast = \int\!d\mu_g\,v(\lb-\dth)w^\ast\,.
\end{align}
 In summary, given $f\in L^2(M)$, there exists a unique $v\in H_0^1(M)$ such that $\dth v=\lb v-f\in L^2(M)$ and  $(\lb-\dth)v=f$, establishing surjectivity.

\medskip

{\itshape Injectivity:} For any element $w\in \mcS':=\set{u\in D(\dth)\given \norm{u}_{L^2}=1}$, 
\begin{align}
\nrm{(\lb-\dth)w}_{L^2(M)}&\geq   \big|\ip{w}{(\lb-\dth)w}_{L^2(M)} \big|
\nonum 
\isa \lambda -\Big[-\sin\theta e(w)-i\cos \theta \ell(w)-i \exx{-i\theta} \int\!d\mu_g\,V|w|^2\Big]\,,
\end{align}
where the second line follows from an integration by parts, and $e(w)$, $\ell(w)$ are defined in \eqref{sec9e}.  This is the difference of two terms,  $\lambda\in \Sigma_{\pi/2+\tilde\theta}$, and the other an element of  $\bbC\setminus \Sigma_{\pi/2+\tilde\theta}$. Then, clearly for  all $w\in \mcS'$
\begin{align}\label{sec9f}
	\big|\bra w|(\lb-\dth)w\rangle\big|\geq \text{dist}(\lb,\,\bbC\setminus \Sigma_{\pi/2+\theta})=:d_\lb(\theta)>0\,,
\end{align}
implying injectivity.

{\itshape Resolvent bounds:} For $\lb\in\Sigma_{\pi/2+\tilde\theta}$,  the bijection $\lb-\dth:D(\dth)\to L^2(M)$ has inverse $(\lb-\dth)\inv$, and \eqref{sec9f}  implies the resolvent bound
\begin{align}\label{sec9f0}
	\nrm{(\lb-\Delta_\theta)\inv}_{\rm op}\leq \frac{1}{d_\lb(\theta)}\,.
\end{align}
When $\tilde{\theta}'\in (0,\tilde{\theta})$, for any $\lb\in \Sigma_{\pi/2+\theta'}$ one has $d_\lb(\theta)\geq {\rm sin}(\tilde{\theta}-\tilde{\theta}')|\lb|$, yielding the uniform bound
\begin{align}
	\nrm{(\lb-\Delta_\theta)\inv}_{\rm op}\leq \frac{1}{{\rm sin}(\tilde\theta-\tilde\theta')}\,\frac{1}{|\lb|}\,,\quad \forall\, \lb \in \Sigma_{\pi/2+\tilde\theta'}\,,
\end{align}
Setting $C_{\tilde\theta'}:={\rm sin}(\tilde{\theta}-\tilde{\theta}')\inv\geq 1$ gives the result \eqref{sec9d}.
On the other hand, restricting to $\lb\in \Sigma_{\tilde\theta'}$, it is easy to see that  $d_\lb(\theta)=|\lb|$. Thus \eqref{sec9f0} yields the result  \eqref{sec9d0}, completing the proof of Lemma \ref{lmsec1}.
\qed

%\newpage
%%%%%%%%%%%%%%%%%%%%%%%%%%%%%%%%%%%%%%%%%%%%%%%%%%%%%%%%%%%%%%%%%%%%%%%%%%%%%%%%%%%%%%%%%

\section{Smoothening and integral kernel} 
\label{Sec3_3}

Two of the key features of the standard heat semigroup associated to the Dirichlet Laplacian on Riemannian manifolds are its smoothening property and its realizability as an integral operator with a smooth kernel. Having shown the existence of the analytic semigroups
generated by $\Delta_{\th}$ for all $\theta\in (0,\pi)$ in the previous section,  we collect a number of results on the semigroup's action in Theorem \ref{thsmooth1}, in particular its smoothening behavior. 
 The second main result of this section, Theorem \ref{thsmooth2}, realizes the semigroup in terms of a unique, smooth integral kernel,
\begin{align}
	&\big(e^{\zeta\dth} \psi\big)(y) 
= \int\! d\mu_g(y')\, 
K^{\th}_\zeta(y,y') \psi(y') \,,
\nonum 
&(\pp_\zeta-\Delta_{\theta,y} )K^{\th}_\zeta(y,y') =0\,,\quad
\lim_{\zeta \ra 0} K_\zeta^{\th}(y,y') =
\delta(y,y')\,.
\end{align}
Recall that $d\mu_g$ is the $\theta$-independent volume form $dy^{1\!+\!d} |g|^{1/2}$,
and the $\delta$-distribution is normalized with respect to it. 
\pagebreak[2]

\begin{theorem}\ \label{thsmooth1}
Let $\theta\in (0,\pi)$ and  $\tilde{\theta}:=\min\set{\theta,\pi-\theta}$. 
\begin{enumerate}[leftmargin=8mm, rightmargin=-0mm, label={\rm (\roman*)}]
  \item For any $\psi\in L^2(M)$ consider the map $\Sigma_{\tilde{\theta}}\ni \zeta \mapsto \exx{\zeta \Delta_\theta}\psi\in L^2(M)$. Then, for each $\zeta\in \Sigma_{\tilde\theta}$, the image $\exx{\zeta \dth}\psi\in L^2(M)$ has a $C^\infty(M)$-representative, denoted by $(\exx{\zeta \dth}\psi)(\cdot)$, and the mapping 
\begin{align}\label{e.sm1}
	\Sigma_{\tilde\theta} \times M \ni (\zeta,y)\mapsto (\exx{\zeta \dth}\psi)(y)\in \bbC
\end{align}
is jointly smooth, and is analytic in $\zeta\in \Sigma_{\tilde\theta}$ for fixed $y\in M$.  Moreover, the mapping \eqref{e.sm1} is a classical solution to the differential equation
\begin{align}\label{e.sm1a}
	\begin{cases}
		(\pp_\zeta -\dth)f(\zeta,y)=0\,, \quad \forall\, (\zeta,y)\in \Sigma_{\tilde\theta}\times M\,,\\[2mm]
		f(0,y)=\psi(y)\,,\quad \text{for a.e. }y\in M\,.
	\end{cases}
\end{align}

\item Consider an arbitrary $\tilde{\theta}'\in (0,\tilde{\theta})$ and $\psi\in L^2(M)$. Then, for any chart $U$ on $M$, and each open set $\Omega\Subset U$, there exist for every $m\in \bbN_0$ constants $c_m>0$ and $\sigma_m\in \bbN$ such that in local coordinates there are the uniform bounds\footnote{The $\nrmm_{C^m(\Omega)}$-norm is defined in the local chart coordinates by $\nrm{u}_{C^m(\Omega)}:=\underset{|\alpha|\leq m}{\sup}\underset{\Omega_0}{\sup}\,|\pp^\alpha u|$.}
\begin{align}\label{e.sm2}
\forall\,\zeta\in \Sigma_{\tilde{\theta}'}\,,\,\forall\,m\in \bbN_0:\quad 	\nrm{\exx{\zeta \dth}\psi}_{C^m(\Omega)}\leq c_m(1+|\zeta|^{-\sigma_m})\nrm{\psi}_{L^2(M)}\,.
\end{align} 
The constant $c_m>0$ depends on $\theta',\,\theta,\,\Om,\,U$, but is independent of  $\psi$ and $\zeta$, and $\sigma_m\in \bbN$ is defined as the smallest integer greater than $m/2+(d+1)/4$.

\item If $\psi\in C_c^\infty(M)$ and $\tilde{\theta}'\in (0,\tilde{\theta})$, then for any chart $U$ on $M$, and each open set $\Omega\Subset U$, in local coordinates
\begin{align}\label{e.sm2a}
\forall\,m\in \bbN_0:\quad 	\nrm{\exx{\zeta \dth}\psi-\psi}_{C^m(\Omega)}\to 0\,,\quad \text{as}\quad \Sigma_{\tilde{\theta}'}\ni \zeta \to 0\,.
\end{align}

\end{enumerate}

\end{theorem}

{\itshape Proof of Theorem \ref{thsmooth1}.} 

(i) Fix an open ball $B_S(\zeta_0):= \set{\zeta\in \bbC\given |\zeta-\zeta_0|<S}\subseteq \Sigma_{\tilde\theta}$, a chart $U$ on $M$, and an open set $\Omega\Subset U$. Integrating in local coordinates in the chart $U$, we have the bound
\begin{align}\label{e.sm3}
	\int_{\Omega}\!dy\,|\Delta_\theta f|^2\leq C^2\int_{\Omega}\!d\mu_g|\,\Delta_\theta f|^2\leq C^2\nrm{\Delta_\theta f}_{L^2(M)}^2\,,
\end{align}
where $C^2:=\sup_{\Omega}(N\sqrt{\wg})^{-1/2}<\infty$, since $\Omega\Subset U$.   

Next, given $\psi\in L^2(M)$, Eq.~\eqref{e.sm3} and the bound \eqref{sec6c2} imply that 
\begin{align}\label{e.sm4}
	\forall\,n\in \bbN_0:\quad \Delta_\theta ^n\,\exx{\zeta_0 \dth}\psi\in L^2_{\rm loc}(U)\,,
\end{align}
where $U$ is regarded as an open subset of $\bbR^{d+1}$ via the chart map, and $L^2_{\rm loc}(U)$ is defined relative to the Lebesgue measure $dy$ on $\bbR^{d+1}$. Further, \eqref{sec6d} entails the Taylor expansion 
\begin{align}\label{e.sm5}
	\Delta_\theta^n \exx{\zeta \dth}\psi=\ssum{k=0}{\infty}\frac{(\zeta-\zeta_0)^k}{k!}\dth^{n+k}\exx{\zeta_0 \dth}\psi,
\end{align}
 converging in $L^2(M)$  for all $\zeta\in B_S(\zeta_0)$ (with $S>0$ sufficiently small), and hence converging in $L^2_\text{loc}(U)$ in local coordinates on the chart $U$. Since in local coordinates, $\dth$ is an operator in $\bbR^{d+1}$ of the type \eqref{apploc1b} considered in Appendix \ref{apploc}, Theorem \ref{apploc_th2} implies that for each $\zeta\in B_S(\zeta_0)$, there exists a $C^\infty(U)$ representative of $\exx{\zeta\dth}\psi$ in the chart, and moreover that the mapping
\begin{align}
	B_S(\zeta_0)\times U\ni (\zeta,y)\mapsto (\exx{\zeta\dth}\psi)(y)\in \bbC
\end{align}
is jointly smooth. Since both the open ball $B_S(\zeta_0)\subseteq \Sigma_{\tilde\theta}$ and the chart $U$ on $M$ was arbitrary, this establishes that the mapping $\Sigma_{\tilde\theta}\times M\to \bbC$  given by $(\exx{\zeta\dth}\psi)(y)$ is smooth. Moreover, for each $y\in M$, the pointwise convergence of \eqref{e.sm5} implies analyticity in $\zeta\in \Sigma_{\tilde\theta}$.
Finally, recall that when regarded as an element of $L^2(M)$, $\exx{\zeta \dth}\psi$ satisfies $(\pp_\zeta-\dth)\exx{\zeta \dth}\psi=0$ and $\lim_{\Sigma_{\tilde\theta}\ni\zeta\to 0}\exx{\zeta\dth}\psi=\psi$. The smoothness of $(\zeta,y)\mapsto \exx{\zeta \dth}\psi(y)$ then clearly entails it satisfies the  differential equation \eqref{e.sm1a}.

\medskip

(ii)
Proceeding next to the local coordinate bound, fix arbitrary $\tilde{\theta}'\in (0,\tilde{\theta})$ and open sets $\Omega\Subset \Omega'\Subset U$ (since $\Omega\Subset U$, it is always possible to find such an open set $\Omega'$). Then by Corollary \ref{apploc_cor1} we have for any $\zeta\in \Sigma_{\tilde{\theta}'}$ the bound
\begin{align}\label{e.sm5a}
	\nrm{\exx{\zeta\dth}\psi}_{C^m(\Omega)}&\leq B\ssum{j=0}{\sigma_m}\nrm{\dth^j\exx{\zeta\dth}\psi}_{L^2(\Omega',\,dy)}\leq B'\ssum{j=0}{\sigma_m}\nrm{\dth^j\exx{\zeta\dth}\psi}_{L^2(M)}
	\nonum
	&\leq B''\big(1+\ssum{j=1}{\sigma_m}(j/|\zeta|)^j\big)\nrm{\psi}_{L^2(M)}
        \leq c_m\big(1+|\zeta|^{-\sigma_m}\big)\nrm{\psi}_{L^2(M)}
\end{align}
where \eqref{sec6c2} is used for the penultimate inequality, and $\ssum{j=1}{N}(j/|\zeta|)^j\leq \ssum{j=1}{N}j^j(1+|\zeta|^{-N})$ for the final bound.

\medskip

(iii) Strong continuity of $(\exx{\zeta\dth})_{\zeta\in \Sigma_{\tilde\theta}\cup \set{0}}$ (c.f.~Definition \ref{asgroupdef}(iii)) entails that for any $\tilde{\theta}'\in (0,\tilde{\theta})$ and $\psi\in L^2(M)$, $|\hspace{-1pt}| \exx{\zeta\dth}\psi-\psi|\hspace{-1pt}|_{L^2(M)}\to 0$ as $\Sigma_{\tilde{\theta}}\ni \zeta \to 0$.
In particular, if $\psi\in C^\infty_c(M)$, then for each $k\in \bbN_0:\,\dth^k\psi\in C^\infty_c(M)$, and hence $\forall\,k\in \bbN_0$
\begin{align}\label{e.sm6}
	\nrm{\dth^k(\exx{\zeta\dth}\psi-\psi)}_{L^2(M)}=\nrm{\exx{\zeta\dth}(\dth^k\psi)-\dth^k\psi}_{L^2(M)}\to 0\quad \text{as }\Sigma_{\tilde{\theta}}\ni \zeta \to 0\,,
\end{align}
where the equality follows from the semigroup commuting with its generator. Corollary \ref{apploc_cor1} then implies the result \eqref{e.sm2a}.
 \qed
 
\bigskip 
 
We  proceed to the kernel theorem.

\begin{theorem} \label{thsmooth2}
Let $\theta\in (0,\pi)$ and  $\tilde{\theta}:=\min\set{\theta,\pi-\theta}$. Then for all $\zeta\in \Sigma_{\tilde\theta}$ there is a unique integral kernel $K_\zeta^\theta\in C^\infty(M\times M)$ such that for all $\psi\in L^2(M)$
\begin{align}\label{e.sm7}
	(\exx{\zeta\dth}\psi)(y)=\int\! d\mu_g(y')\,K_\zeta^\theta(y,y')\psi(y')\,,\quad \forall\,y\in M\,.
\end{align}

This integral kernel has the following properties:
\begin{enumerate}[leftmargin=8mm, rightmargin=-0mm, label={\rm (\roman*)}]
\item (Hermiticity) For all $(\zeta,y,y')\in \Sigma_{\tilde\theta}\times M\times M:\,K_\zeta^\theta(y,y')=K_{\zeta^\ast}^{\pi-\theta}(y',y)^\ast$.

\item (Regularity) The mapping 
\begin{align}\label{e.sm8}
	\Sigma_{\tilde\theta}\times M\times M\ni (\zeta,y,y')\mapsto K_\zeta^\theta(y,y')\in \bbC
\end{align}
 is jointly $C^\infty$-smooth. Moreover for fixed $(y,y')\in M\times M$, $K_\zeta^\theta(y,y')$ is analytic in $\zeta\in \Sigma_{\tilde\theta}$.  
 
 \item (Uniqueness) If for any $\zeta\in \Sigma_{\tilde\theta}$ there exists another integral kernel \\ $J^\theta_\zeta\in C^\infty(M\times M)$ realizing the semigroup action via \eqref{e.sm7}, then $K_\zeta^\theta =  J^\theta_\zeta$.

\item (Semigroup property) For all $y,y'\in M$ and $\zeta_1,\zeta_2\in \Sigma_{\tilde\theta}$
\begin{align}\label{e.sm9}
	K_{\zeta_1+\zeta_2}^\theta(y,y')=\int\!d\mu_g(y'')\,K_{\zeta_1}^\theta(y,y'')K_{\zeta_2}^\theta(y'',y')\,.
\end{align}

\item (Lapse-Wick-rotated heat equation)  For any fixed $y'\in M$, 
\begin{align}\label{e.sm8a}
	(\pp_\zeta-\Delta_{\theta,y})K_\zeta^\theta(y,y')=0 \quad\text{in }\Sigma_{\tilde\theta}\times M\,.
\end{align}
Furthermore if $\psi\in C_c^\infty(M)$ and $\tilde{\theta}'\in (0,\tilde{\theta})$, then for any chart $U$ on $M$, and each open set $\Omega\Subset U$, in local coordinates we have
\begin{align}\label{e.sm10}
	\int\!d\mu_g(y)\,K_\zeta^\theta(\cdot,y)\psi(y)\longrightarrow\psi(\cdot)
\end{align}
in  $C^m(\Omega)$ as $\Sigma_{\tilde\theta'}\ni \zeta \to 0$. 
\end{enumerate}

\end{theorem}

{\bf Remark.} These properties mirror closely those of the heat kernel
$K_s(y,y')$ on a Riemannian manifold $(M, g^+)$. Additionally, the latter
often satisfies $\int \! d\mu_g(y') K_s(y',y) =1$, for all $y \in M$,
referred to as stochastic completeness of $(M, g^+)$. Since this
primarily reflects features of the underlying manifold \cite{Grigorbook}
(not of the operator family $\Delta_{\th}$) we leave this unexplored here.

The proof of this theorem uses the following lemmas.
\begin{lemma}\ \label{lmsmooth1}
	For all $(\zeta,y)\in \Sigma_{\tilde\theta}\times M$ there exists a unique $\kappa_{\zeta,y}^\theta\in L^2(M)$ such that for every $\psi\in L^2(M)$
\begin{align}
	(\exx{\zeta\dth}\psi)(y)=\ip{\kappa_{\zeta,y}^\theta}{\psi}_{L^2(M)}\,,\quad \forall\,(\zeta,y)\in \Sigma_{\tilde\theta}\times M\,.
\end{align}

\end{lemma}

{\itshape Proof of Lemma \ref{lmsmooth1}.} Fixing an arbitrary $ (\zeta,y)\in \Sigma_{\tilde\theta}\times M$, consider the mapping  $L^2(M)\ni \psi\mapsto (\exx{\zeta\dth}\psi)(y)\in \bbC$, which is well-defined by Theorem \ref{thsmooth1} and clearly linear. Moreover, Theorem \ref{thsmooth1}(ii) implies there is a constant $B>0$ (depending on $y,\zeta,\theta$, but independent of $\psi\in L^2(M)$) such that 
\begin{align}
\label{localbound}
  \big|(\exx{\zeta\dth}\psi)(y)\big|\leq B\nrm{\psi}_{L^2(M)}\,,\quad \forall\,\psi\in L^2( M)\,.
\end{align}
Thus, this mapping is continuous, and by the Riesz Representation Theorem is realized by a unique $\kappa_{\zeta,y}^\theta\in L^2(M)$ via $(\exx{\zeta\dth}\psi)(y)=\langle \kappa_{\zeta,y}^\theta|\psi\rangle_{L^2(M)}$, for each $\psi\in L^2(M)$.
\qed

\medskip

Next, as a tool to prove the regularity statement  Theorem \ref{thsmooth2}(ii), we recall the notions of weak and strong differentiability of Hilbert space valued maps. Let $\mcO\subseteq\bbR^n$  be an open set and $h:\mcO\to L^2(M)$. 
\begin{itemize}[leftmargin=6mm, rightmargin=-0mm]
  \item $h$ is said to be {\itshape weakly $C^k,\,k\in \bbN_0$}, if $\forall\,\vp\in L^2(M):\,\ip{\varphi}{h(\cdot)}_{L^2(M)}$ is in $C^k(\mcO)$.
  \item The G\^{a}teaux derivative of $h$ along direction $\hat{e}_i$ in $\Omega$ is 
 \begin{align}
	\pp_ih(z):=\lim_{\varepsilon\to 0}\frac{h(z+\varepsilon \hat{e}_i)-h(z)}{\varepsilon}\,,
\end{align}
with the limit taken in $\nrm{\cdot}_{L^2(M)}$. Then $h$ is said to be {\itshape strongly $C^k,\,k\in \bbN_0$}, if for all multi-indices $\alpha$ with $|\alpha|\leq k$ the G\^{a}teaux derivatives $\pp^\alpha h$ exist throughout $\Omega$ and are strongly continuous. 
\end{itemize}
Then we have the result
\begin{lemma}\ \label{lmsmooth2}
$h:\Omega\to L^2(M)$ is strongly $C^\infty$ iff it is weakly $C^\infty$. 
\end{lemma}
We omit the proof, referring to Corollary 1.42 of \cite{DaviesSemigroup}.

\bigskip

We now proceed to the proof of Theorem \ref{thsmooth2}.

{\itshape Proof of Theorem \ref{thsmooth2}.} We begin with the existence of the pointwise defined integral kernel. Fixing 
arbitrary $\zeta_1,\zeta_2 \in \Sigma_{\tilde\theta}$, $y\in M$ and $\psi\in L^2(M)$, consider
\begin{align}\label{e.sm11a}
  (\exx{(\zeta_1+\zeta_2)\dth}\psi)(y)\isa
  \big(\exx{\zeta_1\dth}(\exx{\zeta_2 \dth}\psi)\big)(y)
  =\ev{\kappa_{\zeta_1,y}^\theta\,\big|\,\exx{\zeta_2 \dth}\psi}_{L^2(M)}
\nonum
\isa \ip{\exx{\zeta_2^\ast  \Delta_{\pi-\theta}}
    \kappa^\theta_{\zeta_1,y}}{\psi}_{L^2(M)}
	= \int\!d\mu_g(y')\,(\exx{\zeta_2^\ast  \Delta_{\pi-\theta}}
        \kappa^\theta_{\zeta_1,y})^\ast(y') \psi(y')
	\nonum
	\isa \int\!d\mu_g(y')\,
       \ip{\kappa_{\zeta_1,y}^\theta}{\kappa_{\zeta_2^\ast ,y'}^{\pi-\theta}}_{L^2(M)}
        \psi(y')\,,
\end{align}
where we use $\exx{\zeta_2^\ast  \Delta_{\pi-\theta}}=(\exx{\zeta_2 \dth})^\ast$ for the third equality on the first line, and Lemma \ref{lmsmooth1} for the last step. In particular, for $\zeta\in \Sigma_{\tilde\theta}$ we have
\begin{align}\label{e.sm11}
	(\exx{\zeta\dth}\psi)(y)\isa\int\!d\mu_g(y')\,\ip{\kappa_{\zeta/2,y}^\theta}{\kappa_{\zeta^\ast/2,y'}^{\pi-\theta}}_{L^2(M)} \psi(y')\,,
\end{align}
so the semigroup has a pointwise defined integral kernel 
\begin{align}\label{e.sm12}
	K_\zeta^\theta(y,y'):=\ip{\kappa_{\zeta/2,y}^\theta}{\kappa_{\zeta^\ast/2,y'}^{\pi-\theta}}_{L^2(M)}\,,\quad \forall\,(\zeta,y,y')\in \Sigma_{\tilde\theta}\times M\times M\,.
\end{align}

(i) The hermiticity property of the kernel follows immediately from the definition \eqref{e.sm12}.

\medskip

(ii) We now turn to its regularity. Let $\mcO$ be a chart of $\Sigma_{\tilde\theta}\times M$ with coordinates  $(\zeta,y)$, and consider the map $(\zeta,y)\mapsto \kappa^\theta_{\zeta,y}\in L^2(M)$. Since $\langle\kappa^\theta_{\zeta ,y}| \psi\rangle_{L^2(M)}=(\exx{\zeta\dth}\psi)(y)$
is smooth in $(\zeta ,y)\in \mcO$ for every $\psi\in L^2(M)$, the map is weakly $C^\infty$, and hence strongly $C^\infty$ by the Lemma \ref{lmsmooth2}. 
Similarly, choosing another chart $\mcO'$ of $\Sigma_{\tilde\theta}\times M$ with coordinates  $(\xi,y')$, define the map
\begin{align}\label{e.sm12a}
 \mcO\times  \mcO'\ni	(\zeta,y;\xi,y')\mapsto \ip{\kappa_{\zeta ,y}^\theta}{\kappa_{\xi^\ast,y'}^{\pi-\theta}}_{L^2(M)}\in \bbC \,.
\end{align}
This is jointly smooth in $(\zeta,y;\xi,y')$ since 
 the maps $(\zeta ,y)\mapsto \kappa^\theta_{\zeta,y}$ and $(\xi,y')\mapsto \kappa^{\pi-\theta}_{\xi^\ast ,y'}$ are strongly $C^\infty$. Since the charts were arbitrary, $(\Sigma_{\tilde\theta}\times M)^2\ni (\zeta,y;\xi,y')\mapsto \langle\kappa^\theta_{\zeta ,y}| \kappa_{\xi^\ast,y'}^{\pi-\theta}\rangle_{L^2(M)}$ is smooth. Restricting to the diagonal in kernel time yields  joint $C^\infty$-smoothness of
 \begin{align}\label{e.sm12b}
	\Sigma_{\tilde\theta}\times M\times M\ni (\zeta,y,y')\mapsto \ip{\kappa_{\zeta/2
	,y}^\theta}{\kappa_{\zeta^\ast/2 ,y'}^{\pi-\theta}}_{L^2(M)}= K_\zeta^\theta(y,y')\in \bbC\,.
\end{align}
Finally, the analyticity in $\zeta$ is a by-product of Theorem \ref{thsmooth1}(i) and equation \eqref{e.sm14} below. 

\medskip
 
(iii)  Next, fixing $\zeta\in \Sigma_{\tilde\th}$, the uniqueness of the kernel $K_\zeta^\theta\in C^\infty(M\times M) $ is implied by the uniqueness (for each $y\in M$) of  $\kappa_{\zeta,y}^\theta\in L^2(M)$ from Lemma \ref{lmsmooth1}. In more detail, fixing $y\in M$, and comparing the realizations of $(\exx{\zeta\dth}\psi)(y)$ in terms of $\kappa_{\zeta,y}^\theta$ via Lemma \ref{lmsmooth1} and the kernel $K_\zeta^\theta$ in \eqref{e.sm7}, yields $\kappa_{\zeta,y}^\theta(y')^\ast =K_\zeta^\theta(y,y')$ for a.e.~$y'\in M$. The same must be true of any other kernel $J_\zeta^\theta\in C^\infty(M\times M) $ satisfying \eqref{e.sm7}, i.e.~for each $y\in M$, they coincide a.e. $y'\in M$. Then smoothness implies they coincide everywhere, i.e. $J_\zeta^\theta=K_\zeta^\theta$.

\medskip

(iv)  To prove the semigroup identity, fix arbitrary $\zeta\in \Sigma_{\tilde\theta}$ and $y,y'\in M$. As discussed in (iii) above, for any $z\in M:\kappa_{\zeta,z}^\theta(z')^\ast =K_\zeta^\theta(z,z')$ for a.e. $z'\in M$. Then, it is   sufficient to prove that for $\zeta_1,\zeta_2\in \Sigma_{\tilde\theta}$ 
\begin{align}\label{e.sm13}
\forall\,y,y'\in M:\,\,	K_{\zeta_1+\zeta_2}^\theta(y,y')=\ip{\kappa_{\zeta_1,y}^\theta}{\kappa_{\zeta_2^\ast,y'}^{\pi-\theta}}_{L^2(M)}\,.
\end{align}
 Indeed, from the definition of the kernel \eqref{e.sm11} and the computation \eqref{e.sm11a}, it is immediate that for each $y\in M:\,K_{\zeta_1+\zeta_2}^\theta(y,y')=\langle\kappa_{\zeta_1,y}^\theta|\kappa_{\zeta_2^\ast,y'}^{\pi-\theta}\rangle_{L^2(M)}$ a.e. $y'\in M$. However, both sides are smooth in $y'\in M$ by the above regularity results \eqref{e.sm12a}, \eqref{e.sm12b}, and hence are equal for all $y'\in M$. This proves \eqref{e.sm13}, and hence the semigroup property.

\medskip

(v) To see that for each $y'\in M$, $(\pp_\zeta-\Delta_{\theta,y})K_\zeta^\theta (y,y')=0 $ in $\Sigma_{\tilde\theta}\times M$, fix an arbitrary $\xi\in \Sigma_{\tilde\theta}$ such that $\zeta-\xi\in \Sigma_{\tilde\theta}$ and consider $f(\zeta,y):=\langle \kappa_{\zeta
	,y}^\theta| \kappa_{\xi^\ast ,y'}^{\pi-\theta}\rangle _{L^2(M)} = (\exx{\zeta\dth}\kappa_{\xi^\ast  ,y'}^{\pi-\theta})(y)$. Then Theorem \ref{thsmooth1}(i) implies  $(\pp_\zeta-\Delta_\theta)f (\zeta,y)=0$, and hence the same holds for $f(\zeta-\xi,y)$ since  $\zeta-\xi\in \Sigma_{\tilde\theta}$. The  semigroup identity \eqref{e.sm13} entails that 
\begin{align}\label{e.sm14}
	f(\zeta-\xi,y)=\ip{\kappa_{\zeta-\xi
	,y}^\theta}{\kappa_{\xi^\ast ,y'}^{\pi-\theta}}_{L^2(M)}=K_\zeta^\theta(y,y')\,,
\end{align}
yielding \eqref{e.sm8a}. Finally, the limit statement \eqref{e.sm10} is an immediate corollary of Theorem \ref{thsmooth1}(iii), completing the proof.
\qed
\bigskip

The above results provide a generalized heat kernel methodology that
remains well defined into the near Lorentzian regime, $1\gg \th >0$. 
We add some comments on the motivation and the relation to the
Hadamard parametrix. 

{\bf Remarks.} 

(i) As mentioned in the introduction, formal series obtained  
from the heat kernel expansion by ad-hoc substitutions
($s \mapsto i \tilde{s}$, $\tilde{s} \in \R$, $g^+ \mapsto g^-$)  
are a widely used computational tool in theoretical physics
\cite{Parkerbook, heatkoffdiag1}. These series are intended to
capture aspects of an evolution group $\R \ni \tilde{s} \mapsto
e^{ -i \tilde{s} \cD_-}$ somehow associated with the heat semigroup
$\R_+ \ni s \mapsto e^{ - s \cD_+}$. Although this lacks
justification on the group level the formal series are not without
indirect rationale: a schematic ansatz for the parametrix of the
putative $e^{ -i \tilde{s} \cD_-}$ group is made in terms of the
Synge function and its derivatives. The Synge function is locally well-defined
on a Lorentzian manifold and the recursion relations for the coefficients
in the ansatz are virtually the same as in the Riemannian case
\cite{heatkoffdiag1}. The solution formulas likewise have no explicit
dependence on the signature parameter. These off-diagonal pseudo-heat kernel
coefficients are moreover in one-to-one correspondence to those occurring
in the (Lorentzian signature) Hadamard expansion \cite{heatkoffdiag1},
so there is little doubt that these coefficients correctly reflect the
short distance behavior of the Hadamard parametrix. While the Hadamard
parametrix can be defined independently of these series expansions
\cite{Wavebook,Synge2},
the use of the inverse Laplace transform to define the corresponding
part of a pseudo-heat kernel is not immediate.

(ii) A rigorous result on a near Lorentzian Hadamard expansion is
\cite{FinsterReg}, where the $i\eps$ part of the Hadamard parametrix is
kept finite and local. Starting from the original, $i\eps$-independent
wave equation the regularization terms are shown to be governed by
recursive relations analogous to the standard ones and the existence of
a deformed parametrix is shown. The relation to a Euclidean 
regime is not discussed. 

(iii) One way to link the Euclidean heat kernel to the Lorentzian signature
Hadamard parametrix is by subsuming both in a setting that invokes complex
analytic metrics \cite{Moretti1}. This allows for a local Wick rotation
(in time), which is however potentially  coordinate dependent. This
framework can be used to prove the symmetry of the coefficients to all
orders by transitioning between different real sections. It also leads to
a natural notion of analytic Hadamard states \cite{AnalyticHad} that extends
beyond the stationary case \cite{WickWrochna}. On the other hand, a
Wick rotation in time may be limited to metrics with a purely electric
Riemann tensor \cite{Wickelectric}.

%%%%%%%%%%%%%%%%%%%
%	New section 
%%%%%%%%%%%%%%%%%%%
\newpage 

\section{Asymptotic expansion of the kernel's diagonal}
\label{Sec5}

Much of the computational uses of the heat kernel rest on
it's asymptotic expansion for small diffusion time. The tabulated
low order coefficients for the diagonal's expansion provide a shortcut
to many otherwise difficult computations on curved backgrounds, see
e.g.~\cite{Avramidibook,Getzlerbook,Waldheatk,Morettiheatk,Parkerbook}.
A justifiable near Lorentzian counterpart of such expansions 
would be desirable (see remark (i) at the end of Section 3)   
and is a by-product of the main result of this section. 
We first present the result and comment on the rationale of the
proof strategy. In section 4.1 we introduce and study the
relevant parametrix before turning to the bounds in Section 4.2.

\begin{theorem}  \label{asymthm}
  The diagonal of the kernel $K_{\zeta}^{\th}$ from Theorem
  \ref{thsmooth2} admits an asymptotic expansion of the form 
\begin{align}
\label{hkasymp}   
K_{\zeta}^{\th}(y,y) \asymp
\frac{ (-i e^{ i\th})^{\frac{d-1}{2}}}{( 4 \pi \zeta)^{\frac{d+1}{2}}} 
\sum_{n \geq 0} A_n^{\th}(y) ( i e^{- i \th} \zeta)^n,
\end{align}
i.e. for all $N\in \bbN$  and $y\in M$
\begin{align}
	(4 \pi \zeta)^{\frac{d+1}{2}}K_{\zeta}^{\th}(y,y)-(-i e^{ i\th})^{\frac{d-1}{2}}\sum_{n = 0}^N A_n^{\th}(y) ( i e^{- i \th} \zeta)^n = O(\zeta^{N+1})
\end{align}
as $\zeta\to 0^+$.
Here, the coefficients $A_n^{\th}(y)$ from (\ref{gpara2a}) are in each 
chart map the standard heat kernel coefficients evaluated on the lapse-Wick-rotated
metric $g^{\th}$.  
\end{theorem}

The proof strategy will have to differ significantly from
standard ones, for reasons we outline now. 
For the heat kernel proper the off-diagonal expansion is usually
formulated in terms of the Synge function (one-half of the square
of the geodesic distance between nearby points). In the present setting the
correctly lapse-Wick-rotated form of the Eikonal equation reads   
\begin{equation} 
\label{weikonal1}
\sigma_{\th}(y,y') = \frac{i e^{- i \th}}{2}  g_{\th}^{\mu\nu}(y)
\dd_{\mu} \sigma_{\th} \dd_{\nu} \sigma_{\th}\,,
\end{equation}
which characterizes the Synge function $\sigma_{\th}$ subject to suitable
boundary conditions. Throughout this section we write $g^{\th}_{\mu\nu}(y)$ for the
components of a lapse-Wick-rotated metric $g^{\th} = g^{\th}_{\mu\nu}(y) dy^{\mu}dy^{\nu}$
and $g_{\th}^{\mu\nu}$ for the components of the inverse metric. In some
reference foliation these components can be read off from (\ref{i1}).     
We omit a derivation as (\ref{weikonal1}) will occur
later as a byproduct of the ansatz for $K_{\zeta}^{\th}(y,y')$ used.
Note that (\ref{weikonal1}) is not symmetric in $y,y'$ while the
solution aimed at is. The relevant boundary conditions are
$\sigma_{\th}(y,y) =0$, $\nabla^{\th}_{\mu} \nabla^{\th}_{\nu}
\sigma_{\th}|_{y = y'} = - i e^{i\th} g^{\th}_{\mu\nu}$,
where $\nabla_{\mu}^{\th}$ is the metric connection built from $g^{\th}$
(acting on the first argument of $\sigma_{\th}$). 
The following proposition summarizes a Wick rotated version of the usual
starting point to establish the existence of an asymptotic expansion. 
\pagebreak[4]

\begin{proposition} Let $\sigma_{\th}(y,y')$ be an exact, symmetric solution
  of the Eikonal equation (\ref{weikonal1}) with boundary conditions
  $\sigma_{\th}(y,y) =0$, $\nabla^{\th}_{\mu} \nabla^{\th}_{\nu}
\sigma_{\th}|_{y = y'} = - i e^{i\th} g^{\th}_{\mu\nu}$. Then 
\begin{align}
\label{hkexp1}   
K_{\zeta}^{\th}(y;y') = \frac{ (-i e^{ i\th})^{\frac{d-1}{2}}}%
    {( 4 \pi \zeta)^{\frac{d+1}{2}}} e^{ - \frac{1}{2 \zeta} \sigma_{\th}(y,y')}
\sum_{n \geq 0} A_n^{\th}(y,y') ( i e^{- i \th} \zeta)^n,
\end{align}
provides a formal solution to the heat equation $(\dd_{\zeta} - \Delta_{\th})
K^{\th}_{\zeta} =0$, iff the coefficients $A_n^{\th}$
satisfy 
\ba
\label{hkexp2} 
&& 2 g_{\th}^{\mu\nu} \dd_{\mu} \sigma_{\th} \dd_{\nu} A_0^{\th}
+ [ \nabla_{\th}^2 \sigma_{\th} + i e^{ i \th} (d\!+\!1)] A_0^{\th} =0\,,
\\[2mm] 
&& 2 g_{\th}^{\mu\nu} \dd_{\mu} \sigma_{\th} \dd_{\nu} A_{n+1}^{\th}
+ \big[\nabla_{\th}^2 \sigma_{\th}
  - \frac{i e^{i \th}}{2} (2 n\!+\!1\!-\!d) \big] A_{n+1}^{\th}
= - e^{ 2 i \th} \Delta_{\th} A_n^{\th} \,, \quad n \geq 0\,.
\nonumber
\ea
The overall normalization is fixed by $A_0(y,y) =1$. Further,  
$\nabla_{\mu}^{\th}$ is the metric connection associated with $g^{\th}$,
and $\nabla_{\th}^2 = g^{\mu\nu}_{\th} \nabla^{\th}_{\mu} \nabla^{\th}_{\nu}$.
\end{proposition} 

{\it Proof.} Direct substitution and comparing powers of $\zeta$. \qed

The recursion (\ref{hkexp2}) generalizes the well-known one for the
off-diagonal heat kernel coefficients $A_n$
(see e.g.~\cite{FullingQFT}, Eq.~9.9a), and specializes to them for
$\th = \pi/2$. For Euclidean signature $A_0$ is expressible
in terms of the Van Vleck-Morette determinant and the recursion
can be solved via the method of characteristics. The construction
masks the symmetry of the coefficients, which has to be established
independently \cite{Moretti1}.

Even with the symmetry of the coefficients ensured, the Euclidean
signature series in (\ref{hkexp1}) may not be properly asymptotic to
the exact heat kernel \cite{Moretti1}. A complete and direct proof
for the existence of a small time asymptotic expansion for the
off-diagonal heat kernel has only been given recently in
\cite{Ludewig}, building on \cite{Kannai} and assuming compactness
of the manifold.  The methodology in
\cite{Ludewig,Kannai} does not make use of the transport equations
(\ref{hkexp2}) and is therefore somewhat decoupled from the way the
(Euclidean signature version of the) ansatz (\ref{hkexp1}) enters  
computations. It is also not immediate that the methodology
would carry over to the lapse-Wick-rotated situation. Since in applications
mostly the diagonal expansion is needed, we aim in the following only
at a small $\zeta$ asymptotic expansion for the diagonal of
our kernel $K_{\zeta}^{\th}(y,y')$. By modifying a proof strategy 
 proposed in \cite{Waldheatk}, and approximating
the solutions of the transport equations via Proposition \ref{syngeseries}
a fairly elementary proof for the existence of the diagonal expansion
can be given that retains contact to (\ref{hkexp1}), (\ref{hkexp2}). This 
is possible because we have independently established the existence
of our kernel $K_{\zeta}^{\th}(y,y')$ with the requisite properties.
In contrast, the approaches \cite{Ludewig,Kannai,Grieser,Waldheatk} aim at  
establishing the kernel's existence through a sequence of
approximants that are asymptotic a-posteriori.  

%%%%%%%%%%%%%%%%%%%%%%%%%%%%%%%%%%%%%%%%%%%%%%%%%%%%%%%%%%%%
\subsection{Local and global parametrix} 

While the Synge function is a natural geometric quantity
it is itself nontrivial to construct, both mathematically \cite{Synge1,Synge2}
and computationally. In our Wick rotated setting the standard local existence
proofs carry over straightforwardly only if the metric is assumed to be
locally analytic. This is somewhat at odds with our framework and
presumably is also not a necessary assumption. We therefore postpone a
detailed investigation of the Wick rotated Synge function to another
occasion and limit ourselves here to an explicitly constructable
asymptotic solution. 
%\pagebreak[4] 

\begin{proposition} \label{syngeseries} Let $g^{\th}$ be a lapse-Wick-rotated
  smooth metric (\ref{i1}) on $M$ and fix a local chart $U \subset M$, which in
  local coordinates can be identified with an open subset of $\R^{d+1}$.
  Define for any $\sigma\in C^\infty(U\times U)$ and $y,y' \in U$,
  $\cE_{\th}[\sigma](y,y') :=
  \sigma(y,y') - \frac{i}{2} e^{- i \th} g_{\th}^{\mu\nu}(y)
  \dd_{\mu} \sigma \dd_{\nu} \sigma$, as well as
  $\Delta y^{\mu} := (y - y')^{\mu}$ and $|\Delta y| := (\delta_{\mu\nu}
\Delta y^{\mu}\Delta y^{\nu})^{1/2}$.
\begin{itemize}
\item[(a)]
Inserting the (truncated) formal series 
\begin{equation}
\label{weikonal2}
\sigma_\theta (y,y') = - i e^{i\th}
\sum_{n \geq 2} \frac{1}{n!} s_{\mu_1 \ldots \mu_{n}}(y)
\Delta y^{\mu_1} \ldots \Delta y^{\mu_n} \,,
\end{equation} 
into $\cE_{\th}[\sigma] \equiv 0$, 
the completely symmetric coefficients $s(y)$ are
uniquely determined differential polynomials in $g_{\mu\nu}^{\th}$
contracted with $g_{\th}^{\rho\sigma}$ at $y$. 
\item[(b)]
There exists a (possibly non-unique) smooth function
$\tilde{\sigma}_{\th}(y,y')$, which in a $\Delta y$ expansion has
Taylor coefficients $-i e^{i \th}s_{\mu_1 \ldots \mu_n}(y)$, such for
all $N \geq 1$
\begin{equation}
\label{weikonal3}
\big| \cE_{\th}[ \tilde{\sigma}_{\th}](y,y') \big| \leq C_N |\Delta y|^N\,,
\end{equation}
for some constants $C_N>0$ and all $y,y'$ in an open neighborhood
$V \subset U$.  
\item[(c)] There exists $\th$ dependent constants
  $0< c_- \leq  c_+ < \infty$
  and an open neighborhood $V \subset U$ such that for all
  $y,y' \in V$
  \begin{eqnarray}
    \label{weikonal4} 
    c_- |\Delta y|^2 \leq {\rm Re}\,\tilde{\sigma}_{\th}(y,y')
    \leq c_+ |\Delta y|^2 \,.
  \end{eqnarray}
\end{itemize} 
\end{proposition}

{\it Proof.} (a) Technically, it is convenient to insert an ansatz
akin to (\ref{weikonal2}) but with coefficients evaluated at $y'$
into into $\cE[\sigma]= 0$. By comparing powers of $\Delta y$ 
then only algebraic recursion relations arise. Moreover, these coefficients
at $y'$ are differential polynomials in $g^{\th}_{\mu\nu}$ with at most $n-2$
differentiations at order $n$, contracted with $g^{\rho\sigma}_{\th}$.
By Taylor expanding the coefficients at $y' = y - \Delta y$ around
$y$ the expansion (\ref{weikonal2}) arises, where each $s(y)$
coefficient is a finite linear combination of the algebraically
determined coefficients at $y'$. 

(b) By the assumed smoothness of the metric there exist a
precompact $V_1 \subset U$ such that $g_{\mu\nu}^{\th}$, $g_{\th}^{\mu\nu}$,
and all its derivatives are bounded on $V_1$. By (a) the coefficients 
$s(y)$ can be bounded in terms of the Taylor coefficients of the
metric at $y$.  For arbitrary $N \geq 2 $, the remainders of the partial sums,
$n \leq N\!-\!1$, can thus be bounded by a constant times
a $O(\Delta y^N)$ contraction. For some open $V \subset V_1$
this gives a bound $C_N |\Delta y|^N$, for all $y,y' \in V$. 
Borel's lemma then provides the existence of $\tilde{\sigma}_{\th}$
satisfying (\ref{weikonal3}). 

(c) We fix again some precompact set
$V_1 \subset U$ in which the components of $g_{\mu\nu}^{\th}$,
$g_{\th}^{\mu\nu}$, and all derivatives thereof are bounded.
Defining
\ba
\label{length} 
&& c_- := \frac{1}{2} \inf_{y \in V_1}  
\inf_{ \Delta y \in \R^{d+1},|\Delta y| =1}
{\rm Re}\Big(\! - i e^{i \th} g_{\mu\nu}^{\th}(y) 
\Delta y^{\mu} \Delta y^{\nu} \Big)\,,
\nonum
&& c_+ := \frac{1}{2} \sup_{y \in V_1} 
\sup_{ \Delta y \in \R^{d+1},|\Delta y| =1}
{\rm Re}\Big(\! - i e^{i \th} g_{\mu\nu}^{\th}(y) 
\Delta y^{\mu} \Delta y^{\nu} \Big)\,,
\ea
one has $0 < c_- < c_+< \infty$. The function $V_1 \times V_1 \ni (y,y')
\mapsto  \tilde{\sigma}_{\th}(y,y') - c_- |\Delta y|^2$ 
is smooth and vanishes on the diagonal.
By (a), (b) one also has $2 {\rm Re}\tilde{\sigma}_{\th}(y,y') =
{\rm Re}(\! - i e^{i \th} g_{\mu\nu}^{\th} 
\Delta y^{\mu} \Delta y^{\nu}) + O(|\Delta y|^3)$ near the diagonal.
It follows that ${\rm Re}\,\tilde{\sigma}_{\th}(y,y')
- c_- |\Delta y|^2 >0$ in some open
neighborhood of the diagonal in $V_1 \times V_1$. This implies that
there is an open neighborhood $V \subset V_1$ such that
${\rm Re}\,\tilde{\sigma}_{\th}(y,y') >
c_- |\Delta y|^2$, for all $y,y' \in V$.
The upper bound in the first equation of (\ref{weikonal4}) is obtained
analogously.
\qed

\medskip

To low orders the coefficients read  
\ba
\label{weikonal5} 
s_{\mu_1 \mu_2} \is g^{\th}_{\mu_1\mu_2}\,, \quad 
s_{\mu_1 \mu_2\mu_3} =
- \frac{3}{2} \dd_{\mu_1} g^{\th}_{\mu_2 \mu_3}\,,
\nonum
s_{\mu_1 \mu_2\mu_3 \mu_4} \is 2 \dd_{\mu_1} \dd_{\mu_2}
g^{\th}_{\mu_3 \mu_4} -
\frac{1}{4} g_{\th}^{\rho\sigma}
\dd_{\rho} g^{\th}_{\mu_1 \mu_2} \dd_{\sigma} g^{\th}_{\mu_3 \mu_4}
\nonum
&+&
g_{\th}^{\rho\sigma}
\dd_{\rho} g^{\th}_{\mu_1 \mu_2} \dd_{\mu_3} g^{\th}_{\sigma \mu_4}
- g_{\th}^{\rho\sigma}
\dd_{\mu_1} g^{\th}_{\rho \mu_2} \dd_{\mu_3} g^{\th}_{\sigma \mu_4}\,,
\ea 
where symmetrization in all indices is understood. An analogous
expansion exists with coefficients $\mathfrak{s}_{\mu_1 \ldots \mu_n}$   
evaluated at $y'$ rather than $y$. A-posteriori one then finds 
$\mathfrak{s}_{\mu_1 \ldots \mu_n} = (-)^ns_{\mu_1 \ldots \mu_n}$, $n \geq 2$,  
indicating that the directly computed coefficients are compatible
with the symmetry of the solution, $\sigma_{\th}(y,y') = \sigma_{\th}(y',y)$.  
For the present purposes symmetry of  
the truncated series (\ref{weikonal2}) will not be needed.

{\bfseries Local parametrix.} For simplicity, we take the heat kernel time
$\zeta$ in the following to be real (and positive) and comment
on the extension of Theorem \ref{asymthm} to
$|{\rm Arg}\zeta| < \tilde{\th}$ at the end. 
Given an arbitrary $y_0\in M$, an open chart neighborhood $U\ni y_0$, and a truncation order $N\in \bbN$,  we initially consider a {\it local} parametrix defined in $U\times U$
of the form 
\begin{align}
\label{hkexp4}   
F_{\zeta}(y,y') :=  
\frac{ (-i e^{ i\th})^{\frac{d-1}{2}}}%
{( 4 \pi \zeta)^{\frac{d+1}{2}}} e^{ - \frac{1}{2 \zeta} s_{\th}(y,y')}
\sum_{n =0}^N A_n^{\th}(y,y') ( i e^{- i \th} \zeta)^n\,.
\end{align}
Later on $s_{\th}$ will be identified with the function
$\tilde{\sigma}_{\th}(y,y')$ from Proposition \ref{syngeseries};
for now we only need $s_{\th}(y,y')$ to be jointly
smooth in $\zeta,y,y'$, for $y,y' \in U$  and such that
$\Re s_{\th} >0$ in a neighborhood
of $y=y'$. Importantly, $s_{\th}$ is neither necessarily symmetric nor
a solution of the Eikonal equation (\ref{weikonal1}).
Similarly, the $A_n^{\th}$ are initially only assumed to be smooth
in $y,y' \in U$ and nonzero in a neighborhood of $y=y'$.  To unclutter the notation we suppress the $N$ and $\th$
dependence in the notation.

To proceed, we act with the heat operator $\dd_{\zeta} - \Delta_{\th}$ on
(\ref{hkexp4}). For convenience we recall the definitions,
$\Delta_{\th} = i e^{- i \th} (\nabla_{\th}^2 - V)$,
$\nabla_{\th}^2 = g_{\th}^{\mu\nu} \nabla^{\th}_{\mu} \nabla^{\th}_{\nu}$. 
A straightforward computation gives
\be
\label{hkexp5} 
\!\!\big[\dd_{\zeta}- \Delta_{\th,y} \big] F_{\zeta}(y,y') = 
\frac{ (-i e^{ i\th})^{\frac{d-1}{2}}}%
{( 4 \pi \zeta)^{\frac{d+1}{2}}} e^{ - \frac{1}{2 \zeta} s_{\th}(y,y')}
%\nonum
%&\times & 
\sum_{n =0}^N ( i e^{- i \th} \zeta)^n
\Big[ \frac{X_n}{( i e^{- i \th} \zeta)^2 }  +
\frac{Y_n}{i e^{- i \th} \zeta }  + Z_n \Big],
\ee
where 
\ba
\label{hkexp6} 
X_n \is \frac{1}{2} (i e^{- i \th})^2 \Big( s_{\th} - \frac{1}{2}
i e^{ - i \th} g_{\th}^{\mu\nu} \dd_{\mu} s_{\th} \dd_{\nu} s_{\th} \Big)
A_n^{\th} \,,
\nonum
Y_n \is \frac{1}{2} i e^{- i \th} \big[ (2 n - (d\!+\!1)) \1 +
\mcL(s_{\th}) \big] A_n^{\th}\,, \quad Z_n = - \Delta_{\th} A_n^{\th}\,,
\nonum
\mcL(s_{\th}) \is i e^{- i \th} \nabla^2_{\th} s_{\th} + 2 i
e^{ - i \th} g_{\th}^{\mu\nu} \dd_{\mu} s_{\th} \dd_{\nu}\,.
\ea 
The structure of the $X_n$ term elucidates the origin of the Eikonal
equation (\ref{weikonal1}). Clearly, $X_n =0$ for all $n \geq 0$
iff (for at least one nonzero $A_{n_0}^{\th}$) $s_{\th} = \sigma_{\th}$ solves
(\ref{weikonal1}). Further, for $X_n =0$, $n \geq 0$, the
remaining terms in the second line reduce to $(i e^{-i\th} \zeta)^N Z_N$
iff $Y_0 =0$ and $Y_{n+1} + Z_n =0$, $n=0, \ldots,N\!-\!1$.
These vanishing conditions are precisely the $n=0, \ldots,N\!-\!1$
transport equations in (\ref{hkexp2}).

The strategy used later on will be to impose {\it approximate}
vanishing conditions for the $X_n$, $n \geq 0$, and $Y_{n+1}
+ Z_n, n =0, \ldots, N\!-\!1$, in an expansion in powers of $\Delta y$.  
This needs to be done in a way that is coordinated with the
order of the expansion in $\zeta$. To do so, we interpret (and
eventually construct) the functions $s_{\th}(y,y'), A_n^{\th}(y,y'),
F_{\zeta}(y,y')$ as truncated series in $\Delta y$ with coefficients
evaluated at $y$. Our convention throughout this section will be that
the first argument indicates the base point at which the coefficients
are evaluated, $f(y,y') = f(y) + f_{\mu}(y) \Delta y^{\mu} +
O(|\Delta y|^2)$, etc.. The remainders in these expansions will be
of $O(|\Delta y|^L)$, for suitable powers $L$.
For the next steps, we assume only that $s_{\th}$ has a leading term
given by the one fixed by Proposition \ref{syngeseries}, i.e.
\begin{equation}
\label{sleading} 
s_{\th}(y,y') = - \frac{i}{2} e^{ i \th}
g^{\th}_{\mu\nu}(y) \Delta y^{\mu} \Delta y^{\nu} + O(|\Delta y|^3)\,,
\end{equation}
 that $A_n^{\th}(y,y')$ is regular and nonzero as $|\Delta y| \ra 0$, and in particular $A_0^{\th}(y,y)=1$.
Before turning to the $\Delta y$ expansion, we prepare a key property
of the local parametrix. All function spaces occurring in the following
are subspaces of $L^2(M)$ and $\Vert \cdot\Vert$ will always
refer to the $L^2(M)$ norm.

\begin{lemma} \label{approxdelta} 
Let $F_{\zeta}(y,y)$, $\zeta >0$,
  be the local parametrix (\ref{hkexp4}) defined in an open chart neighborhood $U$ of the point $y_0\in M$. Then there is a precompact open neighborhood $V\Subset U$ of  $y_0$  such that :
 
\begin{itemize}   
\item[(a)] As a distribution on $C^{\infty}_c(V)$ test functions,
\begin{equation}
\label{lim1}
\lim_{\zeta \ra 0^+} F_{\zeta}(y,y') =
|g|^{-1/2}(y)\, \delta(y-y') \,.
\end{equation}
\item[(b)] Let $\mathbb{F}_{\zeta}$, $\zeta >0$, be the integral operator
  on $L^2(V)$ defined by the kernel $F_{\zeta}$. Then
\begin{equation}
  \label{lim2}
  \lim_{\zeta \ra 0^+} \Vert \mathbb{F}_{\zeta} \psi \Vert
  = \Vert \psi \Vert \,, \quad \psi \in L^2(V)\,. 
  \end{equation}
\end{itemize} 
\end{lemma} 

{\it Proof.} (a) It follows from (\ref{sleading}) that $y_0$ has a precompact open neighborhood $ V\Subset U$ where 
$|2 s_{\th}(y,y') + i e^{i\th} g_{\mu\nu}(y) \Delta y^{\mu} \Delta y^{\nu}|
< c_- |\Delta y|^2$, for all $y,y' \in V$. As in the proof
of Proposition \ref{syngeseries}(c) this implies 
\begin{equation}
\label{lim3} 
\frac{c_-}{4}  |\Delta y|^2 \leq  
\frac{1}{2 \zeta} {\rm Re}
\big[s_{\th}\big(y, y \!-\! \sqrt{\zeta} \Delta y\big) \big]\,,
\end{equation} 
 for all $y,y' \in V$, as needed shortly. We express (\ref{lim1})
 in terms of 
the integral operator $\bbF_{\zeta}$ with kernel $F_{\zeta}(y,y')$.
Changing integration variables
from $y'$ to $\Delta y$, its action on a $C_c^{\infty}(V)$
function $\psi$ can be written as
\ba
\label{lim4}
&& (\bbF_{\zeta} \psi)(y) = 
\frac{ (-i e^{ i\th})^{\frac{d-1}{2}}}{( 4 \pi\zeta)^{\frac{d+1}{2}} }
\sum_{n =0}^N ( i e^{- i \th} \zeta)^n \!
\int\! d^{d+1} (\Delta y) f_{\zeta, n}(y, \Delta y) \,,
\\[2mm] 
&&  f_{\zeta,n}(y, \Delta y) = |g|^{1/2}(y \!-\! \Delta y)
e^{ - \frac{1}{2 \zeta} s_{\th}(y,y-\Delta y)}
A_n^{\th}(y, y\!-\!\Delta y)
\psi(y \!-\! \Delta y) \,. 
\nonumber
\ea
Next, we rescale the integration variable
$\Delta y \mapsto \zeta^{1/2} \Delta y$. In the prefactor
before the sum this cancels the $\zeta^{- \frac{d+1}{2}}$,
leaving only non-negative $\zeta$ powers in front of the
$d^{d+1}\Delta y$ measure. The precompactness of $V$ entails that the  integration ranges over a bounded domain, say $|\Delta y| \leq \delta$. 
After the rescaling one has $|\Delta y| \leq \delta/\sqrt{\zeta}$,
so for $\zeta \ra 0^+$ the integration will effectively
extend over all of $\R^{d+1}$. Since the $A_n^{\th}$
are smooth on $U\times U$, each may be bounded by its supremum over the precompact $V\times V$.
Then (\ref{lim3}) can be used to bound for each $n$ the integrand
by a  constant times $\exp\{ - c_- |\Delta y|^2/4\}$, which is of
course integrable.  
Applying dominated convergence the $\zeta \ra 0^+$ limit can be
brought inside the integral. Pointwise in $\Delta y$ the $n$-th
integrand converges to
\begin{equation}
\label{lim5} 
|g|^{1/2}(y)A_n(y, y) \psi(y) \exp\Big\{
\frac{i}{4} e^{i \th} g^{\th}_{\munu}(y) \Delta y^{\mu} \Delta y^{\nu} \Big\}\,.
\end{equation} 
The $d^{d+1} \Delta y$ integral is a well-defined Gaussian,
as $\th \in (0,\pi)$. Performing it cancels the 
prefactors of the $\sum_{n=0}^N (i e^{- i \th} \zeta)^n$  sum.
When taking the $\zeta \ra 0^+$ limit in (\ref{lim4}) only the $n=0$ term
in the sum contributes and one arrives at $\lim_{\zeta \ra 0^+}
( \bbF_{\zeta} \psi)(y) = \psi(y)$, using $A^{\th}_0(y,y) =1$.  This gives (\ref{lim1}) as a distribution
on $C^{\infty}_c(V)$ test functions. 

(b) Using (\ref{lim4}) and rescaling both difference integration
variables gives
\ba
\label{lim7}
&\nspace & \int\! dy \,|g|^{1/2}(y)\,| \mathbb{F}_{\zeta}\psi|(y)^2 =
\frac{1}{(4\pi)^{d+1}} \sum_{m,n = 0}^{N} \zeta^{n+m}
\int\! d^{d+1} y \,|g|^{1/2}(y) \int\! d^{d+1} \Delta y \int\! d^{d+1} \Delta z
\nonum
&& \quad \times
f_{\zeta,m}(y, \sqrt{\zeta} \Delta z)^* f_{\zeta,n}(y, \sqrt{\zeta} \Delta y) \,.
\ea
The integration in $y$ extends over a bounded domain $V$, while after the rescaling the integration
domain over the difference variables $\Delta y$, $\Delta z$
will tend to $\R^{d+1}$, as $\zeta \ra 0^+$. 
Pointwise in $\Delta y$, $\Delta z$, the integrand converges for
$\zeta \ra 0^+$ to the product of (\ref{lim5}) and a complex
conjugate copy with $n \mapsto m, \Delta y \mapsto \Delta z$. By an argument analogous to that in the proof of part (a) above,  the
integrand is bounded for each $n,m$ by a constant times 
\begin{equation}
\label{lim8} 
|g|^{1/2}(y) |\psi(y)|^2
\exp\{ - c_- |\Delta y|^2/4\}\exp\{ - c_- |\Delta z|^2/4\}\,.
\end{equation}
This is because $\psi$ has compact support, so for fixed $y$ the
$\psi(y - \sqrt{\zeta} \Delta y)$ term can be bounded by its finite
supremum over $\Delta y \in \R^{d+1}$, and similarly for
$\psi(y - \sqrt{\zeta} \Delta z)^*$.  Since the $y$ integration
domain is bounded the expression (\ref{lim8}) is integrable with
respect to the measure in (\ref{lim7}). This allows one to bring the $\zeta
\ra 0^+$ limit inside the triple integral. Taking the $\zeta \ra 0^+$ limit
in (\ref{lim7}) only the $n=m=0$ term is nonzero and results in
\begin{equation}
\lim_{\zeta \ra 0^+}\nrm{\mathbb{F}_{\zeta} \psi}  
= \frac{1}{(4\pi)^{d+1}} \int\! d^{d+1} y |g|^{3/2}(y) |\psi(y)|^2 
\bigg| \int\! d^{d+1} \Delta y \, e^{ \frac{i e^{i\th}}{4}
g^{\th}_{\mu\nu}(y) \Delta y^{\mu} \Delta y^{\nu} } \bigg|^2.
\end{equation} 
The modulus-square of the Gaussian integral evaluates to $(4\pi)^{d+1}/
|g|(y)$, which yields (\ref{lim2}). 
\qed
\medskip

{\bf Global parametrix.} Later on we seek to compare the parametrix
$F_{\zeta}$ to our globally defined kernel $K_{\zeta}$. To this end
an extension of the local parametrix (\ref{hkexp4}) to all of
$M \times M$ is needed; in particular, we want a parametrix that
does not just vanish along the diagonal outside a compact region.

The extension is done using a partition of unity subordinate to
a suitable open cover of $M\times M$. We have already shown through Lemma \ref{approxdelta} that every point in $M$ has an open chart neighborhood on which a local parametrix with properties \eqref{lim1}, \eqref{lim2} can be defined. 
Since $M$ is assumed to have a countable
topological base, it can be covered by countably many such  open chart
neighborhoods  
$(V_l)_{l \in \N}$, where we may assume that no $V_l$ is contained in
any $V_{l'}$ for $l \neq l'$. In fact, we can assume that each
point in $M$ is contained in at most $d+2$ of the $V_l$'s. This is
because on a (smooth, second countable) manifold the Lebesgue
covering dimension coincides with the topological dimension. 
By construction, the charts are such that the local parametrices
$F_{\zeta}^{(l)}(y,y')$, $y,y' \in V_l$, satisfy Lemma \ref{approxdelta}.
Finally, since for fixed $\zeta>0$ the $F_{\zeta}^{(l)}$
  are smooth we may also assume the neighborhoods to be sufficiently small
  such that   
  \begin{align}
\label{gpara0}    
  \int_{V_l}d\mu(y')|F_\zeta^{(l)}(y,y')|\leq
  \frac{C}{(4\pi \zeta)^{\frac{d+1}{2}}}\,,\quad
  \int_{V_l}d\mu(y)|F_\zeta^{(l)}(y,y')|\leq
  \frac{C}{(4\pi \zeta)^{\frac{d+1}{2}}}\,,
\end{align}
for all $y \in V_l$, with a $y$ and $l$ independent constant $C>0$.

The rectangular open neighborhoods $(V_l \times V_l)_{l \in \N}$ then
cover the diagonal of $M \times M$. To get an open cover of $M \times M$
we augment this by $W:= (M \times M)\backslash {\rm diag}(M \times M)$. 
Since ${\rm diag}(M \times M) \subset M \times M$ is closed, $W$ is
indeed open, and
\begin{equation}
\label{gpara1}
\{W\} \cup \{ V_l \times V_l\}_{l \in \N}\,,
\end{equation} 
furnishes an open cover of $M \times M$. Then, there exists a
locally finite partition of unity $(\chi_l)_{l \in \N_0}$ that is
subordinate to this cover. Explicitly,
\begin{itemize}
\item[(i)] $\chi_l \in C^{\infty}(M \times M, [0,1])$, for all
$l \in \N_0$.  
\item[(ii)] ${\rm supp} \,\chi_0 \subset W$, and ${\rm supp} \,\chi_l
  \subset V_l\times V_l$, $l \in \N$.  
\item[(iii)] Every point in $M \times M$ has an open neighborhood
  that intersects only finitely many of the
  $({\rm supp} \,\chi_l)_{l \in \N_0}$. In particular, 
  ${\rm supp}\,\chi_l(y, \,\cdot\,)$ with $y \in M$ fixed
  is nonempty for at most $d+2$ of the $l\geq 1$.     
\item[(iv)] $\sum_{l =0}^{\infty} \chi_l =1$ at every point in $M \times M$. 
\end{itemize}

Using this partition of unity we define the global parametrix by
\begin{equation}
\label{gpara2}
F_{\zeta}(y,y') = \sum_{l \geq 1} \chi_l(y,y') F^{(l)}_{\zeta}(y,y')\,,
\quad y,y' \in M,\;\;\zeta >0\,,
\end{equation} 
where the $l=0$ term does not appear, as the parametrix ought to
be localized around the diagonal. For simplicity we identify points
$p,p' \in V_l$ with their coordinates $y = \vp_l(p)$, $y' = \vp_l(p')$,
in the respective chart map $\vp_l$. By a slight abuse of notation we
extend this to all $p,p' \in M$, with the relevant chart maps
and their consistent coordinatizations implicit.   
Note that for each $(y,y')$ the
sum over $l$ is finite by property (iii) of the partition. 
For simplicity, we continue to write $F_{\zeta}$ for the
resulting bi-scalar parametrix on $M \times M$. It is 
then jointly smooth in $\zeta, y,y'$ and nonzero in  a neighborhood
of the diagonal of $M\times M$. The diagonal $F_{\zeta}(y,y)$ is governed
by the glued diagonal coefficients
\begin{equation}
\label{gpara2a}
A^{\th}_n(y) := \sum_{l \geq 1} \chi_l(\vp_l(p), \vp_l(p))
(A_n^{\th})^{(l)}(\vp_l(p), \vp_l(p))\,,
\end{equation} 
where on the right hand side we indicated the chart maps. By the smoothness
assumption these $A^{\th}_n:M \ra \C$ are globally defined.
The main result needed is the global
counterpart of Lemma \ref{approxdelta}.
\medskip

\begin{proposition} \label{globalapproxdelta}
The global parametrix (\ref{gpara2}) has the property 
(a) from Lemma \ref{approxdelta} but now for $C_c^{\infty}(M)$
test functions. Likewise, property (b) from  Lemma \ref{approxdelta}
holds, but now for all $\psi \in L^2(M)$. 
\end{proposition}  

{\it Proof.} (a) Let $\psi \in C_c^{\infty}(M)$. Then ${\rm supp}\, \psi$
intersects the support of only finitely many $\chi_l(y,\,\cdot\,)
\in C^{\infty}(M, [0,1])$. To see this, consider
$\{ y\} \times {\rm supp}\, \psi \subset M \times M$, which can
be covered by open neighborhoods that intersect only finitely
many of the ${\rm supp}\, \chi_l$ (by property (iii) of the partition).   
Since $\{ y\} \times {\rm supp}\, \psi \subset M \times M$,
is compact, there exist a finite subcover of it by such neighborhoods.
Hence ${\rm supp} \, \psi$ only intersects finitely
many $\chi_l(y,\,\cdot\,)$, as claimed, say
$\chi_{l_1}(y,\,\cdot\,),\ldots, \chi_{l_K}(y,\,\cdot\,)$.

To proceed, let $\mathbb{F}_{\zeta}$ be the integral operator whose
kernel is the global parametrix (\ref{gpara2}). Its action on
$\psi \in C_c^{\infty}(M)$ is given by
\begin{equation}
\label{gpara3}
(\mathbb{F}_{\zeta} \psi)(y) = \sum_{k=1}^K \int_{V_{l_k}}\! d\mu_g(y')
F_{\zeta}^{(l_k)}(y,y') \chi_{l_k}(y,y') \psi(y')\,,
\end{equation} 
using ${\rm supp}\,\chi_{l_k}(y, \, \cdot\,) \subset V_{l_k}$.   
Note in particular, that the image has compact support as well:
$\mathbb{F}_{\zeta} \psi \in C_c^{\infty}(M)$ for
$\psi \in C_c^{\infty}(M)$. Since each $\chi_l(y,\cdot)\in C_c^\infty(V_l)$, Lemma \ref{approxdelta}(a) is applicable to each term in the sum (\ref{gpara3}), giving
\be
\label{gpara4}
\lim_{\zeta \ra 0^+} (\mathbb{F}_{\zeta} \psi)(y) = \sum_{k=1}^K
\chi_{l_k}(y,y) \psi(y)\,.
\ee
By property (iv) of the partition of unity, the sum over all $\chi_l$ is $1$, pointwise in $M \times M$.   
However, as seen above, among all $\chi_l$'s only the $\chi_{l_1},
\ldots, \chi_{l_K}$, are potentially supported at $(y,y)$ with
$y \in {\rm supp}\,\psi$. Further, $\chi_0(y,y) =0$ as
$(y,y) \in {\rm diag}(M \times M)$. Thus, $1 = \sum_{l=0}^{\infty}\chi_l(y,y)
= \sum_{k=1}^K \chi_{l_k}(y,y)$, which gives the assertion (a).

(b) We first use Schur's test (\cite{Laxbook}, p.176) 
to show that (\ref{gpara2}) defines a bounded integral operator
$\mathbb{F}_{\zeta}: L^2(M) \ra L^2(M)$. According to the test, a sufficient
condition for this to be the case is that
\begin{equation}
\label{gpara5}
\sup_{y \in M}\int\! d\mu_g(y') \, |F_{\zeta}(y,y')| < \infty\quad
\mbox{and} \quad
\sup_{y' \in M} \int\! d\mu_g(y) \, |F_{\zeta}(y,y')| < \infty\,,
\end{equation}
where $F_{\zeta}$ is the global parametrix (\ref{gpara2}). 
We start with the first condition and insert (\ref{gpara2}) to find 
\begin{equation}
\label{gpara6}
\int\! d\mu_g(y') \,|F_{\zeta}(y,y')| \leq \sum_{k=1}^{d+2}
\int_{V_{l_k}}\! d\mu_g(y')\, \chi_{l_k}(y,y')
|F_{\zeta}^{(l_k)}(y,y')|\,.
\end{equation} 
Initially, the sum extends over all $l \geq 1$. However, for any fixed $y$
the second part of property (iii) ensures that at most $d+2$ open sets
$V_{l_k}, k =1, \ldots, d+2$, contribute. By definition of the
$\chi_l$, $l \neq 0$, the support of $\chi_{l_k}(y,y')$ is contained
in $V_{l_k} \times V_{l_k}$, even as $y'$ varies. This gives
(\ref{gpara6}). Finally, using property (\ref{gpara0}) the right hand
side of (\ref{gpara6}) can be bounded by  
$C(d\!+\!2)(4\pi \zeta)^{-\frac{d+1}{2}}$, for a $y$ and $l_k$ independent
constant $C$. This gives the first bound in (\ref{gpara5})
for each fixed $\zeta >0$. The flipped version is obtained similarly.
This shows that $\mathbb{F}_{\zeta} :L^2(M) \ra L^2(M)$ is a bounded
operator, for all $\zeta >0$. 

Next, consider the explicit form of $\Vert \mathbb{F}_{\zeta} \psi\Vert^2$,
\ba
\label{gpara8}
\int\! d\mu_g(y) \, |(\mathbb{F}_{\zeta}\psi)(y)|^2 \is 
\int\! d\mu_g(y) \,\int\! d\mu_g(y_1) \,\int\! d\mu_g(y_2) \,
\\[2mm] 
& \times & 
\sum_{l,m\geq 1} \chi_l(y,y_1) \chi_m(y, y_2)
F^{(l)}_\zeta (y,y_1) F^{(m)}_\zeta (y,y_2)^* \psi(y_1) \psi(y_2)^* \,.
\nonumber
\ea 
Anticipating that it is legitimate to take the $\zeta \ra 0^+$ limit inside
the integral one has
\ba
\label{gpara9}
\lim_{\zeta \ra 0^+} \int\! d\mu_g(y) \, |(\mathbb{F}_{\zeta}\psi)(y)|^2 =
\int\! d\mu_g(y) \Big( \sum_{l \geq 1} \chi_l(y,y) \Big)^2 |\psi(y)|^2
=\nrm{ \psi}^2 \,.
\ea
In the penultimate step we used that the $l=0$ term does not contribute to
the partition of unity on the diagonal.

It remains to justify that the limit can be taken inside the integrals.
To this end we initially take $\psi \in C_c^{\infty}(M)$.
By the reasoning before (\ref{gpara3}) then a
variant of (\ref{gpara8}) arises with finite sums and integrals
over compact domains. Specifically, the integrand reads
\begin{equation}
\label{gpara10} 
I(y,y_1,y_2) = \sum_{k,m=1}^K \chi_{l_k}(y,y_1) \chi_{l_m}(y,y_2)
F_{\zeta}^{(l_k)}(y,y_1)
F_{\zeta}^{(l_m)}(y,y_2) \psi(y_1) \psi(y_2)^*\,.
\end{equation}
Here, $\chi_{l_k}$ has support in $V_{l_k} \times V_{l_k}$ and
$\chi_{l_m}$ has support in $V_{l_m} \times V_{l_m}$, which forces $y$
in the $(k,l)$-th term to lie in the intersection $V_{l_k} \cap V_{l_m}$.
In order to take the limit $\zeta \ra 0^+$ inside the triple
integral $d\mu(y) d\mu(y_1) d\mu(y_2)$ with integrand (\ref{gpara10}) 
we show that $I$ can be bounded by an integrable function. As in the
proof of Lemma \ref {approxdelta}a we use $|F_{\zeta}^{(l_k)}(y,y')| \leq C_y^{(l_k)}
\exp\{ - \frac{c_-}{4} |y - y'|^2 \}$, for some constants $C_y^{(l_k)}$.
We can bound the $\chi_{l_k}$ by $1$ to obtain  
\begin{equation}
\label{gpara11}
|I(y,y_1, y_2)| \leq C_y \Big( e^{-\frac{c_-}{4} |y - y_1|^2} \psi(y_1) \Big)
\Big( e^{-\frac{c_-}{4} |y - y_2|^2} \psi(y_2) \Big)^*\,.
\end{equation}
Since the domain of the $y$ integration is also compact (given by
the union of the $V_{l_k} \cap V_{l_m}$) the right hand side is
integrable with respect to $d\mu(y) d\mu(y_1) d\mu(y_2)$ with
integrand (\ref{gpara10}). This justifies the step leading to
(\ref{gpara9}) for all $\psi \in C_c^{\infty}(M)$. 

Finally, we use the fact that $C_c^{\infty}(M)$ is dense in $L^2(M)$.
That is, for a given $\psi \in L^2(M)$ there exists a sequence
$\psi_n, n \in \N$ of $C_c^{\infty}(M)$ functions such that
$\| \psi_n - \psi \Vert \ra 0$, $n \ra \infty$.
We know that $\mathbb{F}_{\zeta} \psi_n \in C_c^{\infty}(M)$ for all $n$
and  $\mathbb{F}_{\zeta} \psi \in L^2(M)$. Since by the first part
$\mathbb{F}_{\zeta}$ is a bounded operator on $L^2(M)$ also
$\Vert \mathbb{F}_{\zeta} \psi_n - \mathbb{F}_{\zeta} \psi \Vert \ra 0$,
$n \ra \infty$ holds. \qed
\medskip

%%%%%%%%%%%%%%%%%%%%%%%%%%%%%%%%%%%%%%%%%%%%%%%%%%%%%%%%%%%%%%%%
\subsection{Bounding the difference kernel}

We now use the global parametrix
$F_{\zeta}(y,y')$ to define a global remainder function
$R_{\zeta}(y,y') := (\dd_{\zeta} - \Delta_{\th,y}) F_{\zeta}(y,y')$,
that is jointly smooth in $\zeta, y,y'$. 
Adapting the strategy in \cite{Waldheatk} we define in terms of it
\be
\label{hkexp10} 
Q_{\zeta}(y,y') := \int_0^{\zeta} \! d\zeta'
\int\! d\mu_g(z) \,K^{\th}_{\zeta - \zeta'}(y,z)
\,R_{\zeta'}(z,y')\,.
\ee
By $\dd_{\zeta}$ differentiation  of (\ref{hkexp10}) one finds
\be
\label{hkexp11}
\dd_{\zeta} \big( F_{\zeta} - Q_{\zeta} \big)(y,y') =
\Delta_{\th,y} \big( F_{\zeta} - Q_{\zeta} \big)(y,y') \,,
\ee
using the heat equation for $\dd_{\zeta}K_{\zeta - \zeta'}^{\th}$
and the definition of $R_{\zeta}$. The difference kernel
$(F_{\zeta} - Q_{\zeta})(y,y')$ therefore satisfies the
same heat equation as $K^{\th}_{\zeta}(y,y')$. Since $Q_{\zeta}$
manifestly vanishes for $\zeta \ra 0^+$ and $F_{\zeta}$ obeys
(\ref{lim1}) by Proposition \ref{globalapproxdelta}, the difference
is also a fundamental solution to the heat equation.

We wish to conclude that $F_{\zeta} - Q_{\zeta}$ therefore
coincides with $K_{\zeta}$. Note, however, that on a noncompact
manifold a fundamental solution even to the standard heat  
equation is not necessarily unique. In order to utilize the
results from Section \ref{Sec3_3} we associate an integral operator
to the difference kernel, defining for $\psi\in L^2(M)$
 \begin{align}
\label{diffk1} 
   ((\mathbb{F}_{\zeta}-\mathbb{Q}_{\zeta})\psi)(y):=
   \int\! d\mu_g(y') (F_{\zeta}-Q_{\zeta})(y,y')\psi(y')\,,
   \quad \zeta >0\,.
\end{align}
As seen above, the image solves the heat equation $(\pp_{\zeta}-\dth)
(\mathbb{F}_{\zeta}-\mathbb{Q}_{\zeta})\psi =0$ for all $\psi \in L^2(M)$.
Further, 
\be
\label{diffk2} 
\lim_{\zeta \ra 0^+} \Vert (\mathbb{F}_{\zeta} - \mathbb{Q}_{\zeta}) \psi \Vert
= \Vert \psi\Vert\,, \quad \psi \in L^2(M)\,. 
\ee
To see this, we first note that $\mathbb{F}_{\zeta} \psi$ and
$\mathbb{Q}_{\zeta} \psi$ are separately in $L^2(M)$. For $\mathbb{F}_{\zeta}
\psi$ this was established before; for $\mathbb{Q}_{\zeta} \psi$
we use the definition and Cauchy-Schwarz to write 
\be
\int\! d\mu(y) |(\mathbb{Q}_{\zeta}\psi)(y)|^2
\leq \Big( \int_0^{\zeta} \! d\zeta'\,  \nrm{e^{ (\zeta- \zeta') \Delta_{\th}} \psi_{\zeta'} }\Big)^2 \,,
\ee
where $\psi_{\zeta}(z) :=\int \! d\mu_g(y') R_{\zeta}(z, y') \psi(y')$. 
Clearly, $\mathbb{Q}_{\zeta} \psi \in L^2(M)$ if $\psi_{\zeta} \in L^2(M)$, where the latter is easily seen. 
To proceed, we anticipate the  (much stronger and independently
established) Lemma \ref{tildeexp} and the ensued bounds on $R_{\zeta}$. Based on this, it can be shown that   $ [0, \zeta] \ni \zeta' \mapsto
\Vert e^{ (\zeta- \zeta')\Delta_{\th}} \psi_{\zeta'} \Vert$ 
is  bounded as $\zeta \ra 0^+$. This implies
$\lim_{\zeta \ra 0^+} \Vert \mathbb{Q}_{\zeta} \psi \Vert =0$. 
The triangle inequality $\Vert (\mathbb{F}_{\zeta}
- \mathbb{Q}_{\zeta}) \psi\Vert - \Vert \mathbb{F}_{\zeta} \psi \Vert \leq
\Vert \mathbb{Q}_{\zeta} \psi \Vert$ and (\ref{gpara9}) then give
the assertion (\ref{diffk2}).

On the other hand, we know that $(\dth, D(\dth))$
generates a strongly continuous semigroup, $\R_+ \ni \zeta \mapsto
 e^{\zeta \dth}$, on $L^2(M)$. As such the solution of the initial value
 problem $\pp_{\zeta}\psi(\zeta)-\dth\psi(\zeta)=0$ with
 $\psi(0)=\psi \in L^2(M)$, is unique; see, Ch.~II, Prop.~6.4
 of \cite{Semigroupbook}. Hence, the semigroups
 $\mathbb{F}_{\zeta}-\mathbb{Q}_{\zeta}$ and $e^{\zeta \dth}$ coincide.
 By Theorem \ref{thsmooth2}(iii) this implies that their kernels
 coincide as well, i.e.
 \be
\label{hkexp12}
K_{\zeta}^{\th}(y,y') - F_{\zeta}(y,y') =  Q_{\zeta}(y,y')\,.
\ee
To complete the proof we now seek to bound $Q_{\zeta}$ appropriately.
Since $Q_{\zeta}$ is defined in terms of $R_{\zeta}$ this amounts to 
rendering $R_{\zeta}$ `small'. So far, mostly the very broad properties
for $s_{\th}$ and $A_n^{\th}$ mentioned after
(\ref{sleading}) entered, and $R_{\zeta}$ will just be given by
the expression (\ref{hkexp5}). We now identify $s_{\th}$
with $\tilde{\sigma}_{\th}$ from Proposition \ref{syngeseries} and make
for $A_n^{\th}(y,y')$ an ansatz in powers of $\Delta y$
to an appropriate order and adjust the coefficients so as to render
all but the term proportional to $(i e^{-i \th} \zeta)^N Z_N$
on the right hand side of (\ref{hkexp5}) `small'. The appropriate
requirement is 
\ba
\label{hkexp8}
X_n \is O\big(|\Delta y|^{2N+8}\big)\,, \;\;\;\, n \geq 0\,,
\quad 
Y_0 = O\big(|\Delta y|^{2N+6}\big)\,, 
\nonum
Y_{n+1} + Z_n \is
    O\big(|\Delta y|^{2N +4}\big)\,,\quad n =0, \ldots, N-1\,,
\ea 
where $N$ is the expansion order in $\zeta$.

We formulate the instrumental fact as a Lemma but postpone its proof. 
\pagebreak[4]

\begin{lemma} \label{tildeexp} For $(y,y') \in V \times V$, some
  open chart neighborhood of the diagonal in $M \times M$, let 
  $s_{\th}(y,y') = \tilde{\sigma}_{\th}(y,y')$
  be given by Proposition \ref{syngeseries} and consider its suitably truncated
  Taylor expansion.  Then, in a polynomial ansatz for
  $A_n^{\th}(y,y')$, $n =0, \ldots, N$, to a suitable $N$-dependent
  order
  in $\Delta y$, the coefficients can be adjusted such that the
  conditions (\ref{hkexp8}) hold, for any given $N \geq (d+1)/2$.
  The adjustment only requires the solution of decoupled linear equations.  
In particular, the resulting diagonal 
  coefficients $A^{\th}_n(y,y)$ are scalar differential
  polynomials in $g^{\th}_{\mu\nu}$ that are smooth in $y$.
\end{lemma}

Based on this Lemma we can bound the failure of $F_{\zeta}$
to furnish a solution of the heat equation. Recall the global remainder function $R_{\zeta}(y,y') =
(\dd_{\zeta} - \Delta_{\th,y}) F_{\zeta}(y,y')$, which may be expressed in terms of  local remainders $R_{\zeta}^{(l)}(y,y')$ through the partition of unity analagously to \eqref{gpara2}. Both are clearly jointly smooth in $\zeta, y,y'$ for all  $\zeta >0$. From  (\ref{hkexp5}) and Lemma \ref{tildeexp} one has
\be
\label{diffk3}
R_{\zeta}^{(l)}(y,y') = 
\frac{e^{-\frac{1}{2\zeta} \tilde{\sigma}_{\th}(y,y')}}{ \zeta^{ \frac{d+1}{2} +2} }
\Big[ \Delta y^{2N+4}\, A(y,\zeta) +
\zeta^{N+2} B(y,y')\Big]  \,.
\ee
Here the first term arises from the approximate vanishing
of the $n=0,\ldots,N\!-\!1$ terms in \eqref{hkexp5}, and the second term is the  $n=N$ contribution proportional to $Z_N$.

Next, it is convenient to reexpress \eqref{hkexp10} in the form 
\be
\label{hkexp16}
Q_{\zeta}(y,y') = \int_0^{\zeta} \! d \zeta'
\Big(\! e^{ (\zeta - \zeta') \Delta_{\th}} \psi_{\zeta', y'}\Big)(y)\,,      
\ee
where $\psi_{\zeta', y'}(z) = R_{\zeta'}(z, y')$ is for fixed
$\zeta', y'$ smooth and of compact support in $z$, and also
the dependence on $\zeta', y'$ is smooth. The integrand may be given a pointwise bound (in $y,y'$)  as follows
\begin{align}\label{diffk2a}
 \big|(e^{ (\zeta - \zeta') \Delta_{\th}} \psi_{\zeta', y'})(y)\big|
 &\leq B_y\ssum{j=0}{\sigma_0}
  \nrm{\Delta_\th^j\exx{(\zeta - \zeta')\Delta_\th}\psi_{\zeta', y'}}=  B_y\ssum{j=0}{\sigma_0}
  \nrm{\exx{(\zeta - \zeta')\Delta_\th}\Delta_\th^j\psi_{\zeta', y'}}
  \nonum
  &
\leq  B_y\ssum{j=0}{\sigma_0}\nrm{\Delta_\th^j\psi_{\zeta', y'}}.
\end{align}
Here, the first inequality follows from
the Sobolev embedding theorem along the lines of  the computation in \eqref{e.sm5a}, and $\sigma_0$ is the smallest integer greater than $(d+1)/4$. In the second   step we commute the operator
$\Delta_\theta$ through the semigroup, and in the final inequality
we use the contractivity of the semigroup.

Thus, it remains to bound the $ \Vert \Delta_\th^j\psi_{\zeta', y'}\Vert $
in \eqref{diffk2a}. To this end we make use of the fact that 
(by the Lebesgue covering dimension) for each $y'$ only
$d+2$ local parametrices contribute to the right hand side of \eqref{diffk2a}. It is therefore
 sufficient to find a bound for only a single local parametrix.
To obtain $\Delta_{\th}^j \psi_{\zeta',y'}$ we need to operate
with $\Delta_\theta^j$ (acting on $y$) on (\ref{diffk3}). The
structure of $\Delta_{\th}^j$ can be inferred from \eqref{hkexp24} below,
it contains between $2j$ and $j$ partial derivatives. Each
derivative acting on the exponential prefactor in (\ref{diffk3})
brings down a power of $\zeta^{-1}$. Each derivative acting on
the term in square brackets produces a term where the order of
$\Delta y$ is reduced and others. After rescaling of the
difference variable $\Delta y = \sqrt{\zeta} z$ this results
in a structure of the following form 
\be
\label{diffk4}
\Delta_{\th,y}^j R_{\zeta}(y,y') \big|_{ \Delta y \mapsto \sqrt{\zeta} z}
= \frac{e^{-\frac{1}{2\zeta} \tilde{\sigma}_{\th}(y,y - \sqrt{\zeta} z)}}%
{ \zeta^{ \frac{d+1}{2} +2} }
\zeta^{N+2 -2 j} P_{\zeta}^{(j)}(y,z) \,,
\ee
where $P_{\zeta}^{(j)}(y,z)$ is a polynomial in $z$ of degree $2N+4$
and also a polynomial in $\zeta$.  
This gives  
\ba
\label{diffk5} 
&&\nrm{\Delta_{\th}^j \psi_{\zeta,y'}}^2 =
\int\! d^{d+1} y |g|^{1/2}(y) \,\big| \Delta_{\th}^j R_{\zeta}(y,y')\big|^2 
\nonum
&& = \zeta^{2 N - 4 j -(d+1)} \int\! d^{d+1} y
\,e^{ - \frac{1}{\zeta} \tilde{\sigma}(y, y - \Delta y)}
|g|^{1/2}(y)\big|P_{\zeta}^{(j)}\big(y, \Delta y/\sqrt{\zeta}\big)\big|^2\,. 
\ea 
The cutoff functions in the partition of unity entail that this integral is over a compact domain, for which
 we may take $|y - y'| < \delta$.
We can bound the $y$-dependence in the non-exponential factors
by its supremuum in this ball and write 
$\tilde{P}^{(j)}(\Delta y/\sqrt{\zeta})$ for the resulting
function of the difference variable. The exponential can be
bounded by $\exp\{- c_- |\Delta y|^2/(2 \zeta)\}$, using \eqref{lim3}.  
The integrand is then a function of $\Delta y$ only and
one can change integration variables to $z = \Delta y/\sqrt{\zeta}$.
This yields
\be
\label{diffk6}
\nrm{\Delta_{\th}^j \psi_{\zeta,y'} }^2
\leq \zeta^{2 N - 4 j - \frac{d+1}{2}} \int\! d^{d+1} z \,
e^{ - \frac{c_-}{2} |z|^2} \tilde{P}^{(j)}_{\zeta}(z) \,.
\ee
After the rescaling the integral is
effectively  over the ball $0< \sqrt{\zeta} |z| < \delta$.
For small $\zeta$ this forces $|z|$ to be large and we can obtain
an upper bound by extending the integration domain to $\R^{d+1}$.
This gives a sum of (polynomially modified) Gaussians in $z$
with a polynomial $\zeta$ dependence. Performing them results
in a polynomial in $\zeta$ (with non-vanishing constant term), which for $0< \zeta < 1$
can be bounded by a constant. The upshot is that there exist a
$C(\delta) >0$ such that
\be
\label{diffk7}
\nrm{\Delta_{\th}^j \psi_{\zeta,y'}}
\leq C(\delta) \zeta^{ N - 2 j - \frac{d+1}{4}}\,.
\ee
Finally,  with (\ref{diffk2a}) one obtains upon performing the $d\zeta'$ integral in the  bound obtained from \eqref{hkexp16}
\be
|Q_{\zeta}(y,y')| \leq C_y \zeta^{N+1-2 \sigma_0 - \frac{d+1}{4}}\,. 
\ee
Used in \eqref{hkexp12}, we arrive at the following result: {\itshape for each $N\in \bbN$ and $y\in M$, for $\zeta\in (0,1)$ there is a constant $C_{y,N}$ such that }
\begin{align}
  |K^\theta_{\zeta}(y,y)-F_\zeta^{N}(y,y)|\leq C_{y,N}\zeta^{N+ 1 -c }
 \end{align}
 {\itshape for $c = 2 \sigma_0 + (d+1)/2 >0$.} Here, we specialized to the diagonal $y=y'$, and denote the the truncation order of the parametrix by a superscript $N$.
 
 For an asymptotic expansion proper, one seeks to have the remainder  of the same order as the first term omitted in the parametrix ansatz. This can be achieved by the following adjustment. Pick an integer $N_0 < N$ and note 
\begin{align}
  F_\zeta^{N}(y,y) = F_\zeta^{N_0}(y,y)
  +\frac{1}{(4\pi \zeta)^{\frac{d+1}{2}}}
  \big\{ A_{N_0+1}^\theta(y)\zeta^{N_0+1}+\ldots+A_{N}^\theta(y)\zeta^{N}\big\}\,.
\end{align}
Then
\begin{align}
|K^\theta_{\zeta}(y,y)-F_\zeta^{N_0}(y,y)| &\leq
|K_{\zeta}^{\th}(y,y) - F_{\zeta}^{N}(y,y)|
+|F_{\zeta}^{N}(y,y) - F_{\zeta}^{N_0}(y,y)|\nonum
& \leq  C_{y,N}\zeta^{N_0 +1 - (c - (N-N_0))} + D_y\zeta^{N_0 +1 - \frac{d+1}{2}}\,.	
\end{align}

The second term dominates when $(N-N_0)+(d+1)/2>c$; with $c= 2 \sigma_0 + (d+1)/2 >0$ and $\sigma_0$ the smallest integer greater than $(d+1)/4$, this can clearly be realized for any given $d$ by adjusting $N-N_0$. 
This establishes Theorem \ref{asymthm} based on Lemma \ref{tildeexp}.

{\bf Remarks.}

(i) The assertion about the coefficients in Theorem \ref{asymthm}
is a by-product of the
proof of Lemma \ref{tildeexp}. Indeed, the proof will show in particular
that in each local chart the $A_{n,0}$ terms in the polynomial ansatz for
$A_n^{c_n}$ below in \eqref{hkexp20} are uniquely
determined differential polynomials in the metric $g^{\th}$ and its inverse.
Since the $X_n,Y_n,Z_n$ coefficients (\ref{hkexp6}) determining them
are $N \mapsto i e^{-i \th} N$ phase rotations of their Euclidean signature
counterparts and only algebraic manipulations are invoked, the
solutions $A_{n,0}$ likewise must be the $N \mapsto i e^{-i \th} N$
phase rotations of the standard diagonal heat kernel coefficients. 
In the special case when for given $p$ the
$(A_n^{\th})^{(l)}(\vp_l(p),\vp_l(p))$ are $l$-independent, i.e.~a single
chart suffices, the expression (\ref{gpara2a}) reduces to the expected
$A_n^{\th}(y,y)$ in that chart.

(ii) As mentioned, the result remains valid for complex $\zeta$ in the
wedge $|{\rm Arg}(\zeta)| < \tilde{\th}$, where the left hand side is
defined by Theorem \ref{thsmooth2}.  
The right hand side is up to a $i e^{- i \th}$ prefactor only a function
of $|\zeta|$ and $\th - {\rm Arg}(\zeta)$, while the asymptoticity
refers to $|\zeta|$. With some additional work one can
also trace the validity of the individual steps in the proof.
\medskip

{\it Proof of Lemma \ref{tildeexp}}. We use power series
ans\"{a}tze of the form
\ba
\label{hkexp20} 
s^L(y',y) \is - i e^{i\th}
\sum_{l= 2}^{L-1} \frac{1}{l!} s_{\mu_1 \ldots \mu_{l}}(y')
\Delta y^{\mu_1}\ldots \Delta y^{\mu_l}\,,
\nonum
A^{c_n}_n(y',y) \is \sum_{k= 0}^{c_n-1} \frac{1}{k!}
A_{n,\nu_1 \ldots \nu_{k}}(y')
\Delta y^{\nu_1}\ldots \Delta y^{\nu_k}\,,
\ea
where the orders $L$ and $c_n$ will be functions of $N$ (the
truncation order in $\zeta$) to be determined. The correction
terms in (\ref{hkexp20}) are understood to be of order $\Delta y^L$
and $\Delta y^{c_n}$, respectively. The base point $y'$ is chosen
so that the partial derivatives $\dd_{\mu} = \dd/\dd y^{\mu}$ only
act on the $\Delta y$ products. To unclutter the notation we
indicate the order in $\Delta y$ directly by the number of indices,
writing $A_{n,\nu_1 \ldots \nu_{k}}(y')$ for $A^{(k)}_{n,\nu_1 \ldots \nu_{k}}(y')$
and $A_{n,0}(y')$ for $A_{n}^{(0)}(y')$. 

Starting with the $X_0$ condition in (\ref{hkexp8})
it is plain that if the coefficients $s_{\mu_1,\ldots, \mu_l}$,
$l =2,\ldots, L\!-\!1$ are adjusted as in Proposition \ref{syngeseries}
the remainder in the Eikonal equation will be of order $|\Delta y|^L$. 
For this to be of order $|\Delta y|^{2 N + 8}$ we take 
\begin{equation}
\label{hkexp21}
L = 2N + 8\,,
\end{equation} 
from here on. Note that this does not constrain $A_0^{c_0}$.
Moreover, the same $L = 2N\!+\!8$ Synge function approximant
$s^L$ will render all $X_n$, $n \geq 1$, also $O(|\Delta y|^{2 N + 4 + d_0})$ 
without constraining the $A_n^{c_n}$ ansatz. Since this is one
or two orders higher than the requirement in the other two conditions
in (\ref{hkexp8}) the latter reduce to 
\begin{equation}
\label{hkexp22}
Y_0 = O(|\Delta y|^{2N + 6})\,,\quad
Y_{n+1} + Z_n = O(|\Delta y|^{2N + 4})\,,
\quad n =0, \ldots, N-1\,.
\end{equation}
In the following we regard $s^L$ with $L =2 N+8$ as a known function
and seek to determine the coefficients in the $A_n^{c_n}$ ans\"{a}tze
such that (\ref{hkexp22}) holds. Irrespective of the details, both
conditions will have an expansion in powers of $\Delta y$.
Consistency with $L = 2N +8$ requires that the order $c_n -1$
of the polynomials used must not exceed $2N+6$, see (\ref{hkexp26}).

For $Y_0$ one needs to take into account that $A_{0,0} =1$;
so the ansatz for $A_0^{c_0}$ only contains $c_0-1$ free parameter functions.
On the other hand, an explicit computation (detailed below) shows that the
constant term in $Y_0$ vanishes, so that only positive
powers of $\Delta y$ occur, $Y_0 = \sum_{p=1}^{c_0-1} (\Delta y)^p Y_{0,p}
+ O(|\Delta y|^{c_0})$. In the absence
of any coincidental cancellations the coefficients $Y_{0,p},
p=1, \ldots, 2N+5$, will have to be made to vanish so as to
render the result
of order $|\Delta y|^{2 N +6}$. Taking $c_0 = 2N+6$
the ansatz for $A_0^{c_0}$
contains $2N+5$ free parameter functions for $2N+5$ linear
conditions to be met. A closer inspection shows that only linear,
decoupled relations arise, which fix the coefficients uniquely.

For $n \geq 1$, the $Z_{n-1}$ term in (\ref{hkexp22}) is
determined by $A_{n-1}^{c_{n-1}}$ and thus constitutes an inhomogeneity
known from the preceding order. The condition $Y_n
+ Z_{n-1} = O(|\Delta y|^{2N+4})$ thus should determine
the $c_n$ parameters in $A_n^{c_n}$.  Assuming $c_n \leq c_{n-1}$ the
expansion will
produce a series of the form $\sum_{p=0}^{c_n-1} (\Delta y)^p
(Y_{n,p} - Z_{n-1,p})$, so that the coefficients $p=0, \ldots, 2N+3$
have to vanish in order to meet the condition. Taking $c_n = 2N+4$
matches the number of parameters to the number of conditions.
Again, the conditions turn out to be linear and decoupled so
that the parameters are uniquely fixed.

In summary, the choice 
\begin{equation}
\label{hkexp23} 
c_0 = 2N +6\,, \quad c_n = 2N+4 \,, \quad n \geq 1\,,
\end{equation}
will uniquely determine the parameters in the ans\"{a}tze,
provided the expansion of the $Y_n$, $n \geq 0$,
terms has an appropriate structure. 

We prepare
\begin{equation}
\label{hkexp24} 
\nabla_{\th}^2 = g_{\th}^{\mu\nu}(y) \nabla_{\mu}^{\th} \nabla_{\nu}^{\th}
= g_{\th}^{\mu\nu}(y) \dd_{\mu} \dd_{\nu} + \gamma_{\th}^{\mu}(y) \dd_{\mu}\,,
\quad \gamma_{\th}^{\mu} := |g|^{-1/2} 
\dd_{\nu} (|g|^{1/2} g_{\th}^{\nu\mu})\,, 
\end{equation}
where again the $\th$-independent density $|g|^{1/2}$ enters. 
For $\mcL(s^L)$ from \eqref{hkexp6} this gives
\ba
\label{hkexp25}
\mcL(s^L) \is g_{\th}^{\mu\nu}(y) g^{\th}_{\mu\nu}(y') + \sum_{l =1}^{L-3}
\frac{1}{l!} \Delta y^{\mu_1} \ldots \Delta y^{\mu_l}
\Big[ g_{\th}^{\mu\nu}(y) s_{\mu\nu \mu_1\ldots \mu_l}(y') 
\nonum
& +& \gamma_{\th}^{\mu}(y) s_{\mu\mu_1\ldots \mu_l}(y') + 2
g_{\th}^{\mu\nu}(y) s_{\mu\mu_1 \ldots \mu_l}(y') \dd_{\nu} \Big] 
\\[2mm] 
& +& \frac{1}{(L\!-\!2)!} \Delta y^{\mu_1} \ldots \Delta y^{\mu_{L-2}}
\Big[   \gamma_{\th}^{\mu}(y) s_{\mu\mu_1\ldots \mu_{L-2}}(y') 
  + 2 g_{\th}^{\mu\nu}(y) s_{\mu\nu \mu_1\ldots \mu_{L-2}}(y') \dd_{\nu}
  \Big]\,.
\nonumber
\ea  
One could now expand the coefficient functions evaluated at
$y' = y - \Delta y$ around $y$ to obtain a local first order linear
differential operator. For our purposes it is better to
expand the coefficients at $y = y' + \Delta y$ around $y'$. This
gives an expression of the form
\ba
\label{hkexp26}
\mcL(s^L) = d\!+\!1 + \sum_{l=1}^{L-2}
\Delta y^{\mu_1} \ldots \Delta y^{\mu_l} \big[ d_{\mu_1 \ldots \mu_{l}}(y')
  + e^{\nu}_{\mu_1 \ldots \mu_l}(y') \dd_{\nu} \big] + O(\Delta y^{L-1})\,,
\ea
where the explicit form of the coefficients is readily inferred from
(\ref{hkexp25}). From here the structure of the $Y_n$ terms follows.
Collecting terms of the same order in $\Delta y$ one finds
\begin{align} 
\label{hkexp27}
& \Delta y^0: 2 n A_{n,0}
\nonum
& \Delta y^1:  [2 n \delta_{\mu_1}^{\nu} + e^{\nu}_{\mu_1} ] A_{n,\nu} +
d_{\mu_1} A_{n,0} \,,
\\[2mm]
& \Delta y^p: \vartheta(c_n\!-1\!-\!p) \bigg\{ 
\Big[\frac{2 n}{p!} \delta_{\mu_1}^{\nu} + \frac{1}{(p\!-\!1)!} e^{\nu}_{\mu_1}
  \Big] A_{n,\nu \mu_2 \ldots \mu_p} +
\nonum
&+ 
\frac{1}{(p\!-\!2)!} e_{\mu_1\mu_2}^{\nu} A_{n, \nu \mu_3 \ldots \mu_p}+ \ldots +
\frac{1}{1!} e^{\nu}_{\mu_1 \ldots \mu_{p-1}} A_{n, \nu \mu_p}\vartheta(L\!-\!p)
\nonum
& + \frac{1}{(p\!-\!1)!} d_{\mu_1} A_{n, \mu_2 \ldots \mu_p}
          + \frac{1}{(p\!-\!2)!} d_{\mu_1\mu_2} A_{n, \mu_3 \ldots \mu_p}
            + \ldots + \frac{1}{1!} d_{\mu_1 \ldots \mu_{p-1}} A_{n, \mu_p}
\vartheta(L\!-\!1\!-\!p) \bigg\}
\nonum          
&+  d_{\mu_1 \ldots \mu_p} A_{n,0} \,\vartheta(L\!-\!2\!-\!p)
      + e^{\nu}_{\mu_1 \ldots \mu_p} A_{n,\nu} \,\vartheta(L\!-\!1\!-\!p) \,.
\nonumber
\end{align}
In the last relation $p \geq 2$ and we use $\vartheta(x) = 1,\,x \geq 0$,
$\vartheta(x) = 0, \,x <0$,  to transcribe the range of the original sums.
Each order in $\Delta y$ will give rise to one condition constraining 
the coefficients in $A_n^{c_n}$. For $n\!=\!0$ there are $2N\! +5$
free coefficients in $A_0^{c_0}$, with $c_0 = 2N\! +6$. The vanishing of
the $p =1, \ldots, p_{\rm max}$, relations will match the
number of parameters for $p_{\rm max} = 2N +5$. The overall step
function in the $\Delta y^p$ condition will then always be unity. 
With $L =2 N +6$ also the $L$-dependent step function is just $1$.
The same happens for $n \geq 1$. By 
(\ref{hkexp23}) there are $2N +4$ coefficients and the step functions limit
the validity of the given expression to $p\leq p_{\rm max} = 2N+3$.
Hence, the $2N+4$ coefficients in the ansatz for $A_n^{2N}$ are 
are subjected to as many equations. Further, for $L =2N +6$ the
$L$ dependent step function is always unity. In summary, with
the choices (\ref{hkexp21}), (\ref{hkexp23}) the relations
(\ref{hkexp27}) have all step functions evaluate to $1$.

The expressions (\ref{hkexp27}) clearly have a triangular structure, the
$A_{n, \nu \mu_2 \ldots \mu_p}$ coefficient with $p$ indices occurs
only once and is contracted with a matrix indicated in square brackets. 
This matrix is always non-singular and is in fact a multiple of
the unit matrix. Indeed, returning to (\ref{hkexp25}) one reads off
$e_{\mu_1}^{\nu} = 2 g_{\th}^{\nu \mu}(y') s_{\mu \mu_1}(y') =
2 \delta_{\mu_1}^{\nu}$. The other terms in the $\Delta y^p$
expression have $A_{n,\#}$ coefficients with a smaller number of
indices, which suggests an iterative procedure to impose 
the conditions (\ref{hkexp22}).

For $n=0$ the vanishing of the $\Delta y^p$, $p =1, \ldots, 2N+5$,
expressions in (\ref{hkexp27}) indeed determines the coefficients
$A_{0,\nu}, A_{0, \nu \mu_2}, \ldots, A_{0, \nu, \mu_2 \ldots \mu_{2N+1}}$,
iteratively. For $n \geq 1$, we solve the relations
$Y_n = - Z_{n-1} +O(|\Delta y|^{2N})$, inductively
in $n$. Since $-Z_{n-1}$ is determined by the $A_{n-1}^{c_{n-1}}$
coefficients, the right hand side is known as a polynomial in
$\Delta y$ from the preceding order. Equating its
coefficients to the ones on (\ref{hkexp27}) again leads to a system
triangular in $p$, which determines the coefficients
$A_{n,\nu}, A_{n, \nu \mu_2}, \ldots, A_{n, \nu \mu_2 \ldots \mu_{2N-1}}$,
uniquely. \qed

%%%%%%%%%%%%%
% New Section
%%%%%%%%%%%%%

%%%%%%%%%%%%%%%%%%%%%%%%%%%%%%%%%%%%%%%%%%%%
\newpage 

\section{Schr\"{o}dinger evolution group from the $\th \ra 0^+$ limit}
\label{Sec6} 

The analytic semigroups $\zeta \mapsto e^{\zeta \Delta_{\th}}$ remain
well-defined in the near Lorentzian regime, as long as $\theta>0$.
In this section, we study the strict
Lorentzian limit $\theta\to 0^+$, where the properties change qualitatively.
Based on the form \eqref{sec8a0} of the generator $\dth$, one might hope that
in some sense a unitary Schr\"{o}dinger group will arise in the limit. The following
theorem specifies a sense in which this indeed holds.

\begin{theorem}\ \label{thunit}
  Let $(M,g^-)$ as in our standing assumptions and assume that the operator
  $\mcD_-=-\nabla_{-}^2+V$ is essentially self-adjoint on $C_c^\infty(M)$, i.e.,
  its closure $\overline{\mcD}_-$ is self-adjoint. 
  Then for all $s\in \bbR_{\geq 0}$ there is the limiting behavior
\begin{align}
	\exx{s\Delta_\theta}\xrightarrow{\theta
    \to 0^+}\exx{-i s\overline{\mcD}_-}\,,\quad \exx{s\Delta_{\pi -\theta}}\xrightarrow{\theta
    \to 0^+}\exx{+i s\overline{\mcD}_-}\,,
\end{align}
converging with respect to the ultra-weak
  (or weak-star) topology on $\mfB (L^2(M))$.
  Here   $\exx{-i s\overline{\mcD}_-}, s\in \bbR,$ is the one-parameter unitary group generated by  $\overline{\mcD}_-$. 
\end{theorem}

{\bf Remarks.}

(i) Recall that $(\mfB(L^2(M)),\nrmm_{\rm op})$ is the Banach space of
continuous endomorphisms on $L^2(M)$, while the ultra-weak topology
on $\mfB(L^2(M))$ is the weak-star topology arising from its pre-dual,
the Banach space of trace-class operators $\mathfrak{B}_1(L^2(M))$
equipped with the trace norm. The convergence in the theorem
therefore amounts to (\ref{i4}) in the introduction. 

(ii) The assumption that $\cD_-$ is essentially selfadjoint indirectly
codes rather subtle properties of the underlying Lorentzian manifold,
see \cite{Nonselfad,taira,taira2,vasyesa}. Under more restrictive assumptions
on $(M,g^-)$ a stronger notion of convergence is feasible \cite{Derez1}.   

(iii) 
We shall also need the product topological space $\mfB(L^2(M))^{\bbR_{\geq 0}}$,  where each component $\mfB(L^2(M))$
is endowed with the ultra-weak topology.
The product topology for an uncountable product is defined as follows. Starting with a family $(X_s)_{s\in \bbR_{\geq 0}}$ of topological spaces, the product space $\prod_{s\in \bbR_{\geq 0}}X_s$ (also denoted $X^{\bbR_{\geq 0}}$ when all $X_s\equiv X$, as is the case here) is the set of all functions $f:\bbR_{\geq 0}\to \bigcup_{s\in \bbR_{\geq 0}}X_s$ such that $\forall\,s\in \bbR_{\geq 0}:f(s)\in X_s$.
Then, the canonical base for the product topology consists of sets of the form $\prod_{s\in \bbR_{\geq 0}}\mcV_s$, with each  $\mcV_s\subseteq X_s$ open, {\itshape and} $\mcV_s=X_s$ for all but {\itshape finitely many} $s\in \bbR_{\geq 0}$.

Throughout this section, we  use the notation
\begin{align}\label{unit2}
\forall\,s\geq 0,\,\theta\in (0,\pi):\quad	T(s,\theta):=\exx{s\Delta_\theta}\,.
\end{align}
Clearly,  for each $\theta\in (0,\pi)$, $T(\cdot,\theta)\in
\mfB(L^2(M))^{\bbR_{\geq 0}}$. We begin by preparing the following lemmas.%
\footnote{We are grateful to K. Taira for discussions related to
Lemma \ref{lmunit2} below.}
\begin{lemma}\ \label{lmunit1}
  Let $(M,g^-)$ be a as in our standing assumptions and
  $\mfB(L^2(M))^{\bbR_{\geq 0}}$ as above. 
Regarding $((0,\pi), \preceq)$ as a
directed set with ordering
\begin{align}\label{unit3}
  \theta_1\preceq \theta_2 \iff \theta_1\geq \theta_2\,,\quad
  \forall\,\theta_1,\,\theta_2\in (0,\pi)\,,
\end{align}
the family $(T(\cdot,\theta))_{ \theta\in (0,\pi)}$ is a net in
$\mfB(L^2(M))^{\bbR_{\geq 0}}$. Then:

\begin{enumerate}[leftmargin=8mm, rightmargin=-0mm, label=(\roman*)]
  \item There exists a subnet
$(T(\cdot,\theta_\alpha))_{\alpha\in \mathscr{A}}$, and a strongly continuous
family of contractive operators  $T(\cdot)\in \mfB(L^2(M))^{\bbR_{\geq 0}}$ such
that $T(0)=\1$, and $T(\cdot,\theta_\alpha)\xrightarrow{\alpha\in \mathscr{A}}T(\cdot)$
wrt. the product topology on  $\mfB(L^2(M))^{\bbR_{\geq 0}}$. In particular, this implies
component-wise convergence,
\begin{align}\label{unit4}
\forall\,s\geq 0:\quad T(s,\theta_\alpha)\xrightarrow{\alpha\in \mathscr{A}}T(s)
\end{align}
in  the ultra-weak topology on $\mfB(L^2(M))$.

\item Consider for each $u\in C^\infty_c(M)$ the mapping 
\begin{align}\label{unit5}
\forall\,s\in \bbR_{\geq 0}:\quad	\mfu(s):=T(s)u\,.
\end{align}
Then $\mathfrak{u}\in C(\bbR_{\geq 0},L^2(M))\cap C^1(\bbR_{> 0},L^2(M))$, and is a distributional solution to the 
   Schr\"{o}dinger equation for each $w\in C_c^\infty(M)$, i.e.
\begin{align}\label{unit6}
	\ip{w}{d_s \mfu(s)}_{L^2(M)}+\ip{i\mcD_{-} w}{\mfu(s)}_{L^2(M)}=0\,,\quad s>0\,,
\end{align}
with initial condition $\mfu(0)=u$.
\end{enumerate}
\end{lemma}

\bigskip

\begin{lemma}
	\ \label{lmunit2}
Let $(M,g^-)$ be as in our standing assumptions and assume that the operator
$\mcD_-=-\nabla_{-}^2+V$ is essentially self-adjoint on $C_c^\infty(M)$.
Then the distributional solution $\mff\in C(\bbR_{\geq 0},
L^2(M))\cap C^1(\bbR_{> 0},L^2(M))$ to the Schr\"{o}dinger equation 
\begin{align}\label{unit7}
	d_s\mff(s)+i\mcD_{-}^\ast \mff(s)=0\,,\quad  s>0\,,
\end{align}
with initial condition $\mff(0)=f\in C_c^\infty(M)$ has a unique solution, namely $\mff(s)=\exx{-is\overline{\mcD}_{-}}f$, where $\overline{\mcD}_{-}$ is the (unique, self-adjoint) closure of $(\mcD_{-},C^\infty_c(M))$.
\end{lemma}

The proof of Lemma \ref{lmunit2} is omitted, referring  to  Ch.VI Theorem 1.7 of \cite{berezanskii}.

\bigskip

{\itshape Proof of Theorem \ref{thunit}.}
Consider from Lemma \ref{lmunit1} the subnet $(T(\cdot,\theta_\alpha))_{\alpha\in \mathscr{A}}$ and its limit $T(\cdot)\in \mfB(L^2(M))^{\bbR_{\geq 0}}$. For every $u\in C_c^\infty(M)$, it follows from \eqref{unit5}, \eqref{unit6} that $\mfu(s):=T(s)u$ is a solution to \eqref{unit7} with initial condition $\mfu(0)=u$. This solution is unique by  Lemma \ref{lmunit2}, and hence $\forall\,s\geq 0:\,T(s)u= \exx{-is\overline{\mcD}_{-}}u$. Thus, the action of the families $(T(s))_{s\geq 0}$ and $(\exx{-is\overline{\mcD}_{-}})_{s\geq 0}$ coincide on $C_c^\infty(M)$. Since $C_c^\infty(M)$ is dense in $L^2(M)$, it follows immediately that the families are identical, and so 
\begin{align}\label{unit8}
	\forall\,s\geq 0:\quad T(s,\theta_\alpha)\xrightarrow{\alpha\in \mathscr{A}}\exx{-is\overline{\mcD}_{-}}\,,
\end{align}
in  the ultra-weak topology on $\mfB(L^2(M))$.

Next, consider an arbitrary subnet $(T(\cdot,\theta_\beta))_{\beta\in \mathscr{B}}$ of the net $(T(\cdot,\theta))_{\theta\in (0,\pi)}$ in $\mfB(L^2(M))^{\bbR_{\geq 0}}$. By the same arguments as in the proof of Lemma \ref{lmunit1} below, it follows that there is a  sub-subnet $(T(\cdot,\theta_{\beta(\kappa) }))_{\kappa\in \mathscr{K}}$ converging to a $\widetilde{T}(\cdot)\in \mfB(L^2(M))^{\bbR_{\geq 0}}$ in the product topology. This limit $\widetilde{T}(\cdot)\in \mfB(L^2(M))^{\bbR_{\geq 0}}$ has the same properties as the $T(\cdot)$ from Lemma \ref{lmunit1}; in particular  $\widetilde{\mfu}(s):=\widetilde{T}(s)u$, for $u\in C_c^\infty(M)$, solves the differential equation \eqref{unit7} with initial condition $u$. As above, this implies that $\forall\,s\geq 0:\,\widetilde{T}(s)=\exx{-is\overline{\mcD}_{-}}$. Having shown that every subnet of $(T(\cdot,\theta))_{\theta\in (0,\pi)}$ has a further sub-subnet that converges to $(\exx{-is\overline{\mcD}_{-}})_{s\geq 0}$, it follows%
\footnote{The proof of this statement for nets is essentially the same
  as the proof of the corresponding result for sequences. Indeed, if the
  original net did not converge, then there would be an open neighborhood
  $V$  of the (sub-subnet) limit point and a subnet that lives in the
  complement of $V$. Then, this subnet cannot have a further sub-subnet
  that converges to the limit point, a contradiction.}
that the original net converges to the same, i.e. $T(\cdot,\theta)\xrightarrow{\theta\to 0^+}\exx{-is\overline{\mcD}_{-}}$ in the product topology. This entails component-wise convergence, i.e.,
\begin{align}\label{unit9}
	\forall\,s\in \bbR_{\geq 0}:\quad T(s,\theta)=\exx{s\Delta_\theta}\xrightarrow{\theta\to 0^+}\exx{-i s\overline{\mcD}_-}\,,
\end{align}
in the ultra-weak topology of $\mfB(L^2(M))$, completing the proof.
\qed

It remains to prove Lemma \ref{lmunit1}.

\bigskip

{\itshape Proof of Lemma \ref{lmunit1}.}

(i) Since for all $s\in \bbR_{\geq 0},\,\theta\in (0,\pi):\,\nrm{T(s,\theta)}_{\rm op}\leq 1$, it follows that $(T(\cdot,\theta))_{\theta\in (0,\pi)}$ is a net in the product topological space $\mcB^{\bbR_{\geq 0}}$, with $\mcB$ the norm-closed unit ball of $\mfB(L^2(M))$ equipped with the (relative) ultra-weak topology inherited from $\mfB(L^2(M))$. Since $\mfB(L^2(M))$ is the dual of the space of trace-class operators $\mfB_1(L^2(M))$, $\mcB$ is compact in the ultra-weak topology by the Banach-Alaoglu theorem, and hence $\mcB^{\bbR_{\geq 0}}$ is compact in the product topology (Tychonoff's theorem). Compactness then entails that there is a subnet $(T(\cdot,\theta_\alpha))_{\alpha\in \mathscr{A}}$ converging to $T(\cdot)\in \mcB^{\bbR_{\geq 0}}$ in the product topology, which in particular implies the component-wise convergence
\begin{align}\label{unit10}
	\forall\,s\geq 0:\quad T(s,\theta_\alpha)\xrightarrow{\alpha\in \mathscr{A}}T(s)\,,
\end{align}
in the ultra-weak topology on $\mcB$, from which \eqref{unit4} follows immediately. 

(ii) To proceed, fix  arbitrary $S>0$ and $u\in C_c^\infty(M)$, and consider the family $(\mfu(s,\theta))_{s\in [0,S] \atop \theta\in (0,\pi)}$ in $L^2(M)$, defined by
\begin{align}\label{unit11}
	\mfu(s,\theta):=T(s,\theta)u\,.
\end{align}
It is clear that for any fixed  $\theta\in (0,\pi)$, $\mfu(\cdot,\theta)\in C([0,S],L^2(M))\cap C^\infty((0,S),L^2(M))$, and in particular for any $k\in \bbN$ we have 
\begin{align}
	d^k_s(\mfu(s,\theta))=(\dth)^kT(s,\theta)u=T(s,\theta)(\dth)^ku\,,
\end{align}
leading to the bound
\begin{align}\label{unit12}
	\nrm{d^k_s(\mfu(s,\theta))}_{L^2(M)}=\nrm{T(s,\theta)(\dth)^ku}_{L^2(M)}\leq \nrm{(\dth)^ku}_{L^2(M)}\leq c_k(u)\,.
\end{align}
Here $c_k(u)>0$ is a constant depending on $k\in \bbN$ and $u\in C_c^\infty(M)$, but independent of $\theta\in (0,\pi)$, and the final bound in \eqref{unit12} follows since $\Delta_\theta$ acts classically on $u\in C_c^\infty(M)$.

It then follows that for each $\theta\in (0,\pi)$, $\mfu(\cdot,\theta)\in \mcH^2([0,S],L^2(M))$, the Bochner-space of $L^2(M)$ valued Sobolev functions on $[0,S]$, which is in particular a Hilbert space.\footnote{ We refer to \cite{yosida, hytonen} for textbook accounts of Bochner integration, and to \cite{evans} for a discussion of the associated Sobolev spaces.} Moreover, the result \eqref{unit12} implies that there is a constant $C_2(u)>0$ such that
\begin{align}\label{unit13}
	\sup_{\theta\in (0,\pi)}\nrm{\mfu(\cdot,\theta)}_{\mcH^2([0,S],L^2(M))}\leq C_2(u)\,,
\end{align}
where the $\mcH^2$-norm is defined by
\begin{align}
	\nrm{\mfh}_{\mcH^2([0,S],L^2(M))}^2:=\ssum{j=0}{2} \int_0^S\!\!ds\,\nrm{d_s^j\mfh(s)}_{L^2(M)}^2\,.
\end{align}

Returning now to the subnet $(T(\cdot,\theta_\alpha))_{\alpha\in \mathscr{A}}$ from earlier in the proof, it follows from \eqref{unit13} that $(\mfu(\cdot,\theta_\alpha))_{\alpha\in \mathscr{A}}$ is a net contained in the $\mcH^2$-norm-closed ball of radius $C_2(u)$. Since $\mcH^2([0,S],L^2(M))$ is a Hilbert space, this ball is weakly compact, and hence $(\mfu(\cdot,\theta_\alpha))_{\alpha\in \mathscr{A}}$ has a subnet $(\mfu(\cdot,\theta_{\alpha(\beta)}))_{\beta\in \mathscr{B}}$ converging weakly to   $\mfu(\cdot)\in \mcH^2([0,S],L^2(M))$. By Sobolev embedding (Theorem 2 of Chapter 5.9.2 of \cite{evans}), $\mfu(\cdot)$ has a $C([0,S],L^2(M))\cap C^1((0,S),L^2(M))$ representative, which we continue to denote by $\mfu(\cdot)$. A further consequence of Sobolev embedding is that for any $w\in L^2(M)$ and $s\in [0,S]$, the mapping
\begin{align}
	\mcH^2([0,S],L^2(M))\ni \mfh\mapsto\ip{w}{\mfh(s)}_{L^2(M)}\in \bbC
\end{align}
is an element of the dual space $\mcH^2([0,S],L^2(M))^\ast$. Then, the weak convergence of the subnet $(\mfu(\cdot,\theta_{\alpha(\beta)}))_{\beta\in \mathscr{B}}$ above  entails that for each $s\in [0,S]$ and $w\in L^2(M)$,
\begin{align}\label{unit14}
	\ip{w}{\mfu(s,\theta_{\alpha(\beta)})}_{L^2(M)}=\ip{w}{T(s,\theta_{\alpha(\beta)})u}_{L^2(M)}\xrightarrow{\beta\in \mathscr{B}}\ip{w}{\mfu(s)}_{L^2(M)}\,.
\end{align}
However, the ultra-weak convergence \eqref{unit10} of the subnet $(T(\cdot,\theta_\alpha))_{\alpha\in \mathscr{A}}$ when restricted to the above sub-subnet $(T(\cdot,\theta_{\alpha(\beta)}))_{\beta\in \mathscr{B}}$ implies
\begin{align}\label{unit15}
	\ip{w}{T(s,\theta_{\alpha(\beta)})u}_{L^2(M)}\xrightarrow{\beta\in \mathscr{B}}\ip{w}{T(s)u}_{L^2(M)}\,.
\end{align}
Since $w\in L^2(M)$ is arbitrary, comparing \eqref{unit14} and \eqref{unit15} yields
\begin{align}
	\forall\,s\in [0,S]:\quad \mfu(s)=T(s)u\,,
\end{align}
and hence $T(\cdot)u\in C([0,S],L^2(M))\cap C^1((0,S),L^2(M))$. Finally, it follows from the density of $C_c^\infty(M)\subseteq L^2(M)$ (and since $S>0$ is arbitrary) that the family of operators $(T(s))_{s\in \bbR_{\geq 0}}$ is strongly continuous. 

To complete the proof of Lemma \ref{lmunit1}, it only remains to be shown that $\mfu(s)=T(s)u$ solves the distributional Schr\"{o}dinger equation \eqref{unit6}. To that end, we return to the weakly converging subnet $(\mfu(\cdot,\theta_{\alpha(\beta)}))_{\beta\in \mathscr{B}}$ in $\mcH^2([0,S],L^2(M))$. Fixing an arbitrary $w\in C_c^\infty(M)$, clearly 
\begin{align}
	\ip{w}{\big(\pp_s-\Delta_{\theta_{\alpha(\beta)}}\big)\mfu(s,\theta_{\alpha(\beta)})}_{L^2(M)}=0\,,\quad s\in (0,S)\,.
\end{align}
On the other-hand, by Sobolev embedding and weak convergence in $\mcH^2([0,S],L^2(M))$,
\begin{align}
	&\ip{w}{\pp_s\mfu(s,\theta_{\alpha(\beta)})}_{L^2(M)}=\ip{w}{d_s\mfu(s,\theta_{\alpha(\beta)})}_{L^2(M)}\xrightarrow{\beta\in \mathscr{B}}\ip{w}{d_s\mfu(s)}_{L^2(M)}\,,
	\\[2mm]
	&\ip{w}{\Delta_{\theta_{\alpha(\beta)}}\mfu(s,\theta_{\alpha(\beta)})}_{L^2(M)}=\ip{\Delta_{\pi -\theta_{\alpha(\beta)}}w}{\mfu(s,\theta_{\alpha(\beta)})}_{L^2(M)}\xrightarrow{\beta\in \mathscr{B}}\ip{i \mcD_{-}w}{\mfu(s)}_{L^2(M)}\,,
	\nonumber
\end{align}
and hence 
\begin{align}
  \ip{w}{d_s \mfu(s)}_{L^2(M)}+\ip{i\mcD_{-} w}{\mfu(s)}_{L^2(M)}=0\,,
  \quad s\in (0,S)\,,
\end{align}
with initial condition $\mfu(0)=u$. Since $S>0$ is arbitrary, it follows that $\mfu(s)=T(s)u$ indeed solves the distributional Schr\"{o}dinger equation \eqref{unit6} with initial condition $\mfu(0)=u$.
\qed

%\newpage

%%%%%%%%%%%%%%%%%%%%%%%%%%%%%%%%%%%%%%%%%%%%%%%%%%%%%%%%%%%%%%%%%%%%%%%%%%
\section{Conclusions} 

Motivated by a Wick rotation in the lapse, rather than in time,
a one-parameter family of analytic semigroups $\zeta \mapsto e^{\zeta \Delta_{\th}}$,
$\th \in (0,\pi)$, was introduced and studied in some detail. The family
interpolates between the heat semigroup at $\th = \pi/2$ and a near
Schr\"{o}dinger semigroup for small nonzero $\th>0$. In a suitable weak
sense (and with a subsidiary condition) the strict Lorentzian limit $\th \ra 0^+$
was seen to give rise to a Schr\"{o}dinger evolution group. In contrast
to similarly motivated approaches the underlying manifolds remain real
and foliated throughout. A more detailed exposition of the rationale
for the Wick rotation in the lapse and its covariance properties
will be given elsewhere \cite{wickfoli}. The main results obtained
for the family of
semigroups are summarized in the introductory tabulation. Here we mention
some further directions. 

The most immediate extension is to non-scalar Laplacians, in
particular of Lichnerowicz type. For Euclidean signature
the associated heat semigroups are widely used to investigate
the quantum theory of gauge fields and gravity, often in combination
with the non-perturbative Functional Renormalization Group. We see no
principle obstruction to such a generalization, which would allow one to
explore the near Lorentzian regime of such computations in an  apples-to-apples
comparison. 

In practice, such computations utilize mainly the coefficients
appearing in a small semigroup time asymptotic expansion of the
kernel's diagonal,
as developed here. Nevertheless, the existence of an asymptotic
expansion for the off-diagonal lapse-Wick-rotated heat kernel would be
of considerable interest. Although the existence of an off-diagonal
expansion can presumably be established along the lines of
\cite{Kannai,Grieser,Ludewig} without reference to the Synge function,
the formulas for the coefficients are normally expressed in terms of it. 
It would therefore be desirable to establish existence results for the
lapse-Wick rotated Synge function under weak assumptions on the metric.  
A related issue left unexplored here are generalizations of
the known exact heat kernel bounds \cite{Daviesheatkbook,Grigorbook},
usually formulated in terms of the Synge function.

The generalized Laplace transform of the Wick rotated heat kernel's off-diagonal
expansion relates to a Wick rotated Hadamard parametrix, and the difference
to the exact induced lapse-Wick-rotated Green's function may be expected to code
`state dependent' aspects. Here the strict Lorentzian limit warrants further
investigation, particularly the relation to boundary values  of the Lorentzian
Hessian's resolvent and associated Feynman propagator \cite{Derez2, Derez1, Derez}. We remark that even on Minkowski space the notion of Wick rotation induced on the propagators by our construction does not coincide with the Feynman $i\epsilon$-prescription \cite{wickfoli}. Rather it is closer to a variant originally introduced by Zimmermann, which has the important consequence of rendering all (power-counting viable) Feynman diagrams to all loop orders {\itshape absolutely convergent}, while still (re-)producing the correct Lorentzian limit as $\epsilon\to  0^+$ \cite{zimm}. One of relevant questions is then whether or not the propagators and associated states of \cite{Derez2, Derez1, Derez} are reproduced by  the $\th \ra 0^+$ limit of the lapse-Wick-rotated resolvent in these cases.
\vspace{1cm} 

{\bfseries Acknowledgments.} We would like to thank N.~Viet Dang, J.~Derezinski,
I.~Khavkine, A.~Strohmaier, R.~Verch, A.~Vasy, and M.~Wrochna for their interest and 
helpful discussions at various stages of this work.
Additionally, we are grateful to K.~Taira and C.~J.~Lennard for
suggestions on the proof strategy in Section \ref{Sec6}.
R.B. thanks the Erwin Schr\"{o}dinger Institute in Vienna for hospitality during the `Spectral Theory and Mathematical Relativity' thematic program, and the Institute Henri Poincar\'{e} for hospitality during the `Quantum and classical fields interacting with geometry' thematic program, where parts of this work were completed and many stimulating discussions took place.  The authors acknowledge support of the Institute Henri Poincar\'{e} (UAR 839 CNRS-Sorbonne Universit\'{e}), and LabEx CARMIN (ANR-10-LABX-59-01).

% ----------------------------------------------------------------------
%		Appendices
% ----------------------------------------------------------------------
\newpage
\appendix

%%%%%%%%%%%%%%%%%%%
%	New appendix 
%%%%%%%%%%%%%%%%%%%

%\newpage

\section{Distributions and Sobolev spaces on manifolds} 
\label{App_dist}

We introduce and summarize the main properties of the function and
distribution spaces used in this paper.  
\medskip

{\bfseries Test functions, distributions and distributional derivatives.}
As a {\itshape set}, the  test functions  on the manifold $M$ is simply
$\mfD(M)= C_c^\infty(M)$, the compactly supported smooth complex-valued functions on $M$ (all functions discussed will be complex valued, and this specification will be henceforth omitted). $\mfD(M)$ is equipped with a  topology arising from a family of seminorms, rendering   it a locally convex topological vector space; we refer to \cite{Wavebook,Friedlander1} for further details. It can be shown that, in order to define distributions, it is sufficient to specify the mode of convergence in $\mfD(M)$. 
\begin{definition}\label{testdef1}
	{\itshape A sequence $(u_j)_{j\in \bbN}$ in $\mfD(M)$ is said to converge to $u\in \mfD(M)$ iff the following holds.
	\begin{enumerate}[leftmargin=10mm, rightmargin=-0mm, label=(\roman*)]
  \item There is a compact set $K\subseteq M$ containing the supports of $u$ and all the $u_j$.
  \item There is a covering of $K$ by finitely many charts $\Omega_1,\ldots, \Omega_k$ such that for each chart in local coordinates, for all multi-indices $\alpha$, $\pp^\alpha u_j \to \pp^\alpha u$  uniformly on the chart.
\end{enumerate}
}
\end{definition}
In order to define distributional gradients and Laplacians we introduce the
{\itshape test vector fields} $\mfD(M,TM)$ and
{\itshape test covector fields} $\mfD(M,T^*\!M)$. As {\itshape sets}
these are, respectively, the compactly supported smooth sections
of the tangent bundle $TM$ and cotangent bundle $T^*\!M$, and the
mode of convergence is defined as the straightforward generalization
of Definition \ref{testdef1}.  
\medskip

Next, the distribution space $\mfD'(M)$ comprises the continuous linear functionals over $\mfD(M)$, where continuity means that for any $\varphi\in \mfD'(M)$ and any sequence $(u_j)_{j\in \bbN}$ converging to zero in $\mfD(M)$ we have $(\varphi,u_j)\xrightarrow{j}0$ (here $(\varphi,u_j)$ denotes the action of the distribution $\varphi$). It is clear that $\mfD'(M)$ is a vector space, and convergence therein is defined as usual: $\vp_j\xrightarrow{\mfD'}\vp$ {\itshape iff} $(\vp_j,u)\xrightarrow{j}(\vp,u)$ for all test functions $u\in \mfD(M)$. Similarly, the spaces of {\itshape distributional vector fields} $\mfD'(M,TM)$
and {\itshape distributional covector fields} $\mfD'(M,T^*\!M)$  are,
respectively,  the duals of $\mfD(M,TM)$ and $\mfD(M,T^*\!M)$,
and are linear spaces with convergence defined analogously to the scalar case. 

The above test function and distribution spaces are  well-defined on any smooth manifold, without reference to a metric structure. We now assume that $M$ carries in addition a smooth metric $g$ of Riemannian or Lorentzian signature, and associated volume form $d\mu_g=dy |g|^{1/2}$. 
Recall that  $f\in C^\infty(M)$ canonically defines a distribution $(\tilde{f},u):=\int\!d\mu_g fu $. Then the one-form  $\nabla f$ defines an distributional vector field  $\widetilde{\nabla f}\in \mfD'(M,TM)$ by an integration by parts,
$(\widetilde{\nabla f},v):=-(\tilde{f},\nabla_\alpha v^\alpha)$ for
$v\in \mfD(M, TM)$, where $\nabla_\alpha v^\alpha=|g|^{-1/2}\pp_\alpha( |g|^{1/2}v^\alpha)$ is the divergence. Generalizing, the {\itshape distributional gradient} maps $\mfD'(M)\to \mfD'(M,TM)$, where for each $\varphi \in \mfD'(M)$, the {\itshape distribution} $\nabla\varphi$ is defined by 
\begin{eqnarray}
  (\nabla \varphi,v):=-(\varphi,\nabla_\alpha v^\alpha)\,,\quad \forall\,
  v\in \mfD(M,TM)\,. 
\end{eqnarray}
The distributional Laplacian $\nabla^2$ is defined analogously:  for $\varphi\in \mfD'(M)$, $\nabla^2 \varphi\in \mfD'(M)$ is defined by 
\begin{eqnarray}
	(\nabla^2 \varphi,u):=(\varphi,\nabla^2 u)\,,\quad \forall\, u\in \mfD(M)\,.
\end{eqnarray}

\medskip

{\bfseries $L^2$- and Sobolev spaces.} The $\sigma$-algebra $\mathfrak{M}(M)$  of measurable sets on the manifold $M$ consists of those $E\subseteq M$ such that for any chart $U$ on $M$, the image of  $E\cap U$ under the chart map is a Lebesgue measurable subset of $\bbR^{1+d}$; in particular,  $\mathfrak{M}(M)$ contains all Borel sets. Then,  measurable functions, vector fields, and covector fields can be defined on $M$ in the standard way.
Next, the   volume form associated to the Lorentzian metric $g_-$  induces a complete, regular measure $\mu_g$ on $\mathfrak{M}(M)$; we refer to  \cite{Grigorbook} for details.
  Then, as usual, $L^2(M)$ consists of (equivalence classes of) measurable functions $f:M\to \bbC$ for which the $2$-norm $\norm{f}_{L^2(M)}:=(\int\!d\mu_g\, |f|^2)^{1/2}<\infty$. This is a Hilbert space with inner product
\begin{eqnarray}
	\ip{f_1}{f_2}_{L^2(M)} :=\int\!d\mu_g \,f_1^\ast f_2 \,,\quad f_1,\,f_2\in L^2(M)\,.
\end{eqnarray}
To discuss $L^2$-spaces of (co)-vector functions, a Riemannian metric $g^+$
with inverse $g_+$ is introduced.
The space of square-integrable covector fields $L^2(M,T^*\!M)$ consists of (equivalence classes of) measurable sections  $\omega:M\to T^\ast M$ for which $(g_+^{\alpha \beta}\omega_\alpha^\ast  \omega_\beta)^{1/2}\in L^2(M)$. The definition of the space of square-integrable vector fields $L^2(M,TM)$ is analogous, and moreover the metric $g_+$ defines  a canonical isometry $\omega_\alpha:=g_{\alpha \beta}^+ v^\beta$ between $L^2(M,TM)$ and $L^2(M,T^*\!M)$. Further, as in
the scalar case, these are  Hilbert spaces with  inner products 
\begin{eqnarray}
  \ip{\omega}{\eta}_{\scriptsize{L^2(M,T^*\!M)}}
  :=\int\!d\mu_g \,g_+^{\alpha \beta}\omega_\alpha^\ast \eta_\beta\,,\quad
  \ip{v}{w}_{\scriptsize{L^2(M, TM)}}
	:=\int\!d\mu_g \,g_{\alpha\beta}^+v^{\alpha\,\ast} w^\beta\,.
\end{eqnarray}
The locally integrable versions of the above spaces consist of measurable sections that are square-integrable over every open set set $\Omega\Subset M$, and are denoted e.g. $L^2_\text{loc}(M)$ in the scalar case; moreover, clearly  $L^2(M)\subseteq L^2_\text{loc}(M)$.

Next, we recall the following characterization of a distribution $F\in \mfD'(M)$ having an   $L^2(M)$ realization. In particular, the right hand side of \eqref{apA1c} below defines a (possibly infinite) $L^2$-``norm'' of a distribution $F\in \mfD'(M)$, without the $L^2$-realization being a-priori well defined.
\begin{proposition}[Lemma 2.15 of \cite{Grigorbook}] \label{propa3} Let $F\in \mfD'(M)$. Then 
\begin{align}\label{apA1a}
  \sup_{u\in C_c^\infty(M)\atop \norm{u}_{\scriptsize{L^2(M)}=1}}
  \big|(F,u)\big|<\infty
\end{align}
iff there exists a unique $f\in L^2(M)$ such that for all $u\in C_c^\infty(M)= \mfD(M)$
\begin{align}\label{apA1b}
	(F,u)=\int\!d\mu_g(y) \,f(y)u(y)\,.
\end{align}
Moreover, 
\begin{align}\label{apA1c}
	\norm{f}_{L^2}=\sup_{u\in C_c^\infty(M)\atop \norm{u}_{L^2(M)}=1}\big|(F,u)\big|\,.
\end{align}
\end{proposition}
 \medskip
 
 Proceeding now to Sobolev spaces, every 
 $f\in L^2(M)$ defines a distribution $(\tilde{f},u):=\int\!d\mu_g\, fu$,
 and hence has a well-defined {\itshape distributional gradient}
 $\widetilde{\nabla f}\in \mfD'(M,TM)$,
\begin{eqnarray}
  (\widetilde{\nabla f}, v)=-\int\! d\mu_g\,f\nabla_\alpha v^\alpha \,,
  \quad \forall v\in \mfD(M,TM)\,.
\end{eqnarray}
The Sobolev space $H^1(M)$ consists of those elements $f\in L^2(M)$ for which the {\itshape distributional gradient} $\widetilde{\nabla f}\in \mfD'(M,TM)$
has an $L^2(M,T^*\!M)$ realization, denoted by $\nabla f$ and called the {\itshape weak derivative of} $f\in L^2(M)$. Clearly for every test vector field
$v\in \mfD(M,TM)$ one has
\begin{align}
	\int\! d\mu_g\,\nabla_\alpha f\,v^\alpha=-\int\! d\mu_g\,f\nabla_\alpha  v^\alpha \,.
\end{align}
Clearly, $C_c^\infty(M)\subseteq H^1(M)$, and moreover 
$H^1(M)$ is a Hilbert space with inner product
\begin{align}\label{sobip1}
  \ip{f_1}{f_2}_{H^1(M)}:=\ip{f_1}{f_2}_{L^2(M)}
  +\ip{\nabla f_1}{\nabla f_2}_{\scriptsize{L^2(M,T^*\!M)}}\,.
\end{align}
Finally, the main Sobolev space used in this paper, $H_0^1(M)$, is defined as the closed subspace of $H^1(M)$ obtained by taking the closure of  $C_c^\infty(M)$ in the $H^1$-norm. In particular, $H_0^1(M)$ is a Hilbert space with inner product \eqref{sobip1}, and  embeds densely into $L^2(M)$.

%\newpage
%%%%%%%%%%%%%%%%%%%
%	New appendix 
%%%%%%%%%%%%%%%%%%%

\section{Local regularity results}
\label{apploc}

In this appendix we summarize the salient aspects of the local regularity theory of the following elliptic operators with complex valued smooth coefficients, defined on a  connected open set $\Omega\subseteq \bbR^n$,
\begin{subequations}
\begin{align}\label{apploc1a}
	L(\cdot)\isa \pp_\mu (a^\munu(x)\pp_\nu(\cdot))\,,
	\\[2mm]\label{apploc1b}
	\bar{L}(\cdot)\isa b(x)\pp_\mu (a^\munu(x)\pp_\nu(\cdot))\,.
\end{align}	
\end{subequations}
Here $b\in C^\infty(\Omega,\bbR)$ and $a\in C^\infty(\Omega,M_{n\times n}(\bbC))$ satisfy:  
\begin{enumerate}[leftmargin=8mm, rightmargin=-0mm, label=(\roman*)]
  \item For all $x\in \Omega:\,b(x)>0$.
  \item For each $x\in \Omega:\,a\in M_{n\times n}(\bbC)$ is symmetric. Further, there exists a continuous map $\ell:\Omega\to \bbR_{>0}$ such that 
\begin{align}\label{apploc2}
	\Re(a^\munu(x)\xi_\mu^\ast  \xi_\nu)\geq \ell(x)\delta^\munu \xi_\mu^\ast  \xi_\nu\,,\quad \forall\,\xi \in \bbC^n\,.
\end{align}
\end{enumerate}

We shall make copious use of  Sobolev embedding:\footnote{Here, and throughout this appendix, all $L^2_{\rm loc}(\Omega)$ and Sobolev spaces $H^l_{\rm loc}(\Omega)$ are defined wrt. the Lebesgue measure on $\Omega$. We refer to \cite{Grigorbook} for detailed definitions.}

{\bfseries Sobolev embedding theorem} (Theorem 6.1 of \cite{Grigorbook}){\bfseries .} {\itshape
Let $\Omega\subseteq \bbR^n$ be open and $k,m\in \bbN_0$ such that  $k>m+\frac{n}{2}$. Then each $u\in H^k_{\rm loc}(\Omega)$ has a $C^m(\Omega)$-representative (also denoted by $u$). Moreover, for any relatively compact open sets $\Omega_0,\,\Omega_1$ such that $\Omega_0\Subset\Omega_1\Subset\Omega$ there exists a constant $C>0$ such that 
\begin{align}
\nrm{u}_{C^m(\Omega_0)}\leq C\nrm{u}_{H^k(\Omega_1)}\,,
\end{align}
and $C$ depends on $\Omega_0,\Omega_1,k,m,n$.}
\medskip

The first  result of this appendix is that when $u,\,\bar{L}u,\ldots,\,\bar{L}^ku\in L^2_\text{loc}(\Omega)$ for sufficiently high $k\in \bbN$, $u$ is classically differentiable. This is the content of 
 Theorem \ref{apploc_th1}, and its Corollary \ref{apploc_cor1}.
\begin{theorem}\ \label{apploc_th1}
  Let $\Omega\subseteq \bbR^n$ be open, and assume that for some
  $k\in \bbN$ we have
\begin{align}\label{apploc2b}
	u,\, \bar{L}u,\ldots,\,\bar{L}^ku\in L^2_{\rm loc}(\Omega)\,, 
\end{align}
where the operator $\bar{L}$ is defined in \eqref{apploc1b}. Then $u\in H^{2k}_{\rm loc}(\Omega)$, and for any open sets $\Omega_0\Subset \Omega_1\Subset\Omega$, there exists $M_k>0$ (depending on $\bar{L},\,\Omega_0,\,\Omega_1$) such that 
\begin{equation}
  \label{apploc11a}
\nrm{u}_{H^{2k}(\Omega_0)}\leq M_k\ssum{j=0}{k}\nrm{\bar{L}^ju}_{L^2(\Omega_1)}\,.
\end{equation}

\end{theorem}
Applying the Sobolev embedding theorem then yields
\begin{thmcorollary}\label{apploc_cor1}
	Given the hypotheses of Theorem \ref{apploc_th1}, if there exists $m\in \bbN_0$ such that $k>m/2+n/4$, then $u\in C^m(\Omega)$ (or more precisely $u\in L^2_{\rm loc}(\Omega)$ has a $C^m$-representative). Moreover, for open sets $\Omega_0\Subset \Omega_1\Subset \Omega$ there is a $B>0$ such that 
\begin{equation}
	\nrm{u}_{C^m(\Omega_0)}\leq B\ssum{j=0}{k}\nrm{\bar{L}^ju}_{L^2(\Omega_1)}\,,
\end{equation}
with $\nrm{u}_{C^m(\Omega_0)}:=\underset{|\alpha|\leq m}{\sup}\underset{\Omega_0}{\sup}\,|\pp^\alpha u|$.
\end{thmcorollary}
The second main result of this appendix concerns analytic maps  from the open ball $B_S(0):=\set{\zeta\in \bbC\given |\zeta|<S}$ to $L^2_{\rm loc}(\Omega)$ defined by  power series of the form
\begin{align}
	F(s):=\ssum{j=0}{\infty}\frac{s^j}{j!}f_j,
\end{align}
and  joint regularity in $(s,x)\in B_S(0)\times \Omega$. Indeed, since for each $s$, $F(s)\in L^2_{\rm loc}(\Omega)$ is determined only up to sets of Lebesgue measure zero, joint regularity is not a well-defined notion for such maps in general. However, with the additional hypotheses that under the action of $\bar{L}$, the $F(s),\,f_j$ satisfy local integrability conditions of the form  \eqref{apploc2b} for sufficiently large $k\in \bbN$,  {\itshape classical joint differentiability}  is enforced.  
\begin{theorem}
	\ \label{apploc_th2}
Let there exist a sequence $(f_j)_{j\in \bbN_0}$ in $L^2_{\rm loc}(\Omega)$ and  $k\in \bbN$ so that
\begin{align}\label{apploc19}
\forall\,j\in \bbN_0:\,\,	f_j,\,\bar{L}f_j,\ldots,\,\bar{L}^kf_j\in L^2_{\rm loc}(\Omega)\,.
\end{align}
Suppose there exists an analytic function $F:B_S(0)\to L^2_\text{loc}(\Omega)$, with the following properties for each $\zeta\in B_S(0)$.
\begin{enumerate}[leftmargin=8mm, rightmargin=-0mm, label=(\roman*)]
\item The power series $F(\zeta)=\ssum{j=0}{\infty}\frac{\zeta^j}{j!}f_j$
converges in $L^2_\text{loc}(\Omega)$.
\item $F(\zeta),\,\bar{L}F(\zeta),\,\ldots,\,\bar{L}^kF(\zeta)\in L^2_\text{loc}(\Omega)$.
\item For each $l\in \set{1,\ldots,k}$
%\begin{align}\label{aps2}
$\bar{L}^lF(\zeta)=\ssum{j=0}{\infty}\frac{\zeta^j}{j!}\bar{L}^lf_j$, 
%\end{align}
with the sum converging in $L^2_\text{loc}(\Omega)$.
 
\end{enumerate}
If there is $m\in \bbN_0$ such that $k>m/2+n/4$,  then for each $\zeta\in B_S(0)$ there exists a $C^m(\Omega)$ representative of $F(\zeta)$, denoted by $F(\zeta,\cdot)$, and the mapping 
\begin{align}\label{aps3}
	B_S(0)\times \Omega\ni (\zeta,x)\mapsto F(\zeta,x)\in \bbC
\end{align}
is jointly $C^m$.
\end{theorem}

The technical input required for the proofs of these results are the
following lemmas.
\begin{lemma}\label{apploc_lm2}
Let $\Omega\subseteq \bbR^n$ be an open set. Then for any $k\in \bbN_0,$ if $u\in H^k(\Omega)$ with compact support in $\Omega$,\footnote{$u\in H^k(\Omega)$ compactly supported in $\Omega$ means there exists $K\subseteq \Omega$ such that $K$ is compact and $u=0$ a.e. on $\Omega\setminus K$. It follows that all weak derivatives up to order $k$ are compactly supported and are zero a.e. on $\Omega\setminus K$.} and $Lu \in H^{k-1}(\Omega)$, then
\begin{align}\label{apploc7}
	L(u\ast \varphi_\varepsilon)\xrightarrow{H^{k-1}}Lu\,.
\end{align}
Here $u\ast \varphi_\varepsilon$ is the convolution with the mollifier $\varphi_\varepsilon$.\footnote{Mollifiers are defined by first specifying a non-negative $\varphi\in C_c^\infty(\bbR^n)$ supported in the unit ball  $B_1(0)\subseteq \bbR^n$ and $\int_{\bbR^n}d^nx\,\varphi =1$. Then for any $\varepsilon\in (0,1)$ one sets $\varphi_\varepsilon(x):=\varepsilon^{-n}\varphi(x/\varepsilon)$, which is supported in $B_\varepsilon(0)$.}

Note: since $u\in H^k(\Omega)$ is compactly supported in $\Omega$, it may be extended to an element of $H^k(\bbR^n)$ by setting it to zero outside $\Omega$, thereby rendering the convolution  $u\ast\varphi_\varepsilon$ well defined.	
\end{lemma}

\begin{lemma}\label{apploc_lm1}
	Let $\Omega,\,\Omega'\subseteq \bbR^n$ be open and $\Omega'\Subset\Omega$. Then for every $k\in \bbN:\,\exists\,C_k>0$ (depending on $L$ and $\Omega'$) such that 
\begin{align}\label{apploc2a}
	\nrm{u}_{H^k(\Omega')}\leq C_k\nrm{Lu}_{H^{k-2}(\Omega')}\,,\quad u\in C_c^\infty(\Omega')\,.
\end{align}
\end{lemma}

These results are (respectively) generalizations to all $k\in \bbN$ of Lemmas 6.6 and 6.7 of \cite{Grigorbook}, suitably adapted to the complex coefficient operator \eqref{apploc1a}. The proofs are by induction, with the base case following the same lines as in \cite{Grigorbook}, and the inductive step straightforward; we omit the details.

\bigskip

We now proceed to the proof of Theorems \ref{apploc_th1} and \ref{apploc_th2}.

{\itshape Proof of Theorem \ref{apploc_th1}.} This is by induction on $k\in \bbN$. 

{\itshape Base case $k=1$}: Assume that $u,\,\bar{L}u\in L^2_\text{loc}(\Omega)$, which implies that $Lu=b\inv \bar{L}u\in L^2_\text{loc}(\Omega)$. Fixing arbitrary open sets $\Omega_1\Subset \Omega_2\Subset \Om$, there is a further open set $\Om'$ such that $\Omega_1\Subset \Omega'\Subset \Om_2$. Next, let $\psi\in C_c^\infty(\Omega')$ be a cutoff function for $\Omega_0$, i.e. $\psi:\Om'\to [0,1]$ and  $\psi|_{\Omega_0}\equiv 1$, and define $v:=\psi u$. Then clearly $v\in L^2(\Omega)$, with compact support contained in $\Omega'$. 

Next, consider
\begin{align}\label{apploc12}
	Lv=\psi Lu + L\psi u+2a^\munu \pp_\mu \psi \pp_\nu u.
\end{align}
Since both $u,\,Lu\in L^2_\text{loc}(\Omega)$, the first two terms are in $L^2_\text{loc}(\Omega)$ as well. The final term, with a single derivative acting on $u$ is {\itshape a priori} only in $H^{-1}_\text{loc}(\Omega)$. Thus, $Lv\in H\inv _\text{loc}(\Omega)$.

Since $v$ is compactly supported in $\Omega'$, there is $\varepsilon_0>0$ such that for all $0<\varepsilon<\varepsilon_0$ we have under mollification $v\ast \varphi_\varepsilon \in C_c^\infty(\Omega')$. Thus, Lemma \ref{apploc_lm1} implies that there is $C_1>0$ such that
\begin{align}
	\nrm{v\ast \varphi_\varepsilon}_{H^1(\Omega')}\leq C_1\nrm{L(v\ast \varphi_\varepsilon)}_{H\inv(\Omega')}\,,\quad \forall\,\varepsilon\in (0,\varepsilon_0)\,.
\end{align}
Furthermore, $L(v\ast \varphi_\varepsilon)\xrightarrow{H\inv}Lv$ by Lemma \ref{apploc_lm2}, so 
\begin{align}\label{apploc13}
	\limsup_{\varepsilon\to 0^+}\nrm{v\ast \varphi_\varepsilon}_{H^1(\Omega')}\leq C_1\nrm{Lv}_{H\inv(\Omega')}\,,
\end{align}
and hence $\liminf_{\varepsilon\to 0^+}\nrm{v\ast \varphi_\varepsilon}_{H^1(\Omega')}<\infty$. This 
 implies $v\in H^1(\Omega')$ with $\nrm{v}_{H^1(\Omega')}\leq C_1\nrm{Lv}_{H\inv(\Omega')}$ (c.f. \cite{Grigorbook} Theorem 2.13). Since $u\equiv v$ on $\Omega_0\Subset \Omega' $, together with the expression for $Lv$ in \eqref{apploc12}, it follows that $u\in H^1(\Omega_0)$ and 
\begin{align}\label{apploc14}
	\nrm{u}_{H^1(\Omega_0)}\leq C_1'\big(\nrm{u}_{L^2(\Omega')}+\nrm{Lu}_{L^2(\Omega')}\big)\,,
\end{align}
with $C_1'>0$ independent of $u$.
Moreover, since $\Omega_1\Subset \Omega'\Subset\Omega$ were arbitrary, $u\in H^1_\text{loc}(\Omega)$. 

To go from $H^1$ to $H^2$, note that (the now established) $u\in H^1_\text{loc}(\Omega)\implies Lv\in L^2_\text{loc}(\Omega)$ above. Then Lemmas \ref{apploc_lm1} and \ref{apploc_lm2} in combination yield
\begin{align}
	\limsup_{\varepsilon\to 0^+}\nrm{v\ast \varphi_\varepsilon}_{H^2(\Omega')}\leq C_2\nrm{Lv}_{L^2(\Omega')}\,,
\end{align}
and hence $v\in H^2(\Omega')$ with the bound $\nrm{v}_{H^2(\Omega')}\leq C_2 \nrm{Lv}_{L^2(\Omega')}$. Then
\begin{align}
	\nrm{u}_{H^2(\Omega_0)}&\leq C_2 \nrm{Lv}_{L^2(\Omega')}\leq C_2'\Big(\nrm{u}_{L^2(\Omega')}+\nrm{Lu}_{L^2(\Omega')}+\nrm{u}_{H^1(\Omega')}\Big)
	\nonum 
	&\leq C_2''\Big(\nrm{u}_{L^2(\Omega')}+\nrm{Lu}_{L^2(\Omega')}\Big)
\end{align}
with the final inequality from the bound \eqref{apploc14} applied to $\nrm{u}_{H^1(\Omega')}$. Thus we conclude that $u,\,\bar{L}u\in L^2_{\rm loc}(\Omega)$ implies $u\in H^2_{\rm loc}(\Omega)$. Moreover, since   $\bar{L}=bL$, there is $M_1>0$ (depending on $\bar{L}$ and the open sets $\Om_0,\,\Om_1$, but independent of $u$) such that 
\begin{align}\label{apploc15}
\nrm{u}_{H^{2}(\Omega_0)}\leq M_k\ssum{j=0}{1}\nrm{\bar{L}^ju}_{L^2(\Omega_1)}\,,
\end{align}
establishing the base case.

\medskip

{\itshape The inductive step  $k$ to $k+1$}: Fixing an arbitrary $k\geq 1$, assume that the result holds for $k$. Next, assume that 
\begin{align}
	u,\,\bar{L}u,\ldots,\bar{L}^ku,\,\bar{L}^{k+1}u\in L^2_\text{loc}(\Omega)\,.
\end{align}
Then $u,\,\bar{L}u,\ldots,\bar{L}^ku\in L^2_\text{loc}(\Omega)$ and $\bar{L}u,\ldots,\bar{L}^k(\bar{L}u)\in L^2_\text{loc}(\Omega)$, which together with the inductive hypothesis implies $u,\,\bar{L}u\in H^{2k}_\text{loc}(\Omega)$ along with the requisite norm bounds (c.f. \eqref{apploc11a}).

It then follows that, upon fixing an arbitrary multi-index $\alpha$ with $|\alpha|=2k$, we have $\pp^\alpha u\in L^2_\text{loc}(\Omega)$ and using $\bar{L}=bL$,\footnote{Clearly the first term is in $L^2_{\rm loc}(\Omega)$, and since $|\beta_2|\leq 2k-1$, $u\in H^{2k}_{\rm loc}(\Omega)$, the second term is an element of $H^{-1}_\text{loc}(\Omega)$.} 
\begin{align}\label{apploc16}
	L(\pp^\alpha u)=b\inv \pp^\alpha (\bar{L}u)-\sum_{\beta_1+\beta_2=\alpha\atop |\beta_1|>0}b\inv \pp^{\beta_1}b\,\pp^{\beta_2}Lu\in H^{-1}_\text{loc}(\Omega)\,.
\end{align}
Applying the argument in the proof of the base case above to $\pp^\alpha u$, one concludes that $\pp^\alpha u\in H^2_{\rm loc}(\Omega)$. Moreover, since $\alpha$ is an arbitrary multi-index with $|\alpha|=2k$, it follows that $u\in H^{2k+2}_\text{loc}(\Omega)$.

Turning now to the bound, having chosen open sets $\Omega_0\Subset\Omega_1\Subset\Omega$, choose two further open sets $\Omega',\,\Omega''$ satisfying $\Omega_0\Subset\Omega'\Subset\Omega''\Subset \Omega_1$. The $k=1$ bound \eqref{apploc15} applied to $\pp^\alpha u$ on  $\Omega_0\Subset\Omega'\Subset \Omega$ is
\begin{align}\label{apploc16a}
	\nrm{\pp^\alpha u}_{H^2(\Omega_0)}\leq M_1\Big(\nrm{\pp^\alpha u}_{L^2(\Omega')}+\nrm{\bar{L}(\pp^\alpha u)}_{L^2(\Omega')}\Big)\,.
\end{align}
Next, it follows from \eqref{apploc16} that 
\begin{align}\label{apploc17}
	\nrm{\bar{L}(\pp^\alpha u)}_{L^2(\Omega')}&\leq C_1'\Big(\nrm{\bar{L}u}_{H^{2k}(\Omega')}+\sum_{|\beta|\leq 2k} \nrm{\pp^\beta u}_{H^1(\Omega')}\Big)
	\nonum 
	&\leq C_1''\Big(\nrm{u}_{H^{2k}(\Omega'')}+\nrm{\bar{L}u}_{H^{2k}(\Omega'')}\Big)\,,
\end{align}
where to arrive at the second inequality one applies the $H^1$-bound \eqref{apploc14} to $\pp^\beta u$ on $\Omega'\Subset\Omega''\Subset \Omega$. 

Finally, applying the inductive hypothesis to $u,\,\bar{L}u\in H^{2k}_{\rm loc}(\Omega)$ on $\Omega''\Subset\Omega_1\Subset \Omega$ yields
\begin{align}\label{apploc18}
	\nrm{u}_{H^{2k}(\Omega'')}\leq \tilde{C}\ssum{j=0}{k}\nrm{\bar{L}^ju}_{L^2(\Omega_1)}\,,\quad \nrm{Lu}_{H^{2k}(\Omega'')}\leq \tilde{C}\ssum{j=0}{k}\nrm{\bar{L}^{j+1}u}_{L^2(\Omega_1)}
\end{align}
with  $\tilde{C}>0$ independent of $u$. This, in combination with \eqref{apploc16a} and  \eqref{apploc17} implies the existence of $M_{k+1}>0$ such that 
\begin{equation}
\nrm{u}_{H^{2k+2}(\Omega_0)}\leq M_{k+1}\ssum{j=0}{k+1}\nrm{\bar{L}^ju}_{L^2(\Omega_1)}\,,
\end{equation}
completing the inductive step, and hence the proof of the theorem.
\qed
\bigskip

{\itshape Proof of Theorem \ref{apploc_th2}.} Assume that there exists
$m\in \bbN$ such that $k>m/2+n/4$. It  follows straightforwardly from
Corollary \ref{apploc_cor1} that for each $j\in \bbN_0$ and $\zeta\in B_S(0)$, $f_j,\,F(s)$ have $C^m(\Omega)$-representatives, to be denoted $f_j(\cdot)$ and $F(\zeta,\cdot)$ respectively. Moreover, it follows from the convergence
of the series in Theorem \ref{apploc_th2} that for each $\zeta\in B_S(0)$ the power series 
\begin{equation}
  \label{aps4}
	F(\zeta,\cdot)=\ssum{j=0}{\infty}\frac{\zeta^j}{j!}f_j(\cdot)
\end{equation}
converges in $C^m(\Omega_0)$ for every open set $\Omega_0\Subset \Omega$.\footnote{Recall that the $\nrm{\cdot}_{C^m(\Omega_0)}$-norm is defined by $\nrm{h}_{C^m(\Omega_0)}:=\sup_{\alpha\leq m}\sup_{\Omega_0}|\pp^\alpha h|$.} The uniformity of this convergence for each $\zeta\in B_S(0)$, together with the basic properties of the convergence of power series, implies that $F(\cdot,\cdot)$ is jointly $C^m$ in $(\zeta,x)$. The detailed argument is by induction.

{\itshape Base case (jointly $C^1$)}: Fix an arbitrary open set $\Omega_0\Subset \Omega$ and let  $\zeta\in B_S(0)$ and $\delta \zeta\in B_r(\zeta)\subseteq B_S(0)$. Since the $C^m(\Omega_0)$ convergence of the power series \eqref{aps4} holds for any $\zeta\in B_S(0)$, the sum $\sum_{j\geq 0}\frac{|\zeta+\delta \zeta|^j}{j!}\nrm{f_j}_{C^1(\Omega_0)}$ is uniformly bounded from above for $\delta \zeta\in B_r(\zeta)$. Then 
\begin{align}
\lim_{\delta \zeta\to 0}F(\zeta+\delta \zeta,\cdot)=\lim_{\delta \zeta\to 0} \ssum{j=0}{\infty}\frac{(\zeta+\delta \zeta)^j}{j!}f_j(\cdot)=F(\zeta,\cdot)\,,
\end{align}
with the limit taken with respect to $\nrm{\cdot}_{C^m(\Omega_0)}$, and the above uniform upper bound allows the limit to be taken inside the sum. This continuity in $\zeta$, uniformly over $x\in \Omega$, implies joint continuity in $\zeta\textit{ and } x$.

Having established joint continuity, consider the following series bounding the difference quotient,
\begin{align}
	\ssum{j=1}{\infty}\frac{1}{j!}\bigg|\frac{(\zeta+\delta \zeta)^j-\zeta^j}{\delta \zeta}\bigg|\,\nrm{f}_{C^m(\Omega_0)}\leq \ssum{j=1}{\infty} \frac{r^{j-1}}{(j-1)!}\nrm{f}_{C^m(\Omega_0)}<\infty
\end{align}
with finiteness following from the $C^m(\Omega_0)$  convergence of \eqref{aps4} over $B_S(0)$. Analogous to the argument for continuity, this implies that the difference quotient power series converges to $\pp_\zeta F(\zeta,\cdot)$  in $C^m(\Omega_0)$ as $\delta \zeta\to 0$. In particular, it yields the power series
\begin{align}\label{aps5}
	\pp_\zeta F(\zeta,\cdot)=\ssum{j=0}{\infty}\frac{\zeta^j}{j!}f_{j+1}(\cdot)\,,
\end{align}
converging in $C^m(\Omega_0)$ for $\zeta\in B_S(0)$. Applying the above continuity arguments to this power series implies that $F(\cdot,\cdot)$ is jointly $C^1$, and for indices $i+|\alpha|=1$,
\begin{align}\label{aps6}
  \pp_\zeta^i\pp_x^\alpha F(\zeta,\cdot)=\ssum{j=0}{\infty}
  \frac{\zeta^j}{j!}\pp_x^\alpha f_{j+1}(\cdot)\,,\quad \zeta\in B_S(0)\,,
\end{align}
converging in $C^{m-|\alpha|}(\Omega_0)$.
\medskip

{\itshape The inductive step}: For $1\leq l<m$, assume that
$F(\cdot,\cdot)$ is jointly $C^l$ on $B_S(0)\times \Omega$, and that
for indices $i+|\alpha|\leq l$ the series (\ref{aps6}) converges  
in $C^{m-|\alpha|}(\Omega_0)$ for any open set $\Omega_0\Subset \Omega$.
To proceed, fix such an open set $\Omega_0$. It is to be shown that for indices $i+|\alpha|= l+1$, the power series expansion in \eqref{aps6} converges in $C^{m-|\alpha|}(\Omega_0)$. When $i\leq l$, this holds as a direct consequence of the induction hypothesis. Thus, {\rm w.l.o.g.} assume that $i=l+1$, and consider 
\begin{align}\label{aps7}
	\pp_\zeta^{l}F(\zeta,\cdot)=\ssum{j=0}{\infty}\frac{\zeta^j}{j!} f_{j+l}(\cdot)\,.
\end{align}
By the inductive hypothesis, this converges in $C^m(\Omega_0)$ for each $\zeta\in B_S(0)$. Applying the arguments of the base case above to \eqref{aps7} then implies  $\pp_\zeta^{l}F(\cdot,\cdot)$ is jointly $C^1$, and in particular
\begin{equation}
	\pp_\zeta^{l+1}F(\zeta,\cdot)=\ssum{j=0}{\infty}\frac{\zeta^j}{j!} f_{j+l+1}(\cdot)\,,\quad \zeta\in B_S(0)\,,
\end{equation}
 converging in $C^m(\Omega_0)$ as desired. Thus, $F(\cdot,\cdot)$ is jointly $C^{l+1}$ in $(\zeta,x)\in B_S(0)\times \Omega$, completing the inductive step, and thereby the proof.
 \qed

% ----------------------------------------------------------------------
%		Bibliography
% ----------------------------------------------------------------------

\end{document}